\documentclass{elsart}

\newcommand{\be}{\begin{equation}}
\newcommand{\ee}{\end{equation}}

\usepackage{natbib}
\usepackage{graphicx}

\usepackage{amssymb}

\begin{document}

\begin{frontmatter}



\title{Pulsar wind zone processes in LS 5039}


\author[1]{Agnieszka Sierpowska-Bartosik \&} 
\author[1,2]{Diego F. Torres}

\address[1]{Institut de Ciencies de l'Espai (IEEC-CSIC)
  Campus UAB, Fac. de Ciencies, Torre C5, parell, 2a planta, 08193
  Barcelona,  Spain. E-mail: agni@ieec.uab.es}
  \address[2] {Instituci\'o Catalana de Recerca i Estudis Avan\c{c}ats (ICREA), Spain. E-mail: dtorres@ieec.uab.es }

\begin{abstract}

Several $\gamma$-ray binaries have been recently detected by the High-Energy Stereoscopy Array (H.E.S.S.) and the Major Atmospheric Imaging Cerenkov (MAGIC) 
telescope. In at least two cases, their nature is unknown.
In this paper we aim to provide the details of a theoretical model of close $\gamma$-ray binaries containing a young energetic pulsar as compact object, earlier presented in recent Letters.
This model includes a detailed account of the system geometry, the angular dependence of processes such as Klein-Nishina inverse Compton and $\gamma\gamma$ absorption in the anisotropic radiation field of the massive star, and a Monte Carlo simulation of leptonic cascading. We present and derive the used formulae and give all details about their numerical implementation, particularly, on the computation of cascades.  In this model, emphasis is put in the processes occurring in the pulsar wind zone of the binary, since, as we show, opacities in this region can be already important for close systems.
We provide a detailed study on all relevant opacities and geometrical dependencies along the orbit of binaries, exemplifying with the case of LS 5039. This is used to understand the formation of the very high-energy lightcurve and phase dependent spectrum. For the particular case of LS 5039, we uncover an interesting behavior of the magnitude representing the shock position in the direction to  the observer along the orbit, and analyze its impact in the predictions.
We show that in the case of LS 5039, the H.E.S.S. phenomenology is matched by the presented model, and explore the reasons why this happens while discussing future ways of testing the model.

\end{abstract}


\end{frontmatter}


\section{Introduction}

Very recently, a few massive binaries have been identified as variable very-high-energy (VHE) $\gamma$-ray sources. They are
PSR B1259-63 (Aharonian et al. 2005a), LS 5039 (Aharonian et al. 2005b, 2006), LS I +61 303 
(Albert et al. 2006, 2008a,b), and Cyg X-1 (Albert et al. 2007). 
The nature of only two of these binaries is considered known: PSR B 1259-63 is formed with a pulsar whereas Cyg X-1 is formed with a black hole compact object. The nature of the two remaining systems is under discussion. 
The high-energy phenomenology of 
Cyg X-1 is different from that of the others. It has been detected just once in a flare state for which a duty cycle is yet unknown. The three other sources, instead, present a behavior that is fully correlated with the orbital period. The latter varies from about 4 days in the case of LS 5039 to several years in the case of PSR B1259-63: this span of orbital periodicities introduces its own complications in analyzing the similarities among the three  systems. 


LS I +61 303 shares with LS 5039 the quality of being the only two known microquasars/$\gamma$-ray binaries that are spatially coincident with sources above 100 MeV listed in the 
Third Energetic Gamma-Ray Experiment (EGRET) catalog (Hartman et al. 1999). 
These sources both show low X-ray emission and variability, and no signs of emission lines or disk accretion. For LS I +61 303, extended, apparently precessing, radio emitting structures at angular extensions of 0.01-0.05 arcsec have been reported by Massi et al. (2001, 2004); this discovery 
has earlier supported its microquasar interpretation.  
But the uncertainty as to what kind of compact object, a black hole or a neutron star, is 
part of the system (e.g., Casares et al. 2005a), seems settled for many after the results presented by Dhawan et al. (2006). These authors have presented 
observations from a July 2006 VLBI campaign in which rapid 
changes are seen in the orientation of what seems to be a cometary tail at periastron. This tail is 
consistent with it being the result of a pulsar wind. 
Indeed, no large features or high-velocity flows were noted on any of 
the observing days, which implies at least its non-permanent nature. The changes within 3 hours were found to be insignificant, so the velocity 
can not be much over 0.05$c$. 
Still, discussion is on-going (e.g. see Romero et al. 2007, Zdziarski et al. 2008). 
New campaigns with similar radio resolution, as well as new
observations in the $\gamma$-ray domain have been obtained since the Dhawan's et al. original results (Albert et al. 2008b). A key aspect  in these high-angular-resolution campaigns is the observed maintenance in time of the morphology of the radio emission of the system: the changing morphology of the radio emission along the
orbit would require a highly unstable jet, which details are not expected to be reproduced orbit after
orbit as indicated by current results (Albert et al. 2008b). The absence of accretion signatures in X-rays in Chandra and XMM-Newton observations (as reported by Sidoli et al. 2006,
Chernyakova et al. 2006, and Paredes et al. 2007) is another relevant aspect of the discussion about the compact object companion.
Finally, it is interesting to note that neutrino detection or non-detection with ICECUBE will shed light on the nature of the $\gamma$-ray emission irrespective of the system composition (e.g.,  Aharonian et al. 2006b, Torres and Halzen 2007).


For LS 5039, a periodicity in the $\gamma$-ray flux, consistent with the orbital timescale as determined by Casares et al. (2005b), was found with amazing precision (Aharonian et al. 2006). Short timescale variability displayed  on top of this periodic behavior, both in flux and spectrum, was also reported. It was found that the parameters of power-law fits to the $\gamma$-ray data obtained in 0.1 phase binning already displayed significant variability. Current H.E.S.S. observations of LS 5039  ($\sim 70$ hours distributed over many orbital cycles, Aharonian et al. 2006)
constitute one of the most detailed datasets of 
high-energy astrophysics. 
Similarly to LS I +61 303, the discovery of a jet-like radio structure in LS 5039 and the fact of it being the only
radio/X-ray source co-localized with a mildly variable (Torres et al. 2001a,b) EGRET detection, prompted a microquasar interpretation (advanced already by Paredes et al. 2000). 
However, the current mentioned findings at radio and VHE $\gamma$-rays in the cases of LS I +61 303 (Dhawan et al. 2006, Albert et al. 2006, 2008b) or PSR B1259-63 (Aharonian et al. 2005), gave the perspective that all three systems are different realizations of the same scenario: a pulsar-massive star binary. Dubus (2006a,b) has studied these similarities. He provided simulations of the extended radio emission of LS 5039 showing that the features found in high resolution radio observations could also be interpreted as the result of a pulsar wind. Recently, Rib\'o et al. (2008) provided VLBA radio observations of LS 5039 with morphological and astrometric information at milliarcsecond scales. They showed that a microquasar scenario cannot easily explain the observed changes in morphology. All these results, together with  the assessment of the low X-ray state (Martocchia et al. 2005) made the pulsar hypothesis tenable, and the possibility of explaining the H.E.S.S. phenomenology in such a case, an interesting working hypothesis. 
 

High energy emission from pulsar binaries has been subject of study for a long time (just to quote a non-exhaustive list of references note the works of Maraschi and Treves 1981; Protheroe and Stanev 1987, Arons and Tavani 1993, 1994;  Moskalenko et al. 1993; Bednarek 1997, Kirk et al. 1999, Ball and Kirk 2000, Romero et al. 2001,  Anchordoqui et al. 2003, and others already cited above).
LS 5039 has been recently subject of intense theoretical studies (e.g., Bednarek 2006, 2007; which we comment on in more detail below, Bosch-Ramon et al. 2005; B\"ottcher 2007; B\"ottcher and Dermer 2005; Dermer and B\"ottcher 2006; Dubus 2006a,b; Paredes et al. 2006; Khangulyan et al. 2007; Dubus et al. 2007).

In the penultimate paper mentioned in the list above, Khangulyan et al. (2007), and contrary to the assumption here, authors assumed a jet structure   perpendicular to the orbital plane of the system. The energy spectrum and lightcurves were computed, accounting for the acceleration efficiency, the location 
of the accelerator along the jet, the speed of the emitting flow, the inclination angle of the system, 
as well as specific features related to anisotropic inverse Compton (IC) scattering and 
pair production. Different magnetic fields, affecting Synchrotron emission, and the losses they produced, were also tested given a large model parameter space. Authors found a good agreement between H.E.S.S. data for some of  their models. 

In the last of these papers, Dubus et al. (2007) computed the phase dependent lightcurve and spectra expected from inverse Compton interactions from electrons injected close to the compact object, assumed as a likely rotation-powered pulsar. 
Since the angle at which an observer sees the binary and propagating electrons 
changes with the orbit (see below), a phase dependence of the spectrum is 
expected, and anisotropic inverse Compton is needed to compute it. 
In general, they found that the  lightcurve is a 
good fit to the observations, except at the phases of maximum attenuation where pair cascade emission plays a role.  Dubus et al. (2007) do not consider cascading in their models, as we do here. Without cascading, zero flux is expected at a broad phase around periastron, which is not found. This lack of cascading in their model also affects the spectra, which are not reproduced well, particularly at the superior conjunction broad phases of the orbit. They mentioned that both, cascading and/or a change in the slope of the power-law injection for the interacting electron distribution  could be needed to explain the spectrum in these phases, what we explore in detail in this work. 

In order to compute inverse Compton emission from LS 5039, we use, as in previous works, leptons interacting with the star photon field. Geometry is described there with different levels of detail, what influence the results. In general, cascading processes were not taken into account, and the goodness of fitting the H.E.S.S. data is arguable in most cases, both for the lightcurve and spectrum.  

In none of the papers mentioned above, the theoretical predictions for the short timescale spectral variability found by H.E.S.S. in 0.1 phase binning was shown and compared with data.  We discuss these results from our model below.


In recent Letters (Sierpowska-Bartosik and Torres 2007, 2008) under the assumption that LS 5039 is composed by a pulsar rotating around an O6.5V star in the $\sim 3.9$ day orbit,  we
presented the results of a leptonic (for a generic hadronic model see Romero et al. 2003) theoretical modeling for the high-energy phenomenology observed by  H.E.S.S. These works studied the lightcurve, the spectral orbital variability in both broad orbital phases and in shorter (0.1 phase binning) timescales and have found  
a complete agreement between H.E.S.S. observations and our predictions. We have also analyzed how this model could be tested by Gamma-ray Large Area Space Telescope (GLAST), and how much time would be needed for this satellite in order to rule the model out in case theory significantly departs from reality. But many details of implementation which are not only useful for the case of LS 5039 but for all others close massive $\gamma$-ray binaries, as well as many interesting results concerning the binary geometry, wind termination,  opacities to different processes along the orbit of the system, and further testing at the highest energy $\gamma$-ray domain were left without discussion in our previous works. Here, we provide these details,  together with benchmark cases that are useful to understand the  formation of the very high-energy lightcurve and phase dependent spectra.


The rest of this paper is organized as follows. Next Section introduces the model concept and its main properties. It provides a discussion of geometry, wind termination, and opacities along the orbit of the system (we focus on LS 5039). 
An accompanying Appendix provides mathematical derivations of the formulae used and useful intermediate results that are key for the model, but too cumbersome to include them as part of the main text. It also deals with numerical implementation, and describes in detail the Monte Carlo simulation of the cascading processes. 
The results follow: Section 3 deals with a mono-energetic interacting particle population, and Section 4, with power-law primary distributions. Comparison with H.E.S.S. results is made in these Sections and details about additional tests are given. Final concluding remarks are provided at the end.

\section{Description of the model and its implementation}

\begin{figure*}
   \includegraphics[width=0.45\textwidth,angle=0,clip]{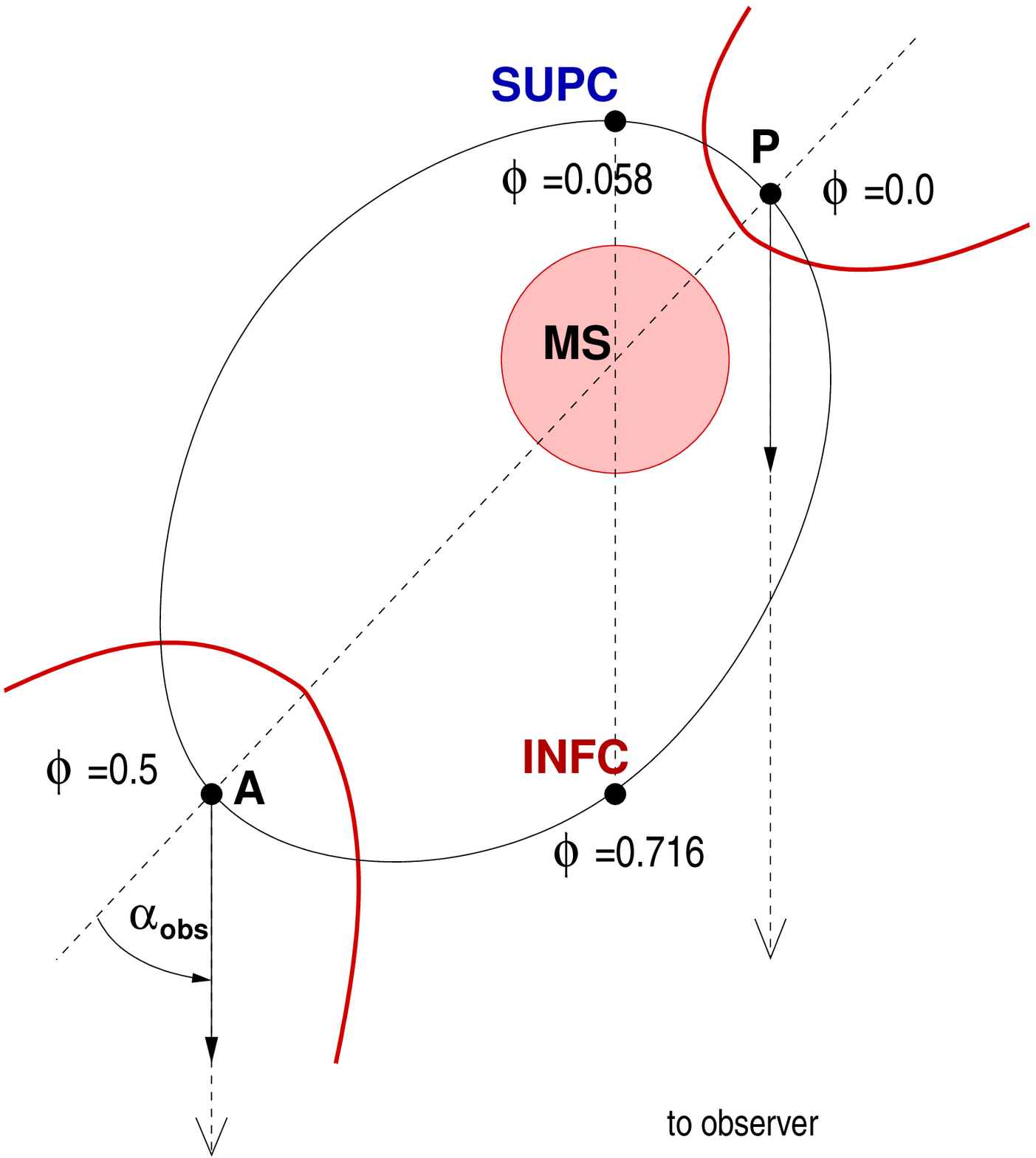}
   \includegraphics[width=0.45\textwidth,angle=0,clip]{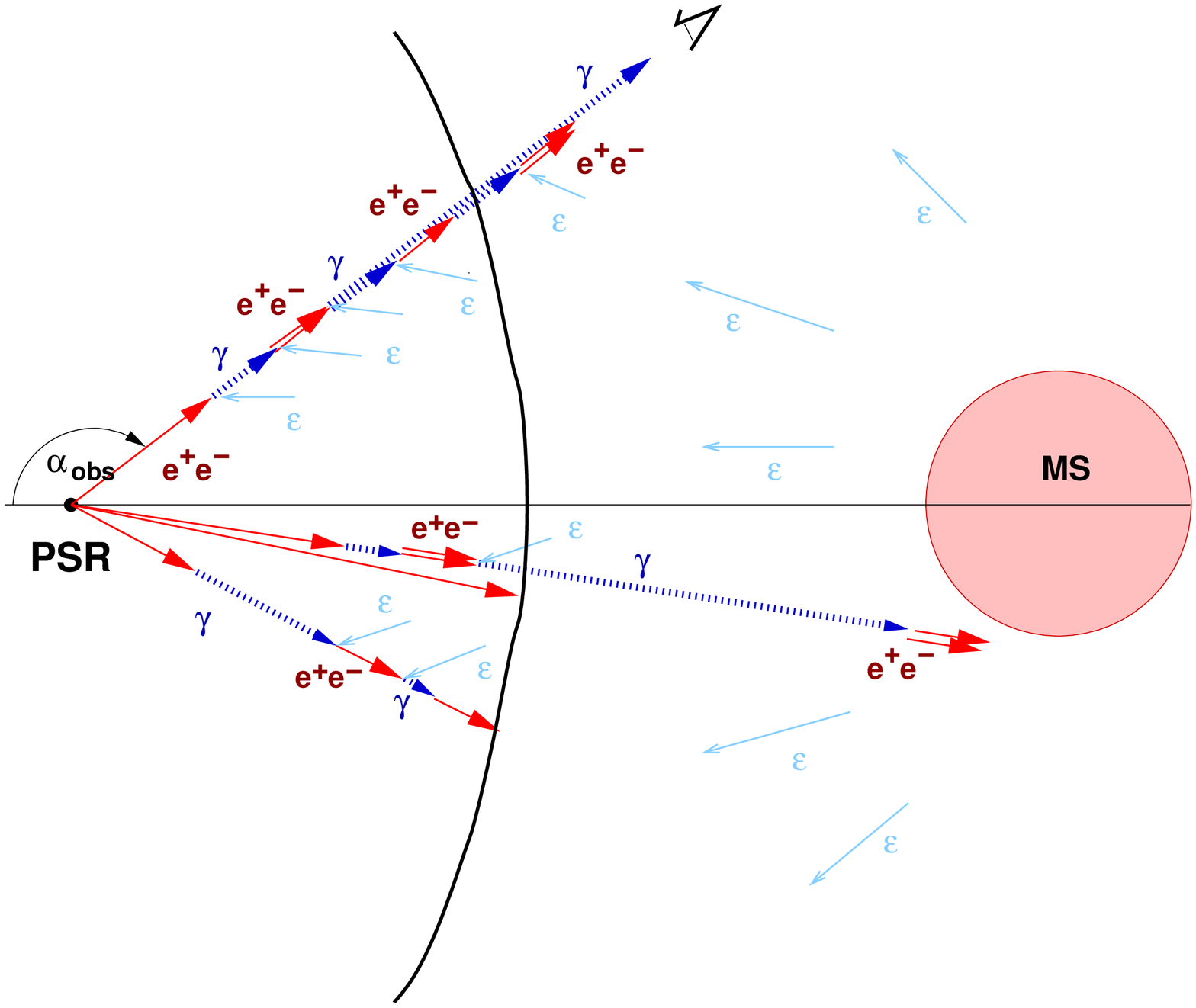}
\caption{\label{fig:general} Left: Sketch of a close binary, such as the LS 5039 system. P stands for the orbital position of periastron, A for apastron, INFC  for inferior conjunction, and SUPC  for superior conjunction. The orbital phase, $\phi$, and the angle to the observer, $\alpha_{obs}$, are marked. The orbital plane is inclined with respect to the direction of the observer (this angle is not marked). The termination shock created in the interaction of the pulsar and the massive star winds is also marked for two opposite phases --periastron and apastron-- together with the direction to the observer for both phases. Right: The physical scenario for high-energy photon production in the PWZ terminated by the shock. $e^+e^-$ are injected by the pulsar or by a close-to-the-pulsar shock and travel towards the observer, producing Inverse Compton photons, $\gamma$, via up-scattering  thermal photons from the massive star, $\varepsilon$. $\gamma$-photons can initiate IC cascade due to absorption in the same thermal field. The cascade is developing up to the termination shock. Electron reaching the shock are trapped there in the local magnetic field, while photons propagate further and escape from the binary or are absorbed in massive star radiation field. The cascade develops radially  following the initial injection direction given by $\alpha_{obs}$. }
\end{figure*}

Under the assumption that the pulsar in the binary is energetic enough to prevent matter from the massive companion from accreting, a termination shock is created in the interaction region of the pulsar and donor star winds. This is represented in Fig. \ref{fig:general}. We focus on the specific case of the binary LS 5039, which we use as a testbed all along this paper. The volume of the system is separated by the termination shock, which structure depends on features of the colliding winds: it may be influenced by the anisotropy of the winds themselves, the motion of the pulsar along the orbit, turbulences in the shock flow, etc. For simplicity it is assumed here that the winds are radial and spherically symmetric, and that the termination of the pulsar wind is an axial
symmetric structure with negligible thickness. In this general picture there are three regions of different properties in the binary: the pulsar wind zone (PWZ), the shock (SR), and the massive star wind zone (MSWZ). 


The energy content in the interaction population of particles is assumed as a fraction of pulsar spin-down power $L_{sd}$. In case of young energetic pulsars, this power is typically $\sim 10^{36} - 10^{37}\, \rm erg\, s^{-1}$. Propagating pairs up-scatter thermal photons from the massive star due to inverse Compton process. For close binaries, the radiation field of hot massive stars (type O, Be or WR, having typical surface temperatures in the range $T_s \sim 10^4 - 10^5\, \rm K$ and linear dimension $R_s \sim 10 R_{\odot}$) dominates along the whole orbit over other possible fields (e.g., the magnetic field or the thermal field of the neutron star). This thermal radiation field is anisotropic, particularly for $e^+e^-$ injected close to the pulsar (the radiation source is misplaced with respect to the electron injection place). 

The high-energy photons produced by pairs can initiate cascades due to subsequent pair production in absorption ($\gamma\gamma$) process with the same radiation field (as sketched in Fig. \ref{fig:general}). We assume that these cascades develop along the primary injection direction, i.e., in a one-dimensional way, which is certainly justified based on the relativistic velocity of the interacting electrons.
This process is followed up to the termination shock unless leptons lose their energy before reaching it. Those leptons which propagate to the shock region are trapped there by its magnetic field. Radiation from them is isotropised. The photons produced in cascades which reach the shock can get through it and finally escape from the binary or be absorbed in the radiation field close to the massive star,  some may even reach the stellar surface.

In the shock region leptons move along it with velocity $\sim c/3$. 
They could be re-accelerated and produce radiation via synchrotron (local magnetic field from the pulsar side) or inverse Compton scattering (ICS, thermal radiation field from the massive star) processes. However, as they are  isotropised in the local magnetic field, photons are produced in different directions and their directionality towards the observer is lost. It was already shown by Sierpowska and Bednarek (2005) that in compact binary systems (as an example, the parameters of Cyg X-3 were taken by these authors) the radiation processes in the shock region do not dominate: the energy carried by $e^+e^-$ reaching the shock is a small fraction of total injected power. Furthermore, we will show that for the parameters relevant to the LS 5039 scenario, the PWZ is relatively large with respect to the whole volume of the system for the significant range of the binary orbit.

\subsection{Hydrodynamic balance }

Assuming that both, the pulsar and the massive star winds are spherically symmetric, and based on the hydrodynamic equilibrium of the flows, the geometry of the termination shock is described by parameter 
$
\eta = {\dot{M}_i V_i}/{\dot{M}_o V_o} ,  
$
where $\dot{M}_i V_i$ and  $\dot{M}_o V_o$ are the loss mass rates and velocities of the two winds (Girard and Wilson, 1987). The shock will be symmetric with respect to the line joining two stars, with a shock front at a distance $r_s  $ from the one of the stars:
\begin{equation}
r_s  = D \frac {\sqrt{\eta}}{(1+\sqrt{\eta})}. 
\label{r0}
\end{equation}
The surface of the shock front can be approximated then by a cone-like structure with opening angle given by 
$
 \psi = 2.1 \left(  1 - \frac{\overline{\eta}^{2/5}}  {4} \right) \overline{\eta}^{1/3},
$
where $\overline{\eta} = min(\eta, \eta^{-1})$ takes the smaller value between the two magnitudes quoted. This last expression was achieved under the assumption of non-relativistic winds in the simulations of the termination shock structure in Girard and Wilson (1987), albeit it is also in agreement with the relativistic winds case (e.g., Eichler and Usov, 1993; Bogovalov et al., 2007). If one of the stars is a pulsar of a spin down luminosity $L_{sd}$ and the power of the massive star is $\dot{M_s} V_s$, the parameter $\eta$ can be calculated from the formula
$
\eta = \frac{L_{sd}}{c(\dot{M_s} V_s)}
$
(e.g., Ball and Kirk 2000). Note that for $\eta < 1$, the star wind dominates over the pulsar's and the termination shock wraps around it. Note also that for $\eta = 1$, the shock is at equal distance, $d/2$, between the stars. In the case of LS 5039, for the assumed (nominal) spin-down luminosity $L_{sd}$ discussed below, the value of $\eta$ is between $0.5$ (periastron) and $0.3$ (apastron).

The massive star in the LS 5039 binary system is of O type, which wind is radiation driven. The velocity of the wind at a certain distance can be described by classical velocity law  
$
V_s(r) = V_0 + (V_{\infty} - V_0)\left(1- \frac{R_s}{r}\right)^{\beta}, 
\label{windy}
$
where $V_{\infty}$ is the wind velocity at infinity, $R_s$ the hydrostatic radius of the star, $V_0$ is the velocity close to the stellar surface, and $\beta = 0.8-1.5$ (e.g., Cassinelli 1979, Lamers and Cassinelli 1999) and we assume $\beta = 1.5$. As can be seen by plotting the velocity law, the influence of this parameter is minor.
Typical wind velocities for O/Be type stars are $V_{\infty} \approx (1-3)\times \, 10^3 \,\rm km\,\rm s^{-1}$. For LS 5039 we have $V_{\infty} = 2.4\times 10^3 $ $\rm km\, s^{-1}$ and $V_0 = 4$ $\rm km\, s^{-1}$ (Casares et al. 2005). The assumed value for the star mass-loss rate is $ 10^{-7}$ M$_\odot$ yr$^{-1}$, while the typical values for O/Be stars are  $10^{-6} - 10^{-7}$ M$_\odot$ yr$^{-1}$.

\subsection{The PWZ and the interacting lepton population}

The magnetization parameter $\sigma= B^2/4\pi \gamma n mc^2$,  (e.g., Langdon et al. 1988) is defined as the ratio of Poynting flux to relativistic particle energy flux. The magnetic field in $\sigma$ is that of the upstream shock propagating with bulk Lorentz factor $\gamma$ and $n$ being the relativistic particle density. The processes establishing the effective change of $\sigma$ along the PWZ are 
the central issue in the discussion of dissipation mechanisms in relativistic plasma flows, exemplified with the Crab pulsar wind (e.g., Kennel and Coroniti, 1984), where variations seem notable. The Crab wind is originally Poynting-dominated ($\sigma \sim 10^4$ close to the neutron star, e.g. Arons 1979); but it is kinetic-dominated near the termination shock ($\sigma \sim 10^{-3}$, e.g., 
Kennel \& Coroniti 1984).  The change in $\sigma$ is produced as a result of dissipative plasma processes in the PWZ, 
which is characterized by high bulk Lorentz factor (e.g., Melatos 1998).
%
Dissipation (plasma processes engaged in the conversion of electromagnetic towards particle kinetic energy) in Poynting flux dominated plasma flows can be in the form of stochastic/non-stochastic and 
adiabatic/non-adiabatic processes, thermal heating/non- 
thermal particle generation, and isotropic adiabatic 
expansion/directed bulk acceleration of the plasma flow (Jaroschek et al. 2008). 
The microphysical details are decisive when looking for the type of particle energization and their spectra.

The existence of wisps in the inner structure of the Crab nebula
has been discussed by, e.g., Lou (1998). The interesting fact of some of them being close 
to the pulsar, apparently well inside the PWZ, was interpreted as being produced by slightly inhomogeneous wind streams, demonstrating that reverse fast MHD shocks at various spin latitudes can appear quasi-stationary in space when their propagation speeds relative to the pulsar wind are comparable to the relativistic outflow.
The possibility of an inhomogeneous wind stream is not implausible. 
Successive radio pulses from a pulsar indeed vary in their shapes, eventhough the average pulse is stable.  A slower wind stream will be eventually caught by a faster one to trigger forward and reverse fast MHD shocks inside the PWZ. In this zone, charged particles can be further accelerated by these turbulences and magnetic reconnection.
Lou proposed that an isotropised power-law like energy distribution of the electrons thus produced help to understand the properties of the changing and brilliant inner nebula.

A mono-energetic assumption for the distribution of leptons in the PWZ can be considered as a first approach to the problem.
On one hand, the magnetization parameter may be a function of angle, and although must be very small in the equatorial part of the wind (e.g., Kirk 2006),  simulations do not favor an angle-independent low value (Komissarov \& Lyubarsky 2004). 
On the other hand, Contopoulos \& Kazanas (2002) already showed that the Lorentz factor of the outflowing plasma could increase linearly with distance from the light cylinder 
(implying that $\sigma$ decreases inversely proportional to the distance).
Contopoulos \& Kazanas (2002) mentioned that this specific radial dependence of the pulsar winds 
Lorentz factor is expected to have additional observational 
consequences: e.g., Bogovalov \& Aharonian (2000) computed the 
Comptonization of soft photons to TeV energies in the 
Crab through their interaction with the expanding MHD 
wind, while Tavani \& Arons (1997) and Ball \& Kirk (2000) 
computed the corresponding radiation expected by the 
radio-pulsar Be star binary system PSR B1259-63 through 
the interaction of the relativistic wind with the photon field 
of the companion in much the same way we do here. Indeed, the details in these predictions would be modified, as shown by Sierpowska \& Bednarek (2004, 2005) should the linear acceleration model be adopted.
Hibschman and Arons (2001) discussed the creation of electron-positron cascades in the context of pulsar polar cap acceleration 
models. They computed the spectrum of pairs that would be produced outflowing the magnetosphere. They found that the pair spectra should be described by a power-law. 

One possibility for  the dissipative conversion is established by magnetic reconnection 
processes between anti-parallel magnetic stripes during 
outwards propagation in the PWZ (Lyubarsky and Kirk, 
2001; Kirk and Skjaeraasen, 2003). 
Kirk (2004) considered acceleration in relativistic current sheets (large magnetization parameter, with Alfven speed $v_A=c\sqrt(\sigma/(\sigma+1)$ close to c).
Recently, Jaroschek et al. (2008, and see references therein for related work)  addressed the problem of interacting relativistic current sheets in self-consistent kinetic plasma simulations, identifying the generation of non-thermal particles and formation of a stable power-law shape in the particle energy distributions $f(\gamma) d\gamma  \propto \gamma^{-s} d\gamma$. Depending on the dimension of the simulation, spectral index from 2 (1D, attributed to a stochastic Fermi-type acceleration) to 3-4 (recognized as a rather universal index of relativistic magnetic reconnection in previous 2D and 3D kinetic simulations, see Jaroschek et al. 2004, Zenitani and Hoshino 2005) were found.
Lyubarsky and Liverts (2008) also studied the compression driven magnetic reconnection in the relativistic pair plasma, using 2.5D (i.e., 2D spatial, 3D velocity) simulations, finding
that the spectrum of particles was non-thermal, and a power law was produced. It seems a power-law distribution for the leptons inside the PWZ is then a plausible assumption.

All in all, to find an a-priori dissipation solution for pulsars, and in particular, for the assumed pulsar in the LS 5039 which is the one we focus, is beyond the scope of this work (and actually, for the latter particular case, such solution is beyond what is by definition possible for a pulsar that we do not know exists). In general, we note that an additional difficulty resides in the fact that the PWZ of pulsars in binaries may be subject to conditions others than those found in isolated pulsars. It is not implausible that close systems may trigger different phenomenology within the PWZ, ultimately affecting particle acceleration there.
Nevertheless, it is relevant for this paper to assume a particular particle's energy spectra with which we compute high energy processes in the PWZ, e.g. the up-Comptonization of the stellar field. We will assume two cases, a mono-energetic spectra -as a benchmark- (e.g., see  Bogovalov \& Aharonian 2000) and a generic power-law spectra (that could itself  
be subject to orbital variability). Both act as a phenomenological assumption in this paper, which goodness is to be assessed a posteriori, by comparison with data.

 As discussed, initial injection could come directly from the pulsar (the interacting particle population can of course be later affected by the equilibrium between this injected distribution and the losses to which it is subject, just as in the case of shock-provided electron primaries). The more compact the binary is, the more these two settings (shock and pulsar injection, equilibrated by losses) are similar to each other. 
Given the directionality of the high Lorentz factor inverse Compton process, photons directed towards the observer are generally those coming from electrons moving in the same direction. 
Opacities to processes such as inverse Compton and $\gamma\gamma$ absorption are high in close $\gamma$-ray binaries, cascades can develop, and high-energy processes can already happen, as we explicitly show below, in the pre-shock region. 

\begin{figure*}
  \centering
   \includegraphics[width=0.49\textwidth,angle=0,clip]{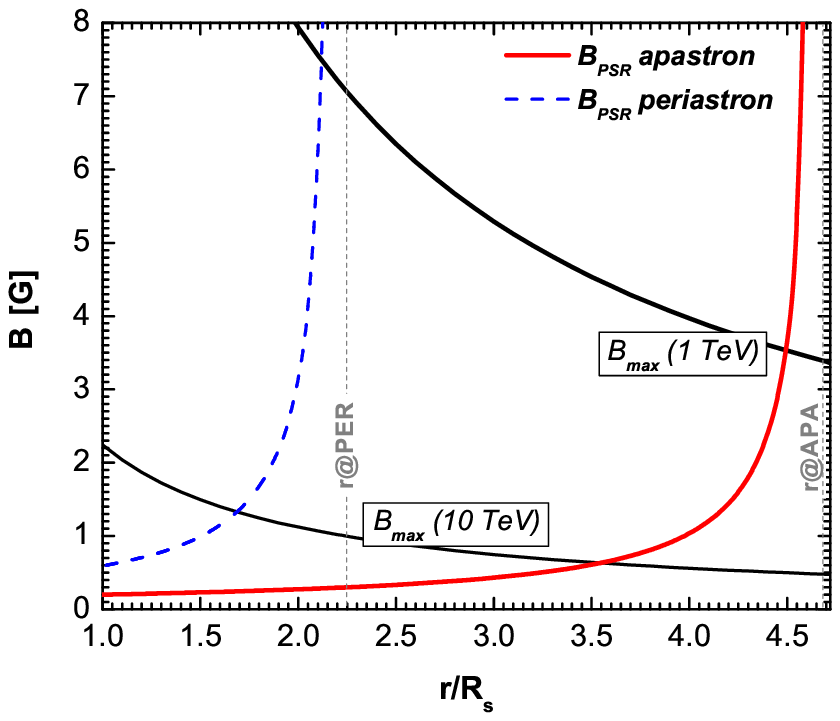}
   \includegraphics[width=0.49\textwidth,angle=0,clip]{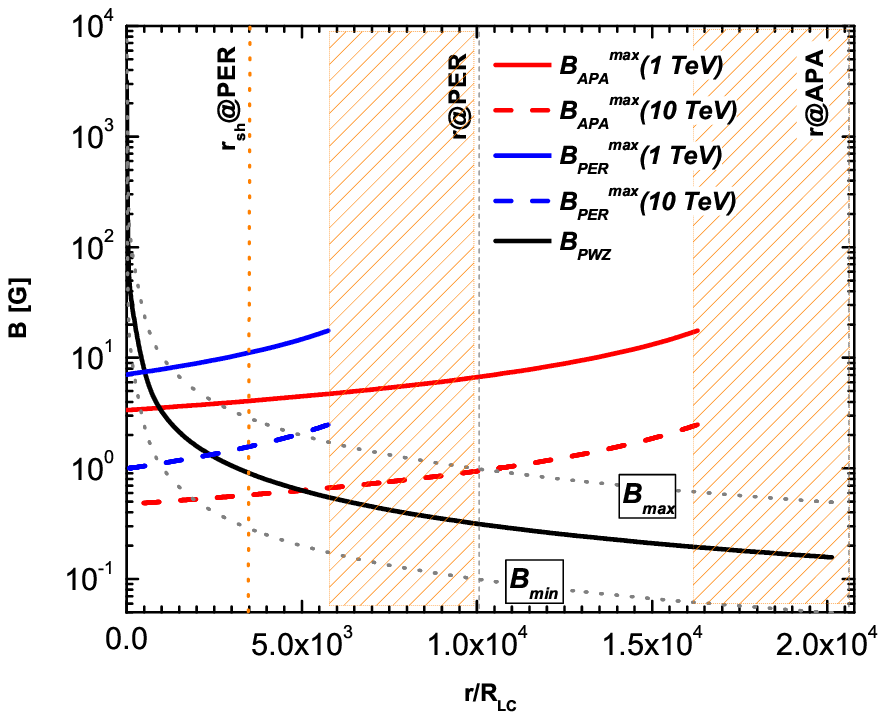}
\caption{\label{fig:locB} Conditions upon the local magnetic field in the pulsar wind zone (PWZ). Left Panel: The black lines represent the maximum magnetic field, $B_{max}$,  for which IC losses dominate over synchrotron ones as a function of distance $r/R_s$ from the massive star for $1$ TeV electrons (thick curve) and $10$ TeV (thin curve). The magnetic field in the PWZ, also as a function of radius from the massive star, is shown with a red line (for the pulsar at the apastron position) and with the blue line (for the pulsar at periastron). The system separation radius at periastron and apastron are also marked (gray lines). The magnetic field decays in the PWZ from its value close to the light cylinder. Right Panel: A different way of showing these conditions. PWZ magnetic field for different magnetization parameters ($\sigma=10^{-2}$ (labeled $B_{max}$) 10$^{-3},$ 10$^{-4}$ (labeled $B_{min}$) as a way of artificially introducing uncertainty in the real value. the magnetization parameter is expected to change within the PWZ (e.g., as in Crab). Maximum magnetic field for IC domination at periastron and apastron are again shown for different electron energies. The size of the star is represented by a shadow rectangle. The linear scale in the x-axis is in units of the light cylinder. The shock radius at periastron, and the separation at periastron and apastron are all shown.}
\end{figure*}

{  In the model where $e^\pm$ pairs are injected as monoenergetic particles with the energy corresponding to the bulk Lorentz factor of the pulsar wind, they are frozen in the B-field. Under this assumption we neglect here the synchrotron losses, since there are none.

When the injected $e^\pm$ pairs distribution is given by a power law spectrum the situation is more complex, since not all of them may be frozen in the PWZ field. The synchrotron cooling time is given by equation:
\be
 t_{syn} = 400 B_G^{-2} E_{TeV}^{-1} \, \rm s,
\ee 
where $B_G$ is the local magnetic field and is given in Gauss, and $E_{TeV}$ is electron energy given in TeV. 
For IC scattering, the timescale is instead given by  
\be
t_{ic} = 7\times 10^3 \omega_0^{-1} E_{TeV}^{0.7}\, \rm s
\ee 
which gives good approximation for electron energy loss time for $E \gtrsim 1$ TeV (e.g., Khangulyan et al. 2008). 
Timescales approach when 
\be
B_G^{max} \approx 0.24 \omega_0^{0.5} E_{TeV}^{-0.85}\, \rm G.
\ee
For a star with effective temperature $T_s = 3.9\times 10^4$ K, the thermal field density at certain point at distance $r$ from the massive star center is given by:
\be
\omega_0 = 4 \sigma T^4/c \times (R_s/2r)^2 \approx 1.75 \times 10^4 (R_s/2r)^2 \, \rm erg\, cm^{-3}.
\ee
Thus, the local magnetic field in the pulsar wind region is given by:
\be
B_G^{max} \approx 31.75 (R_s/2r )E_{TeV}^{-0.85}\, \rm G. 
\ee

Applying this condition at periastron ($r = 2.25 R_s$), the local magnetic field at the assumed injection place results in $B_{max-per} \sim 7$ G. At apastron ($r = 4.72 R_s$), it results in an stronger condition $B_{max-apa} \sim 3.4$ G. That is, the magnetic field should be less than these values in order for IC to dominate over synchrotron losses at that particular position (the light cylinder) in the PWZ. Figure \ref{fig:locB} shows this in detail.\\

The magnetic field at the light cylinder distance $R_{LC}$ is given by the dipole formula $B_{LC} = B_0 (R_{psr}/R_{LC})^3$. In the PWZ, the magnetic field is decreasing with distance as $B(r) \sim \sqrt{\sigma/(1+\sigma)} \, B_{LC} (R_{LC}/r)$. For a millisecond pulsar we get $R_{LC} \sim 5\times 10^7$ cm and $B_{LC} \sim 10^5$ G  (assuming $B_0 = 10^{12}$ G and $P = 10$ ms) up to $R_{LC}\sim 5\times 10^8$ cm and $B_{LC} \sim 8 \times 10^4$ G  (assuming $B_0 = 10^{13}$ G and $P \sim 100$ ms), where $R_{psr} \approx 10$ km. The results for these different parameters are similar as there were obtained with fixed pulsar power in the model $L_{sd} = 10^{37}$ erg s$^{-1}$ and they are related by the standard formula $L_{sd} = B_0^2 R_{psr}^6 c / 4 R_{LC}^4 $. We also assume here that the magnetization parameter is $\sigma = 0.001$, but have explored other values of this and other parameters as well, with similar results (see Figure \ref{fig:locB}). \\

Given our results (see Fig. \ref{fig:locB}) where we show the local magnetic field for which IC dominates and the magnetic field in the PWZ as a function of the distance from the light cylinder we can conclude that the injection for the model have to (generically) occur at some distance from the light cylinder, or/and, if  closer to it, the synchrotron losses can be important.   However, as the separation
of the binary is $\sim 10^{12}$ cm, several orders of magnitude larger than the light cylinder (e.g., $R_{LC} \sim 10^8$ cm), the change of the injection place within e.g. $\sim 1\% - 5\%$ already gives the initial injection distance at $\sim (100-500) R_{LC}$. Thus,  no significant effect in the PWZ photon spectra and lightcurve is produced. }

\subsection{Normalization of the relativistic particle power}

The fraction of the pulsar spin-down power ending in the $e^+e^-$ interacting pairs can then be written as:
\begin{equation}
\beta L_{sd} = \int N_{e^+e^-}(E) E dE.  
\label{norma}
\end{equation}
Assuming that the distance to the source is $d = 2.5$ kpc, 
the normalization factor for electrons traveling towards Earth is $ A = N_{e^+e^-}/ 4\pi d^2.$  The specific normalization factors in the expression of the injection rate $N_{e^+e^-}(E)$ will be given together with the results for two models in the corresponding sections below. In the models presented here, only a small fraction
($\sim $1\%) 
of the pulsar's $L_{\rm SD}$ ends up in relativistic leptons. 
This is consistent with ions carrying much of the wind energetics.
In the case of mono-energetic lepton distribution, where the energy of the primaries is fixed at $E_0=10$ TeV, we have $N_{e^+e^-}(E) \propto \delta(E-E_0)$. In the case of a power-law in energy, that may be constant or vary along the orbit, we have $N_{e^+e^-}(E) \propto E^{-\alpha_i}$. 

\subsection{On parameter interdependencies}

\begin{figure}
  \centering
 \includegraphics[width=0.5\textwidth,angle=0,clip]{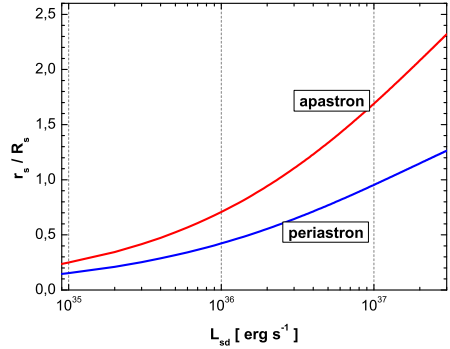}
\caption{\label{fig:r0} Dependence of the distance to the shock form the pulsar side, $r_s$, as a function of the spin-down power, $L_{sd}$, for fixed values of star mass loss rate, $\dot{M} = 10^{-7}$ M$_\odot$ yr$^{-1}$, and terminal star wind velocity, $V_{\infty} = 2.4\times 10^3 $ $\rm km\, s^{-1}$. }
\end{figure}

The $\eta$ expression represents the realization of a specific scenario fixed by different values of the pulsar and massive star parameters. For instance, assuming $L_{sd} = 10^{37}$ $\rm erg\, s^{-1}$ and $\dot{M_s} V_s$ as given in the Table \ref{orb-param} below,
we get $\eta \approx 0.5$ for the periastron phase. 
But we would achieve the same value of $\eta$ with a smaller $L_{sd}$, say 10$^{36}$ $\rm erg\, s^{-1}$, if within uncertainties we adopt a smaller value of $\dot{M_s} V_s$. Clearly, if only one of the quantities, $L_{sd}$ or $\dot{M_s} V_s$ changes, the shock position moves, what results in a correspondingly smaller or larger PWZ  in which cascading processes are set. A key parameter is then given by $r_s$, instead of $\eta$, since this is what defines the real size of the PWZ. The mild dependence of $r_s$ on $\eta$ makes possible to accommodate a large variation in the latter maintaining  similar results.  Fig. \ref{fig:r0} shows this dependence for a relevant range of $L_{sd}$.

In addition, it is obvious that $L_{sd}$ and $\beta$ are linearly (inversely) related. 
We note that a fixed value of $\eta$ does not imply a specific value for $L_{sd}$ as it is combined with parameters of the massive star wind, subject to uncertainty in their measurements for the specific case of binary treated (e.g., for LS 5039), if at all known. 
In the end, the same results can be obtained for different sets of parameters. There is a degeneracy between the shock position, which determines the extent of the PWZ, and the injected power in relativistic electrons. For a smaller PWZ (e.g., if we assume a smaller $L_{sd}$ for the same product of $\dot{M_s} V_s$)
a larger amount of injected power compensates the reduced interacting region.
We found that to get the similar results when the $\eta$ parameter is 
smaller, the $\beta$ parameter have 
to be increased roughly by the same factor.

\subsection{Wind termination}

\begin{figure}
  \centering
   \includegraphics*[width=0.5\textwidth,angle=0,clip]{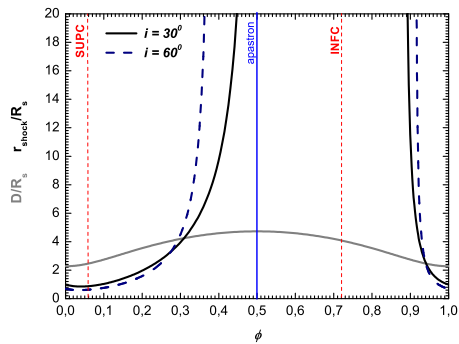}
\caption{\label{fig:shock_sep} {The distance from the pulsar  to the termination shock in the direction to the observer (in units of stellar radius $R_s$), for the two different inclination angles, $\textit{i}$, analyzed in this paper. INFC, SUPC, periastron, and apastron phases are marked. Additionally, the gray line shows the separation of the binary (also in units of $R_s$) as a function of phase along the orbit.}
}
\end{figure}

Very interesting in the context of LS 5039 model properties is the dependence of the distance from the pulsar to the termination shock in the direction to the observer as a function of the orbital phase. This is shown in Fig. \ref{fig:shock_sep}. 
%
We find that for both inclinations, the wind is unterminated for a specific range of phases along the orbit, i.e., the electron propagating in the direction of the observer would find no shock.
The PWZ would always be limited in the observer's direction only if the inclination of the system is close to zero, i.e.,  the smaller the binary inclination the narrower the region of the wind non-termination viewed by the observer.  In the discussed scenarios for LS 5039, the termination of the pulsar wind is limited to the phases $\phi \sim 0.36 - 0.92$ for inclination $i=60^o$ and to $\phi \sim 0.45 - 0.89$ for $i=30^0$, what coincides with the phases where the emission maximum is detected in the very high-energy photon range (see below the observed lightcurve obtained by H.E.S.S.). 
The unterminated wind is viewed by the observer at the phase range between apastron and INFC, while a strongly limited wind appears from periastron and SUPC. 
Note that the important differences in the range of phases in which the unterminated wind appears for distinct inclinations (e.g., the observer begins to see the unterminated pulsar wind at $\sim 0.36$ for $i=60^o$ compared to  $\sim 0.45$ for  $i=30^0$) produce a distinguishable feature of any model with a fixed orbital inclination angle.

From the geometrical properties of the shock surface, there can also be specific phases for which the termination shock is directed {\sl edge-on} to the observer.
These are phases close to the non-terminated wind viewing conditions:
slightly before and after those specific phases for which the wind becomes non-terminated, e.g., for the case of LS 5039 and $i=30^0$, $\phi \sim 0.4$ and $0.93$; whereas for $i=60^0$, it appears at $\phi \sim 0.32$ and $0.93$.
Thus, we find that the condition for lepton propagation  change significantly in a relatively short phase period, when the termination shock is getting further and closer to the pulsar (phases periods $\sim 0.2-0.45$ and $\sim 0.9 - 0.96$).
In Table \ref{tab:orbit} we show the differences in these geometrical parameters for a few characteristic orbital phases. Note that even when they are essential to understand the formation of the lightcurve and phase dependent high energy photon spectra, all the above features are based on the simple approximation of the geometry of the termination shock. In a more realistic scenario, the transition between the terminated and unterminated wind, and its connection with orbital phases and inclination are expected to be more complicated yet.

\begin{table}
\centering
\caption{\label{tab:orbit} Geometrical properties for specific phases along the orbit. $p_1$ and $p_2$ stand for the phases at which the angle to the observer is $90^o$; $r_1$ and $r_2$ stand for the phases between which the shock in the observer direction is non-terminated. The separation $d$ and the shock distance $r_{s}$ are given in units of star radii $R_s$.}
\vspace{0.2cm}
\centering
\begin{tabular}{ l | l  | llll | llll}
\hline
& & \multicolumn{4}{c}{$i = 30^0$} & \multicolumn{4}{c}{$i = 60^0$} \\
\hline
 & $\theta\, [^o]$ & $\phi$ & $d$ & $\alpha_{obs}$ & $r_{s}$ & $\phi$ & $d$ & $\alpha_{obs}$ & $r_{s}$ \\
\hline
Periastron      & 0  & 0.0 & 2.25 & $111^0$ & 0.8 & 0.0 & 2.25 & $130^0$ & 0.6 \\
SUPC       & 44  & 0.06 & 2.45 & $120^0$ & 0.8 & 0.06 & 2.45 & $150^0$ & 0.6 \\ 
$p_1$         & 134  & 0.27 & 4.06 & $90^0$ & 3.2 & 0.27 & 4.06 & $90^0$ & 3.2 \\
$r_1$         & 171   & 0.45 & 4.7 & $72^0$ & \ldots & 0.36 & 4.5 & $73^0$ & \ldots \\
Apastron      & 180   & 0.5 & 4.72 & $69^0$ & \ldots & 0.5 & 4.72 & $43^0$ & \ldots \\
INFC      & 224  & 0.72 & 4.1 &  $60^0$ & \ldots& 0.72 & 4.1 & $30^0$ & \ldots \\
$r_2$       & 285   & 0.89   & 2.8 & $76^0$ & \ldots & 0.92 & 2.6 & $80^0$ & \ldots\\
$p_2$     & 314   & 0.94  & 2.47 & $90^0$ & 2.4 & 0.94 & 2.47 & $90^0$ & 2.4 \\
\hline
\end{tabular}
\end{table}

\subsection{Opacities along the LS 5039 orbit}

\begin{figure*}
  \centering
   \includegraphics[width=0.49\textwidth,angle=0,clip]{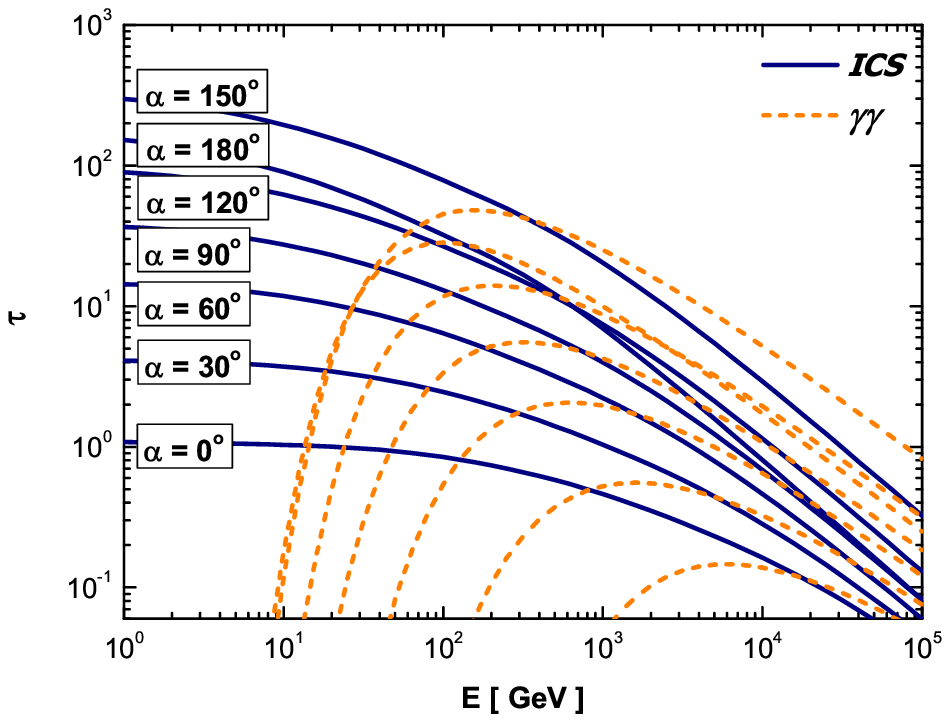}
   \includegraphics[width=0.49\textwidth,angle=0,clip]{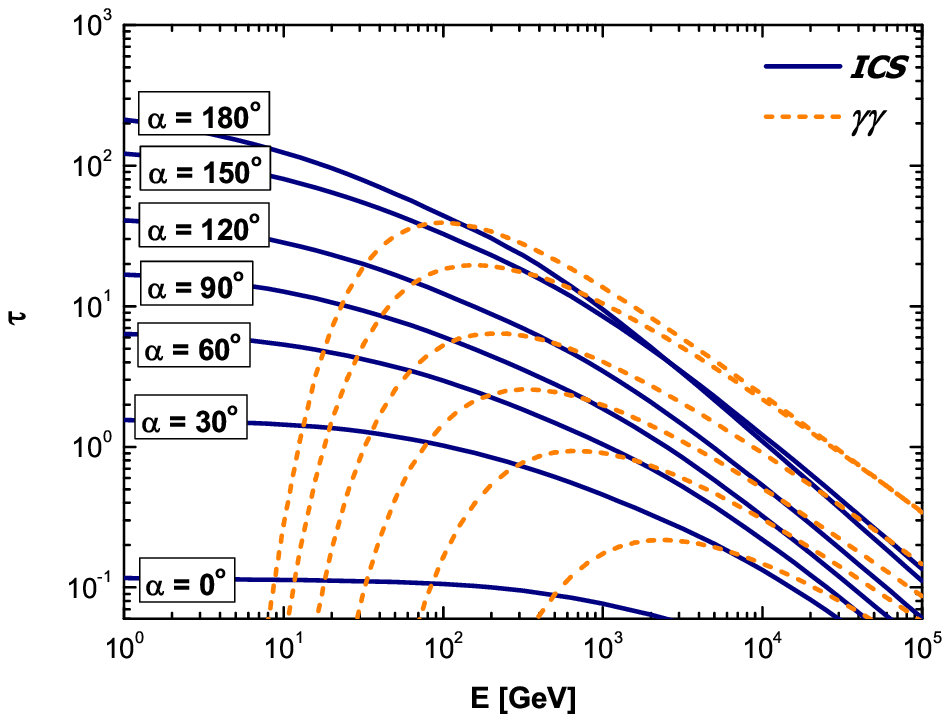}
\caption{\label{fig:tau_gen} {Opacities to ICS for electrons and pair production for photons as a function of electron/photon energy. Opacities were calculated up to infinity for injection at the periastron (left panel) and apastron (right panel) pulsar distance for different angles of propagation with respect to the massive star. }}
\end{figure*}

The conditions for leptonic processes for this specific binary can be discussed based on optical depths to ICS and $\gamma\gamma$ absorption. 
The target photons for IC scattering of injected electrons and for $e^+e^-$ pair production for secondary photons are low energy photons of black-body spectrum with temperature $T_s = 3.9 \times 10^4$, which is the surface temperature of the massive star. This is an anisotropic field as the source of thermal photons differ from the place of injection of relativistic electrons, which for simplicity is assumed to be at the pulsar location for any given orbital phase.
For fixed geometry parameters, the optical depths change with the separation of the binary (in general with the distance to the massive star), the angle of injection (the direction of propagation with respect to the massive star), and the energy of injected particle (the electrons or photons for $\gamma$ absorption).

To have a first handle on opacities, we have calculated the optical depths adopting the binary separation at periastron and apastron. In case of the optical depths for photons we had to do an additional simplifying assumption to allow for a direct comparison with the optical depth for electrons. In the discussed model of this paper, photons are secondary particles and they do not have the same place of injection as electrons, but because the path of electrons up to their interaction is much smaller than the system separation, we assume now the same place of injection for photons and electrons: for a first generation of photons, this is sufficient for comparison. 

\begin{figure*}
  \centering
   \includegraphics[width=0.49\textwidth,angle=0,clip]{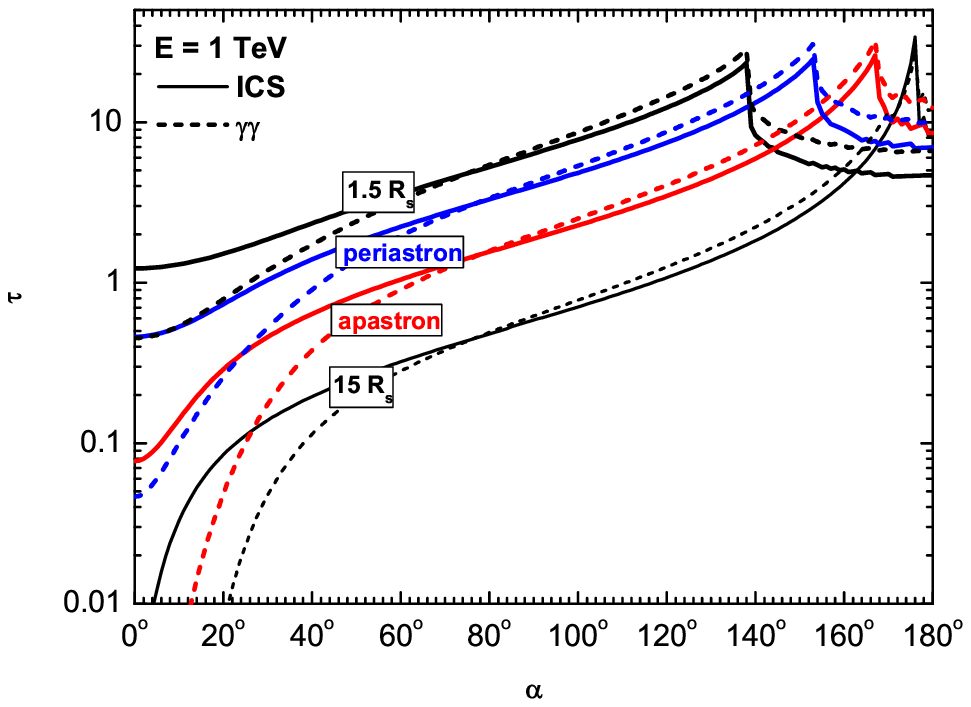}
   \includegraphics[width=0.49\textwidth,angle=0,clip]{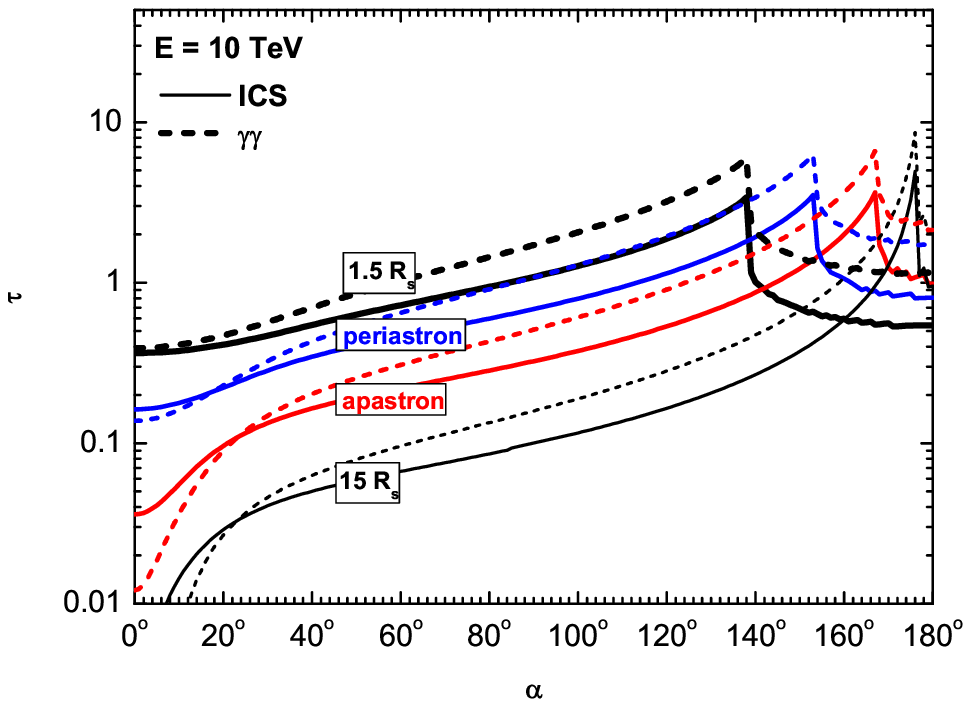}
\caption{\label{fig:tau_alpha} {Opacities to ICS for electrons and pair production for photons as a function of the injection angle, $\alpha$, in the LS 5039 system, for different initial energies of the particles (left panel $E = 1 $ TeV, right panel $E = 10$ TeV) and distance to the massive star (including periastron and apastron separation). Opacities were calculated up to infinity. The injection angle is provided in the plot and for pair production process they are given in the same order as for ICS. }}
\end{figure*}

Results are shown in Fig. \ref{fig:tau_gen} and the needed formulae for its computation are given in the Appendix. 
The optical depths for ICS can be as high as $\sim $ few $100$ for significant fractions of the orbit, and are still above unity for energies close to $ 10 $ TeV. With respect to the injection angle, the opacities are above $\sim 1$ for all propagation angles, except the outward directions at the apastron separation. (Recall that outward --inward-- directions 
refer to $\alpha$-angles close to 0$^o$ and 180$^o$, respectively, as shown in Fig. \ref{fig:general}).
In the case of $\gamma\gamma$ 
pair production, the interactions are limited to specific energies of the photons and are strongly angle-dependent. The most favorable case for pair creation is represented by those photons with energy in the range $0.1$ to few TeV, propagating at $\gtrsim 50^o$. 
The highest optical depths to pair production are found for directions tangent to the massive star surface: for propagation directly towards the massive star pair production processes are limited by the star surface itself (for periastron the tangent angle is at $\sim 154^o$; for apastron, it is given at an angle of $\sim 168^o$). This is better shown in Fig. \ref{fig:tau_alpha}.\footnote{The range of parameters depicted in this figure is just to emphasize that there could also be propagation and interactions outside the innermost parts of the system, in general, although of course with a reduced opacity. The actual range of angles towards the observer allowed in a particular system depend on the orbit inclination.}

The opacities for pair production are characterized by maxima which change with respect to the angle of photon injection. For inward directions (in general for angles larger then $90^o$), the peak energy is higher, $ \sim 10^2 \; \rm GeV - 3 \times 10^2 \; \rm GeV $, than for outward directions, where we find $ \sim 3 \times 10^2 \; \rm GeV - 5 \; \rm TeV$. For energies up to the {\sl peak} for pair production, the IC process dominate for the same angle of particle injection, but for higher energies the opacities for $\gamma$-absorption become slightly higher than for ICS.

Cascading processes are effective when the interaction path for electrons and photons are short enough so that a cascade can be initiated already inside the pulsar wind zone. As we can see in Fig. \ref{fig:tau_gen}, the opacities at high-energy range (Klein-Nishina) are comparable to the maximal opacities for $\gamma \gamma$ absorption. This means similar probability of interaction for both photons and electrons. If the electrons scatter in the Klein-Nishina regime, the photon produced, with comparable energy to that carried by the initiating electron, is likely to be absorbed, and a $e^+e^-$ cascade is produced.

\begin{figure*}
  \centering
   \includegraphics[width=0.49\textwidth,angle=0,clip]{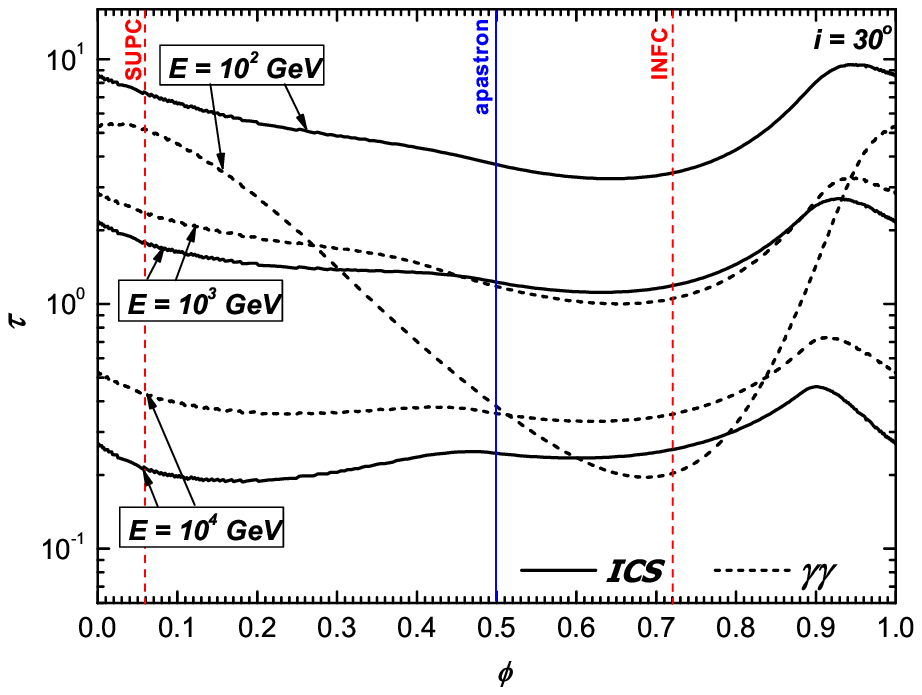}
   \includegraphics[width=0.49\textwidth,angle=0,clip]{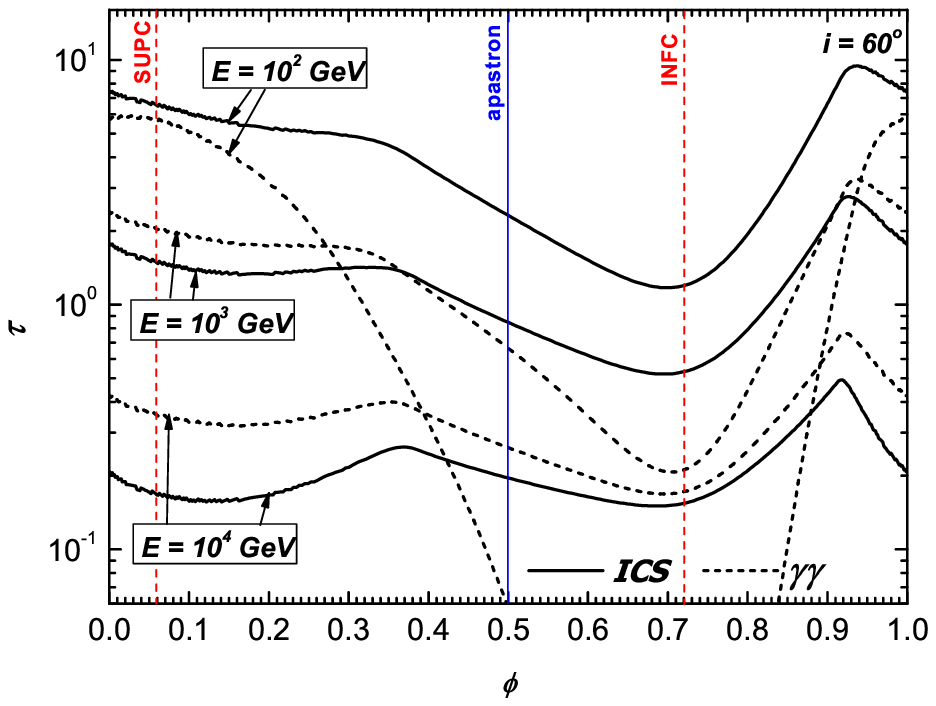}
\caption{\label{fig:tau_phi} {Opacities to ICS for electrons and pair production for photons as a function of the orbital phase, $\phi$ in the LS 5039 system, for different inclination of the binary orbit (left panel $i=30^0$, right panel $i=60^0$) and energy of injected particle. Opacities are calculated for specific phases 
up to the termination shock in direction to the observer. }}
\end{figure*}

\begin{figure} 
  \centering
   \includegraphics*[width=0.49\textwidth,angle=0,clip]{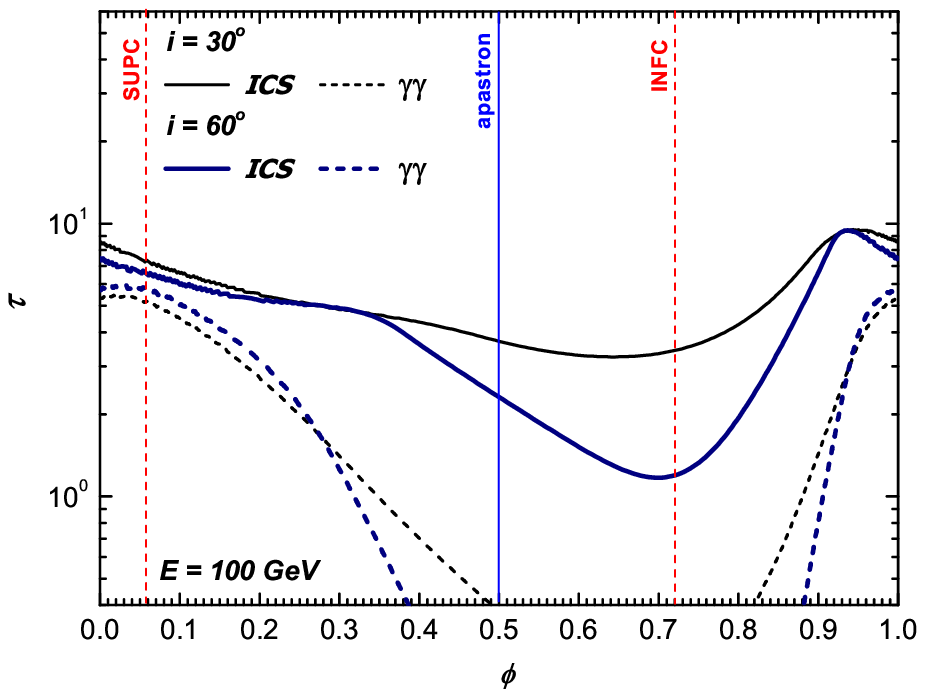}
   \includegraphics*[width=0.49\textwidth,angle=0,clip]{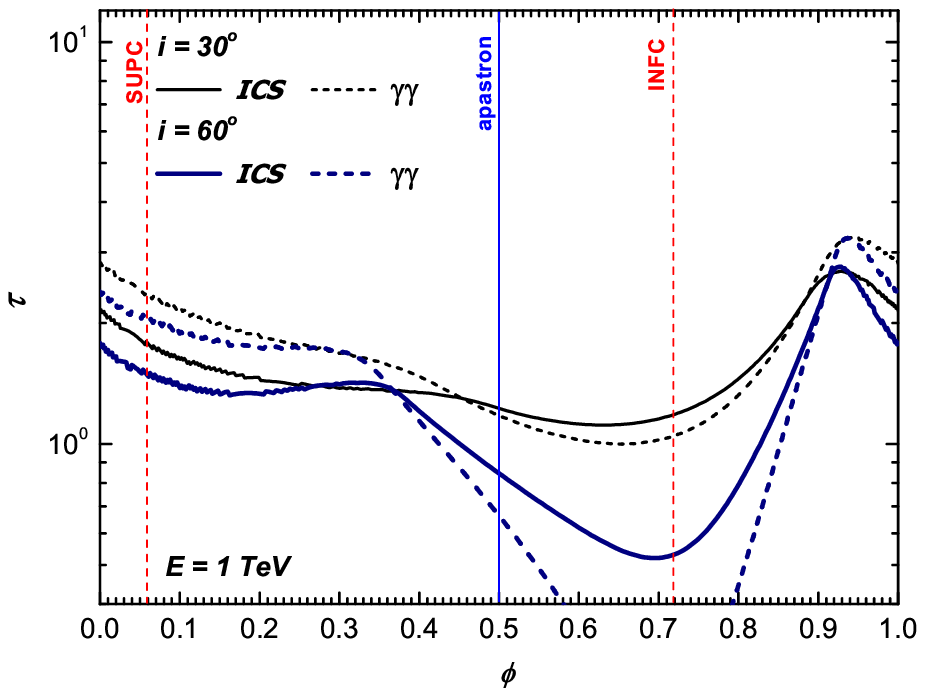}
   \includegraphics*[width=0.49\textwidth,angle=0,clip]{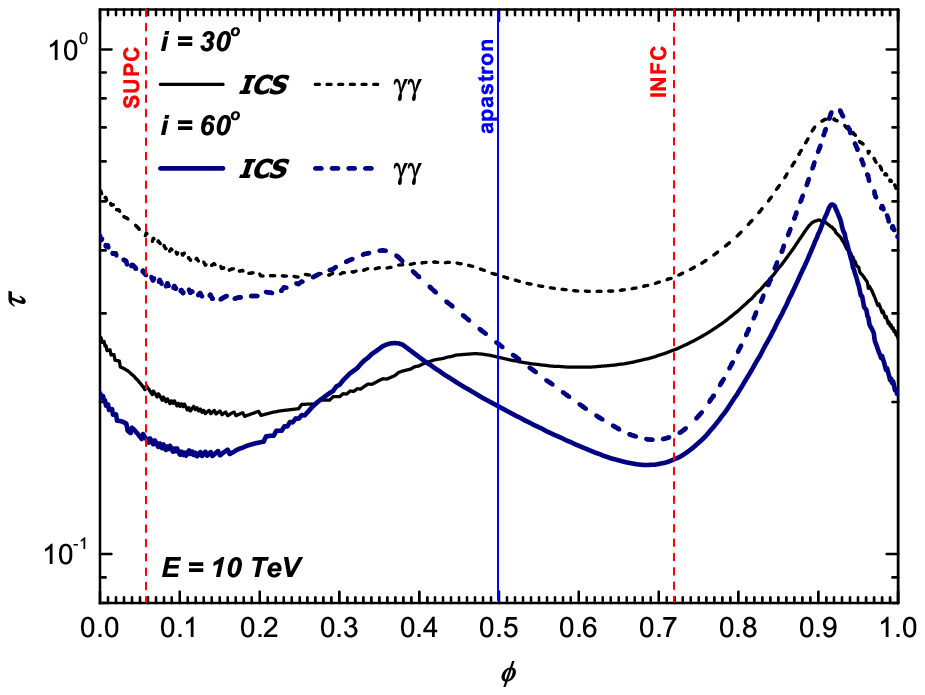}
\caption{\label{fig:tau_E} {Comparison of the opacities to ICS for electrons and pair production for photons along the orbit of LS 5039, for fixed particle energy but different inclination angles ($i=30^0$, $i=60^0$). }}
\end{figure}

The dependence of the optical depths upon the orbital phase for specific parameters of the LS 5039 is shown in Fig. \ref{fig:tau_phi}. These opacities are calculated up to the termination shock in the observer direction. The presence of the shock (at a distance $r_s$) limits the optical depths. Since this parameter is highly variable along the orbit (see Fig. \ref{fig:shock_sep}), its influence on the optical depth values is not minor. 
As the angle to the observer, which defines the primary injection angle, vary in the range $(90-i, 90+i)$ (for INFC and SUPC, respectively), we found that there is a range of orbital phases for which the PWZ is non-terminated. In that case the cascading process --which develops linearly-- is followed up to electron's complete cooling (defined by the energy $E_{min} = 0.5\, \rm GeV$). The optical depths in the non-terminated pulsar wind are thus comparable to those presented in Fig. \ref{fig:tau_gen}, taking into account the differences between the separation and angle of injection. 

\begin{figure*}
  \centering
   \includegraphics[width=0.49\textwidth,angle=0,clip]{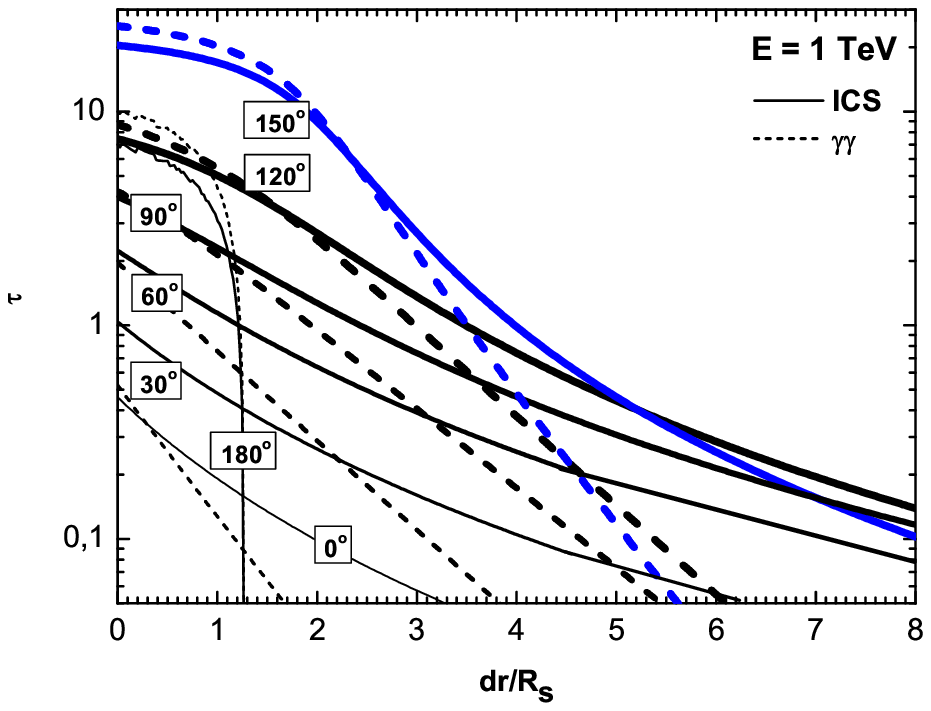}
   \includegraphics[width=0.49\textwidth,angle=0,clip]{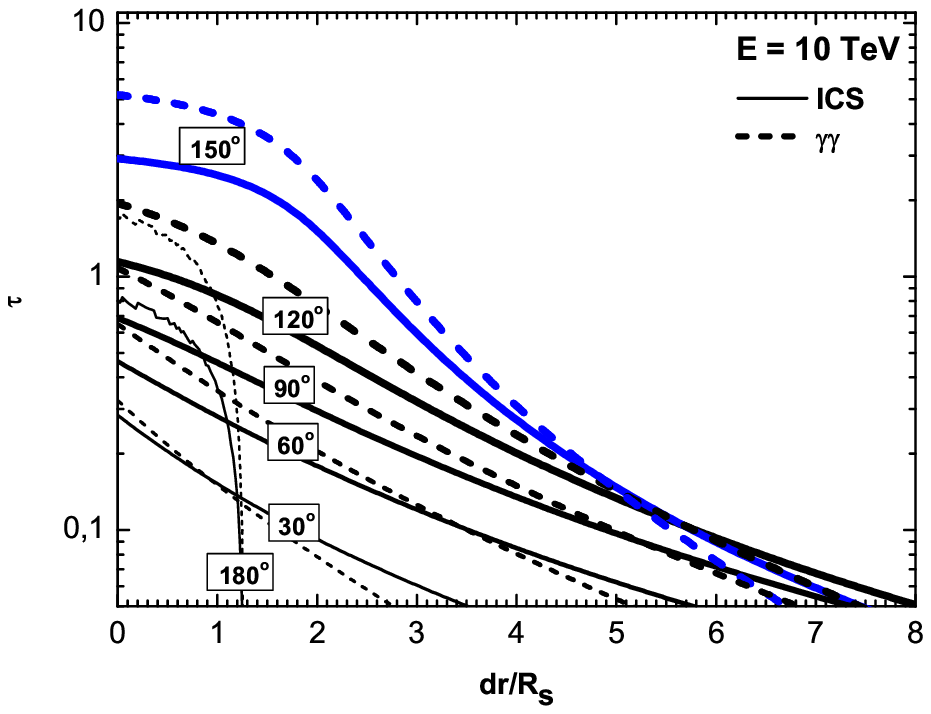}
\caption{\label{fig:tau_dr} {Opacities to ICS for electrons and pair production for photons as a function of the distance from the injection place along the propagation path from the pulsar side, $dr$, in the LS 5039 system. Opacities are calculated for two values of the particles initial energies,  $E = 1 $ TeV (left panel), and  $E = 10$ TeV (right panel), and different angles of injection. }}
\end{figure*}

Apart from the condition for electron cooling, in the case of the terminated wind the cascade is followed up to the moment when the electron reaches the shock region, whatever happens first. As can be seen from comparison of Fig. \ref{fig:tau_gen}  and \ref{fig:tau_phi}, the opacities are significantly smaller when the propagation is limited. For instance, as a comparison we can choose the electron injection at periastron, the angle $\alpha_{obs} = 110^o$ ($130^o$ for $i=60^o$), and fix the energy to $\sim 1$ TeV. Then, the opacities along the orbit up to the terminated shock are $\sim 2-3$ while for the unterminated processes the opacity are $\sim 10-20$. A direct comparison for specific energies of injected particles and the two inclinations considered is shown in Fig. \ref{fig:tau_E}.
Moreover, the opacities for both processes decreases along the propagation path, see Fig. \ref{fig:tau_dr}, as the angle to the observer also decreases. This indicate that even if the photons propagate very close to the massive star, the probability for absorption can be much smaller than that at its injection place, and finally become less then $1$ at the distance from the pulsar $\sim 3-4 \, R_s$, for an energy of $10$ TeV and $1$ TeV respectively.

\section{Mono-energetic electrons in the PWZ}



Following the normalization formula given by Eq. (\ref{norma}), for the primary lepton energy $E = 10$ TeV and the nominal value of pulsar spin-down power $L_{sd} = 10^{37} \rm erg\, s ^{-1}$,  and assuming a distance to the source $d = 2.5$ kpc we get $ N_{e^+ e^-} = \beta \times 6.24 \times 10^{35}\, \delta(E-10 {\rm TeV})\, \rm s^{-1}.$ The normalization factor for electrons traveling towards Earth is $A = N_e/4\pi d^2 = \beta \times 8.83 \times 10^{-10} \, \rm s^{-1} \, cm^{-2}$. As a first normalization of the simulation results below, we have fixed $\beta = 10^{-2}$.


\subsection{Lightcurve and broad-phase spectra}

\begin{figure*}
  \centering
  \includegraphics[width=0.49\textwidth,angle=0,clip]{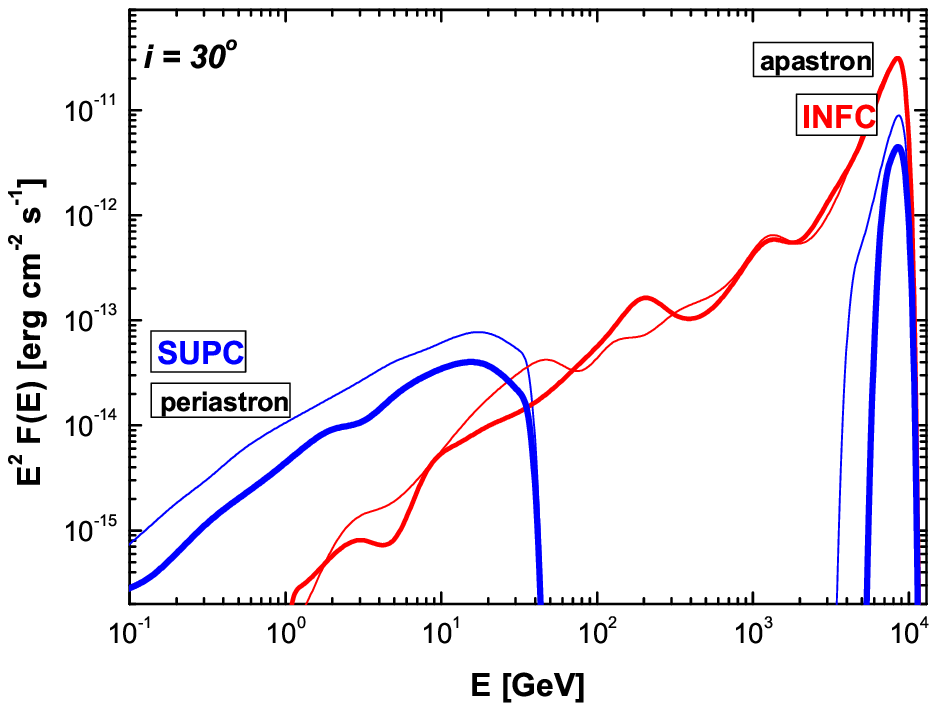}
  \includegraphics[width=0.49\textwidth,angle=0,clip]{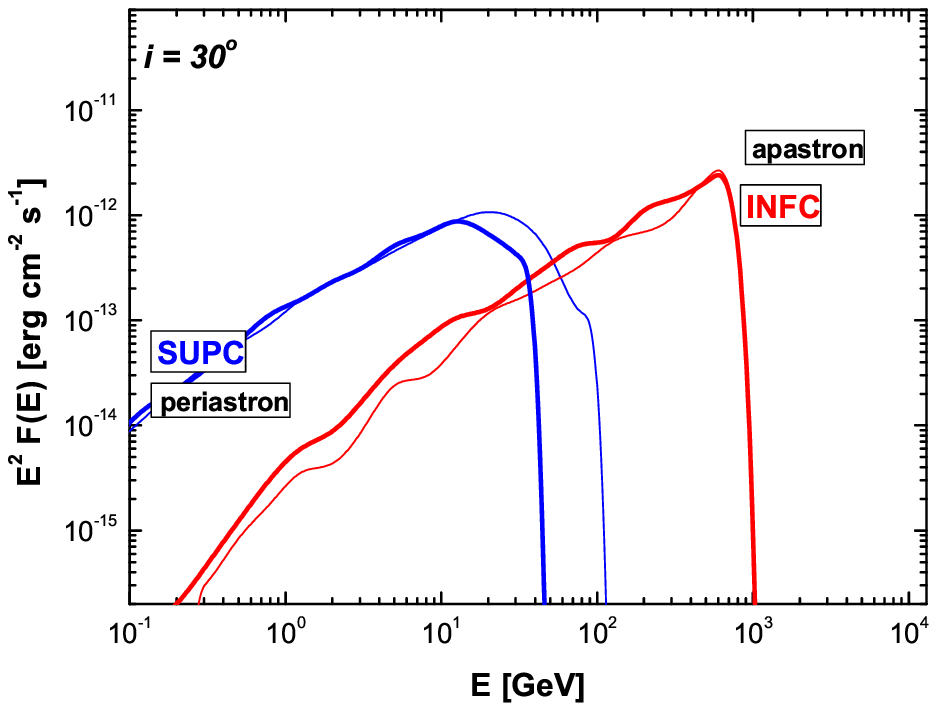}\\
  \includegraphics[width=0.49\textwidth,angle=0,clip]{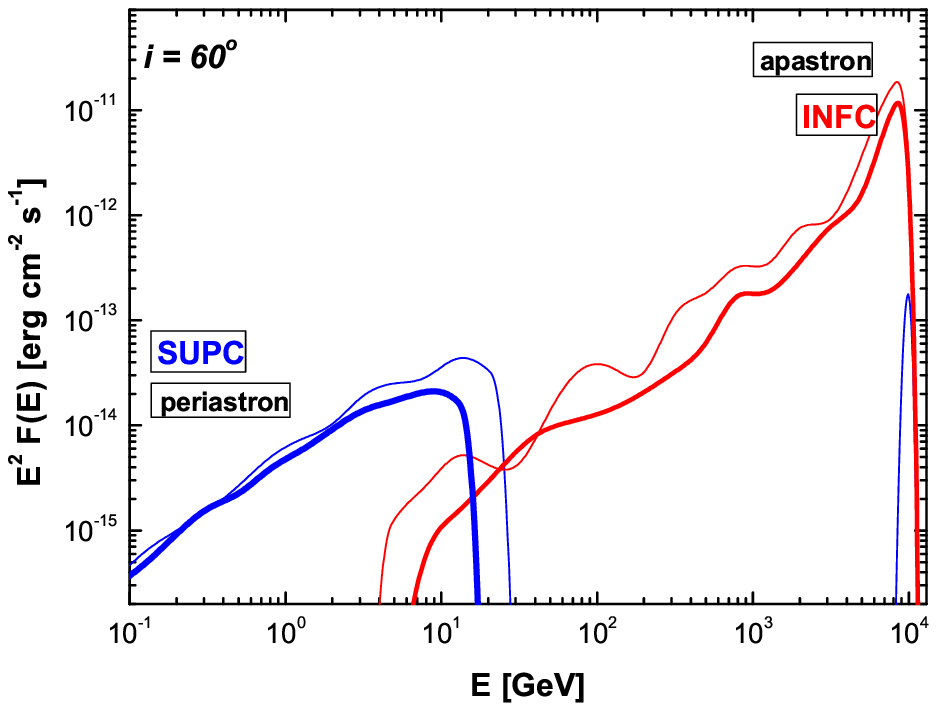}
  \includegraphics[width=0.49\textwidth,angle=0,clip]{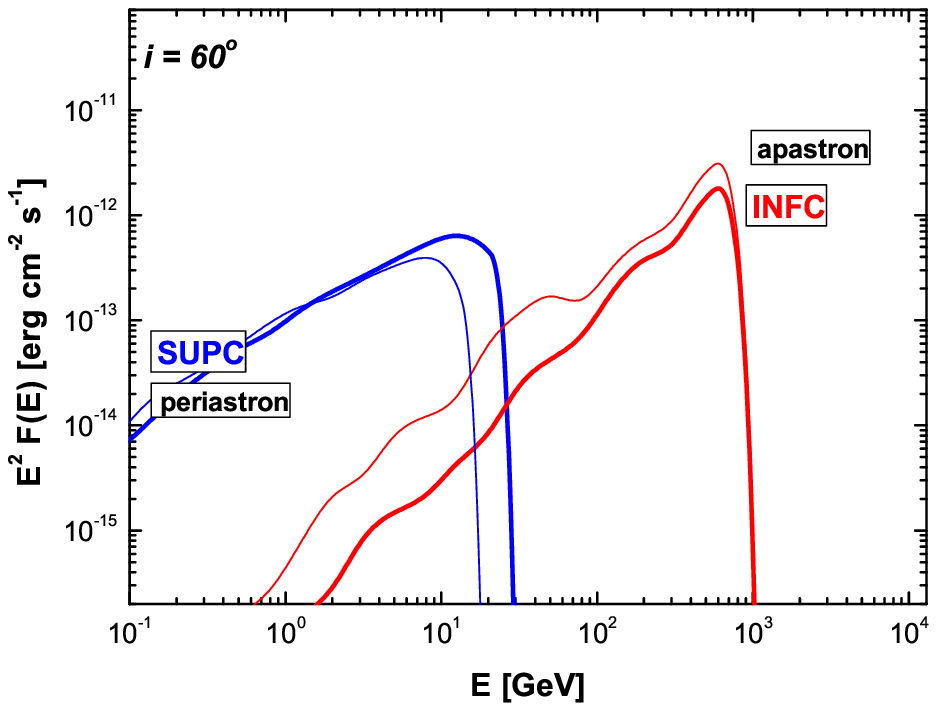}
\caption{\label{fig:mono_spectra} {VHE photon spectra for specific phases, at periastron, apastron (thin lines), INFC, and SUPC (bold lines), for two inclination angles, $i=30^o$ (top) and $i=60^o$ (bottom). The spectra were calculated for two different primary energy of electrons, $E=10$ TeV (left) and $E=1$ TeV (right). The few free parameters involved in the mono-energetic model and their assumed values are given in Table \ref{mono-values}. For direct comparison, the same normalization factor was used in case of each injection energy. Discussion on the non-uniqueness of model parameters is given in Section 2.}}
\end{figure*}

\begin{table}
\centering
\caption{\label{tab:par_mono} Model parameters for the example of mono-energetic injection considered }
\vspace{0.2cm}
\begin{tabular}{lll}
\hline
{Meaning} & {Symbol} & {Adopted value}  \\
\hline
Spin-down power of assumed pulsar & $L_{\rm sd}$ & $10^{37}$ erg s$^{-1}$\\
Fraction of $L_{SD}$  in relativistic electrons & $\beta$ & $10^{-2}$  \\  
The energy of injected pairs  & $E_0$ & 10 TeV \\
\hline
\end{tabular}
\label{mono-values}
\end{table}

\begin{figure*}[t]
  \centering
  \includegraphics*[width=0.49\textwidth,angle=0,clip]{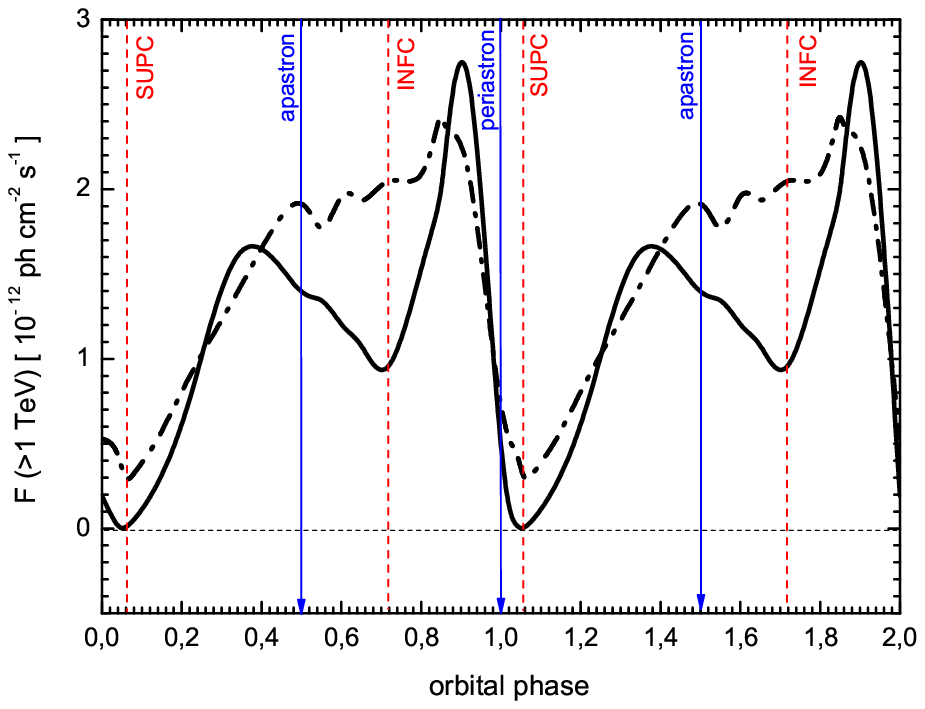}
  \includegraphics*[width=0.49\textwidth,angle=0,clip]{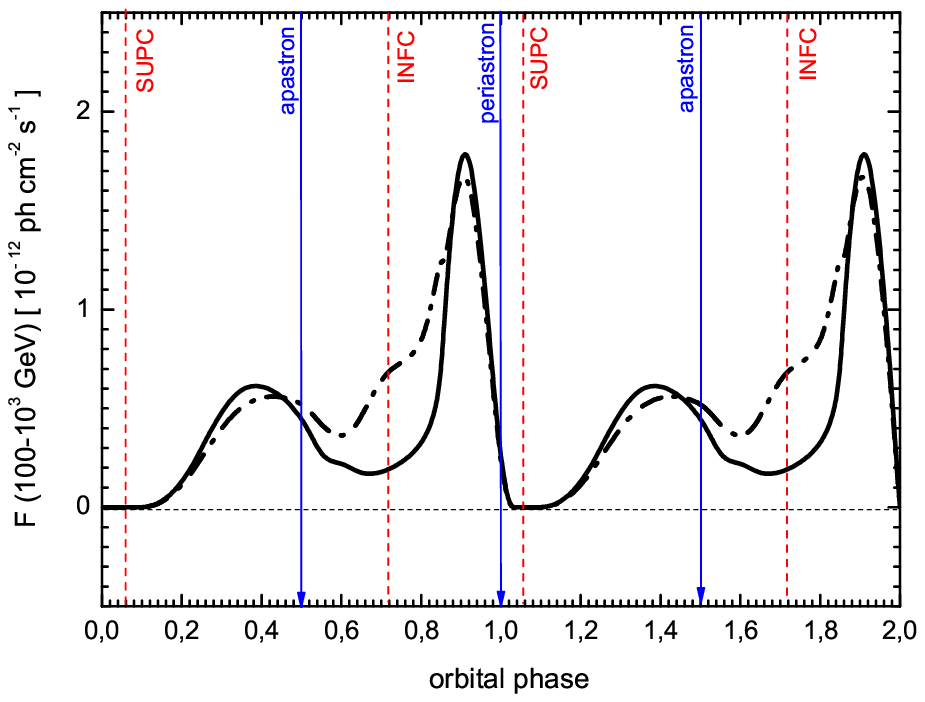}\\
  \includegraphics*[width=0.49\textwidth,angle=0,clip]{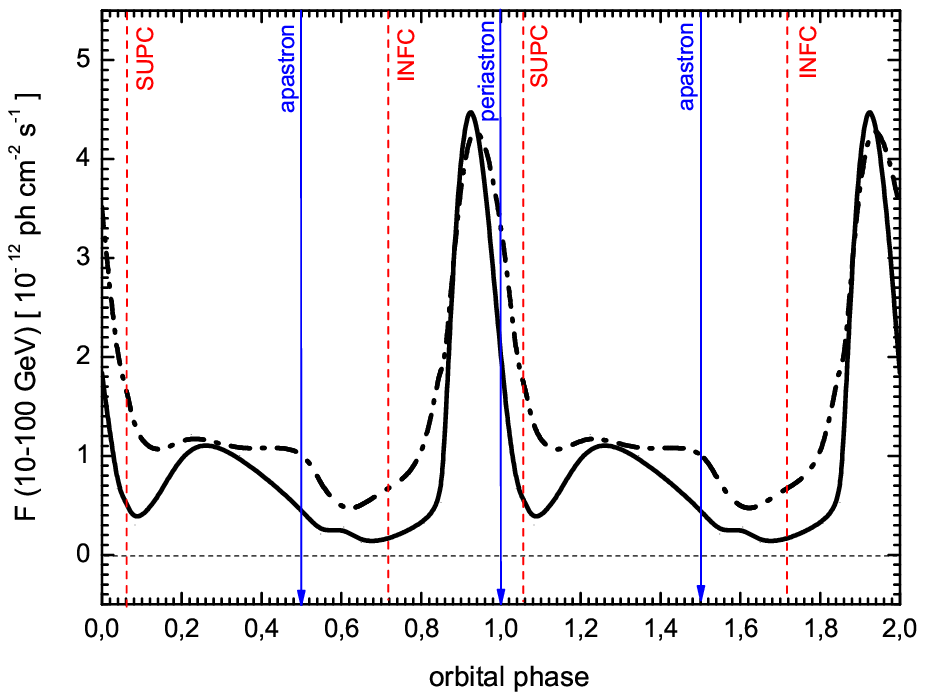}
  \includegraphics*[width=0.49\textwidth,angle=0,clip]{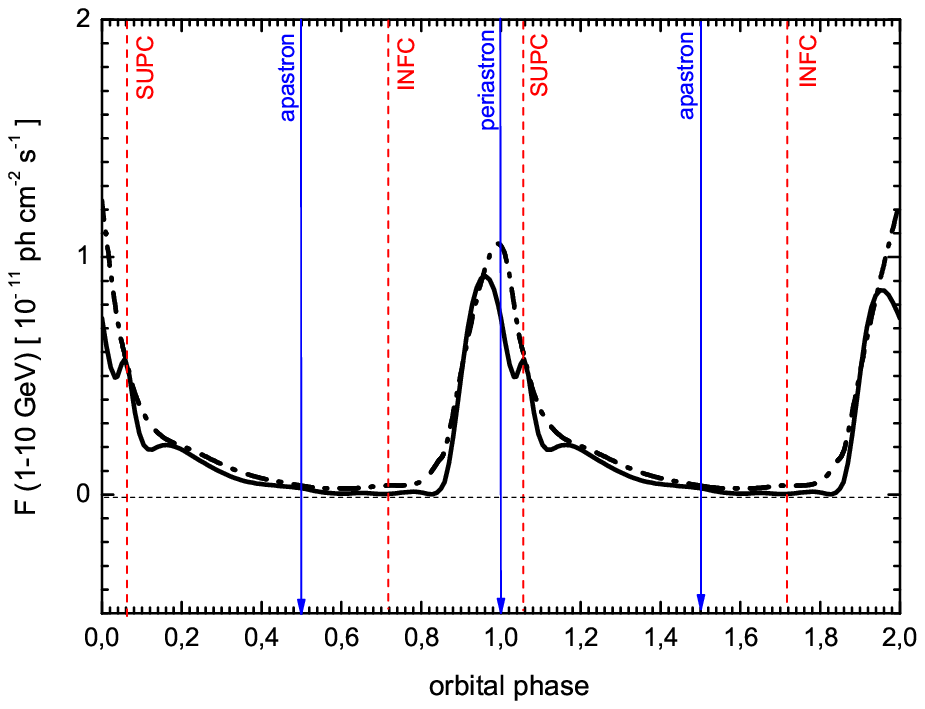}
\caption{\label{fig:mono_LC} {Theoretical lightcurves for mono-energetic injection and two different inclination angles. As before: i=60$^0$, solid; i=30$^0$, dot-dashed. }}
\end{figure*}

Based on the simulations for the specific phases along the orbit, the photon spectra and lightcurves in different energy ranges ($10-100$ GeV, $10-10^3$ GeV, and above $1$ TeV) were calculated for two assumed inclinations of the orbit of the system, $i=30^o$ and $i=60^o$. The highest energy range corresponds to the data presented by H.E.S.S. (Aharonian et. al. 2006). To allow for a comparison, the photon spectra were calculated also for a primary electron energy of $1$ TeV for specific orbital phases (periastron, apastron, SUPC and INFC).

The photon spectra (SED) are presented in Fig. \ref{fig:mono_spectra}. All the spectra for primary electron energy $E=10$ TeV are hard, presenting a photon index $< 2$, and they have significantly higher flux at the highest energy range, close to the initial primaries' $\sim 10$ TeV, as the ICS of primary pairs mainly occur in Klein-Nishina range. Below this peak the spectra are well represent by the power-law with photon index close to $\sim -1$, albeit one can also notice the change in the photon index from SUPC to INFC (harder spectrum) and the notably different energy cutoffs. These features are discussed below.

While for phases around SUPC the PWZ is limited in the direction to the observer, for the opposite phases the PWZ is unterminated. On the other hand, for INFC the angle to the observer is at its minimum (and also the separation of the system is larger) what causes the decreasing of optical depth to both processes considered. In contrary to the INFC phases, those at SUPC
present strong absorption features at an energy range from $\sim 0.1$ to few TeV, which cause the spectra to be cut at lower energies. 
This is due to the dependence of the optical depth to $\gamma\gamma$ absorption on the energy of photon and the influence of geometry in defining the photon path towards the observer. The larger the angle to the observer, the stronger the effects of absorption are, what can be seen from comparison of the results for two inclination angles (SUPC and periastron phases). Notice, that for $i = 60^o$ the angle to the observer attains the largest value of the discussed examples, $\alpha_{obs} = 150^o$. The absorption for SUPC phases takes place already in the PWZ, what can initiate the cascading processes. Photons which pass through the shock region can be also absorbed in the MSWZ. 
The opacities for $\gamma\gamma$ absorption also strongly depend on photon energy (see Fig. \ref{fig:tau_E}). High energy photons, with energy from few GeV up to few TeV, are the most likely to be absorbed, as the opacities decrease with energy for all shown angles of the photon injection. 
Indeed, the number of the photons absorbed in MSWZ is about $\sim 10 - 20$ \% of the number of photons reaching the shock region and the effect is significant in the final spectra. On the other hand, the cascading in PWZ produce lower energy leptons and photons (in the cooling process of secondary pairs) and this causes a higher photon flux at energies up to few 10 GeV. The processes in PWZ are limited by the presence of the shock, what influence the photon production rate. For INFC phases the absorption of $\gamma$-rays is minor, as the photons propagate outwards of the massive star; once produced, most photons can escape from the system (also because of the threshold to pair production). Higher energy photons are produced mostly in the first IC interaction of the primary pairs, while further electron cooling, not limited by the termination shock, supply the spectrum in lower energy photons. This flux is not as high as in the case of SUPC phases, where most of the cascading takes place. 

Similar dependencies, both for INFC and SUPC phases, are present in the spectra obtained from the simulations for primary energy of pairs $E = 1$ TeV. As the optical depths to IC scattering are higher in that case, the processes are more efficient and the number of produced photons are higher. To give an example, most photons of energy $>$ 20-30 GeV at SUPC-periastron phase are absorbed in MSWZ.

Comparing the photon spectra produced at SUPC and INFC, an anticorrelation between GeV and TeV photon fluxes is evident, at least comparing the spectra at energies below and above $\sim 10$ GeV. All these effects play a role in the formation of the $\gamma$-ray lightcurve, shown in Fig. \ref{fig:mono_LC}. 
The lightcurve in the highest energy range ($>1$ TeV) has a broad minimum around SUPC, $0.96 < \phi < 0.25$. From the comparison of the position of the termination shock along the orbit, the opacities to ICS and $\gamma\gamma$ absorption we see that this minimum is formed despite being at phases with the highest opacities to both processes considered and so having an effective cascading in the PWZ region. 
This minimum is mainly due to the absorption of the photons which get through the shock and propagate into the massive star wind region. The broad maximum in the high-energy lightcurve is formed at opposite phases around INFC, $0.4 < \phi < 0.9$. The maximum for higher inclination is also characterized by a local minimum close to the INFC phase, what is the result of the IC opacity dependence on the propagation angle. 
The range of INFC phases corresponds to the unterminated pulsar wind in the observer direction. The opacities for both inclinations have local peaks in this range, what was discussed in the previous paragraph. The local minimum in photon flux within the INFC range reflects a similar behavior of the opacities. So, as far as the propagation of the particles in the PWZ is unterminated, the high-energy lightcurve formation is in agreement with the dependence of the optical depths. Additionally, we can also see that the first local peak in this lightcurve is formed earlier in phase for inclination $i=60^o$ ($\phi \sim 0.35$), what is also in agreement with the dependence shown by the distance to the termination shock. Note that the second peak in the broad TeV lightcurve maximum, at $\sim 0.85$, is higher than the first one, at $\sim 0.4$ what is the result of a change in the separation of the system together with the angle to the observer.

\subsection{Peaks and dips in the lightcurve}

The geometrical conditions for leptonic processes change significantly along the orbit, and they are more efficient for phases around SUPC. However, these processes, as we can already see from the opacities dependence on the orbital phases, is limited by the presence of the termination shock. 
This causes that the best conditions for very high-energy photon production, and finally escaping from the system, occur for an specific combination of the angle to the observer, the separation of the system, and the distance to the shock from the pulsar side. From Figs. \ref{fig:tau_phi} and \ref{fig:shock_sep} we can see that this happens at the phases  $\sim 0.3 - 0.4$ and $\sim 0.9$ what reflects in the TeV lightcurve. 
At phases close to INFC photons are produced mainly in the primary $e^+e^-$ pairs cooling when the propagating electron undergo frequent scatterings. Together with the fact that there is no efficient absorption of photons once they are produced, they finally escape from the system, what yields to the broad maximum in the TeV lightcurve. 
On the other hand, at phases around SUPC, high-energy photons are absorbed when propagating through the system in the MSWZ, what causes a dip in the lightcurve. For these phases many more lower energy photons are produced in cascades in the PWZ, what yields to a maximum in the GeV lightcurve for SUPC, anticorrelated with the behavior at TeV energies.


\section{A power-law electron distribution}

For further exploration of the $\gamma$-ray production model we set a new assumption: the energy distribution of the interacting $e^+e^-$ pairs is given by a power-law spectrum.  We will additionally assume that the power-law may be constant or vary along the orbit.
Motivated by the different observational behavior found, we have assumed that two different spectral indices correspond to the two broad orbital intervals proposed by HESS (Aharonian et. al., 2006). For direct comparison we have specified the interval around the inferior conjunction (with the apastron phase): $0.45 < \phi < 0.90$, and around superior conjunction (including the periastron phase): $\phi < 45$ and $\phi > 0.90$ as being bathed by different electron distributions. 
The results for power-laws distribution were already summarily presented in our earlier work Sierpowska-Bartosik and Torres (2007a,b) and we refer to these works for further details. The agreement in both spectra and lightcurve is notable (particularly for the case of a variable lepton spectrum along the orbit), as can be seen in Figs. \ref{fig:spec_var} and \ref{fig:lc_var} discussed below.

\begin{table}
\centering
\caption{\label{tab:param} Model parameters for interacting electrons described by power-laws}
\vspace{0.2cm}
\begin{tabular}{lll}
\hline
{Meaning} & {Symbol} & {Adopted value}  \\
\hline
Spin-down power of assumed pulsar & $L_{\rm sd}$ & $10^{37}$ erg s$^{-1}$\\
VHE cutoff of the injection spectra & $E_{\rm max}  $ & $50$ TeV \\
\hline
{\bf Constant lepton spectrum along the orbit}\\
\hline
 Fraction of $L_{\rm sd}$ in leptons & $\beta$ & $10^{-2}$ \\  
 Slope of the power-law  & $\Gamma_e$ & $-2.0$ \\
 \hline
{\bf Variable lepton spectrum along the orbit}\\
\hline
 Fraction of $L_{\rm sd}$ in leptons at INFC  interval & $\beta$ &  $8.0 \times 10^{-3}$ \\  
 Slope of the power-law at INFC  interval& $\Gamma_e$ & $-1.9$ \\
  Fraction of $L_{\rm sd}$ in leptons at SUPC  interval & $\beta$ & $2.4 \times 10^{-2}$ \\  
Slope of the power-law  at SUPC  interval & $\Gamma_e$ & $-2.4$ \\
\hline
\end{tabular}
\label{spectrum-values}
\end{table}

\begin{figure*}
  \centering
  \includegraphics*[width=0.49\textwidth,angle=0,clip]{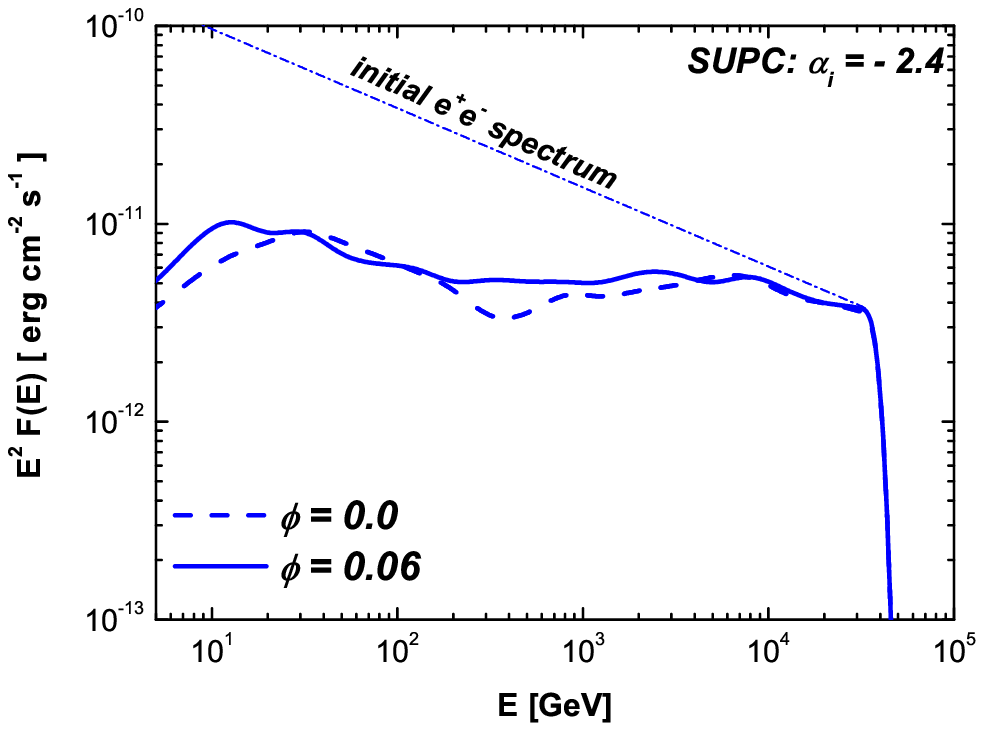}
  \includegraphics*[width=0.49\textwidth,angle=0,clip]{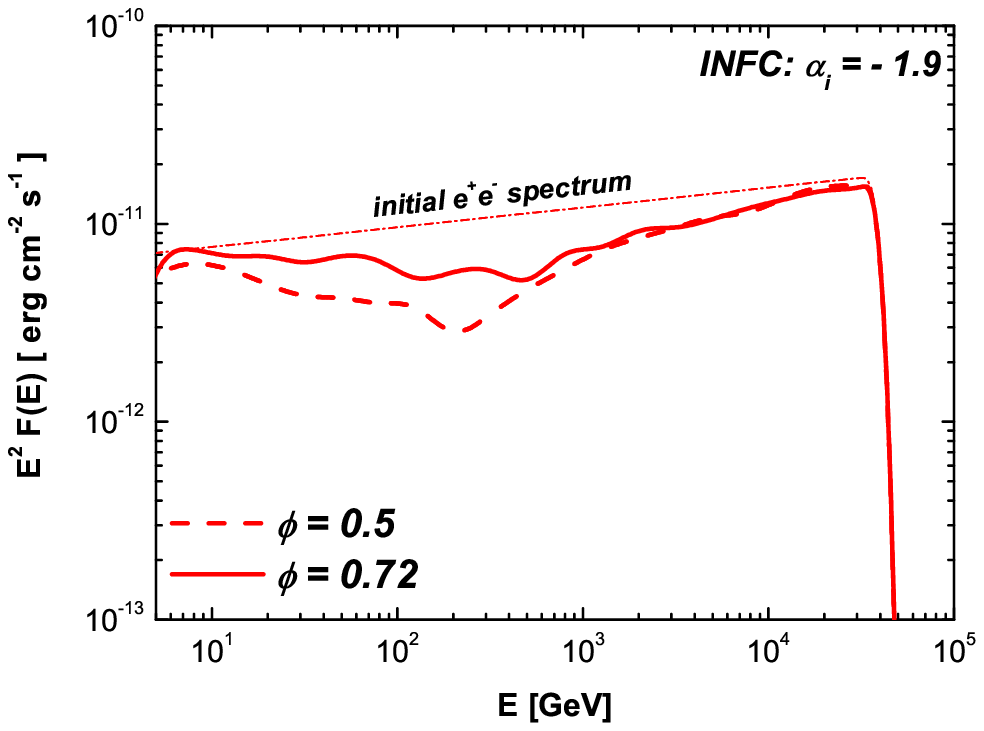} \\
\caption{\label{fig:e_spec_var} {Comparison of the spectra of electrons injected in the PWZ with the corresponding spectrum of electrons which reach the shock region (SUPC, periastron) or leave the innermost part of the PWZ after propagation in it (INFC, apastron) in case of non-terminated pulsar wind. Left: the electron spectrum injected at the SUPC and periastron (PER) with the initial index $\Gamma_e = -2.4$ (dot-dashed line) and spectrum for electrons reaching the shock: at SUPC (solid line) and PER (dashed line). Right: the electron spectrum injected at the INFC and apastron (APA) with the initial index $\Gamma_e = -1.9$ and spectrum for electrons after interaction in the local radiation field (the shock for this phases is not terminated in the direction to the observer): at INFC (solid line) and APA (dashed).} }
\end{figure*}

{ 
As we could have already noticed from Fig. \ref{fig:tau_gen}, the optical depth for high energy electrons is below unity. In that case, part of the initial electrons will be interacting in the PWZ less efficiently and finally will reach the shock region (we remind that in the direction of the observer, there is not always a shock in the electron's propagation). In Fig. \ref{fig:e_spec_var} we show the spectra of initial electron distribution and corresponding electron spectra after propagation in the PWZ. The electrons which reach the termination shock are isotropised there. The termination shock shape is specific for each phase and it is limited in space. The electrons at the shock, locally re-accelerated,  become the initial spectra for the next generation of photons (not only radiative but adiabatic losses have to be taken into account). 

The MSWZ can play a role (although we believe it will not be too important in such close binaries like LS 5039, because of the high opacities already encountered in the PWZ along most of the orbit and  the loss of directionality --consequently, of random photon emission-- of the electron population). 
An estimation of the contribution of the cascades in MSW is not trivial, especially from the normalization point of view, as the magnetic filed in this region causes isotropisation of produced photons. In addition, the magnetic field of the massive star is ordered, what could cause additional effects e.g. focusing the propagating electrons in some regions close to the massive star (where the magnetic field is dipolar) as was shown in Sierpowska \& Bednarek (2005). The impact of the MSWZ could make the need for a change in the injection index of relativistic particles less severe.  A 3D cascading code is needed for such estimates, and we expect to report on that in the future.
}

\begin{figure*}
  \centering
  \includegraphics*[width=0.49\textwidth,angle=0,clip]{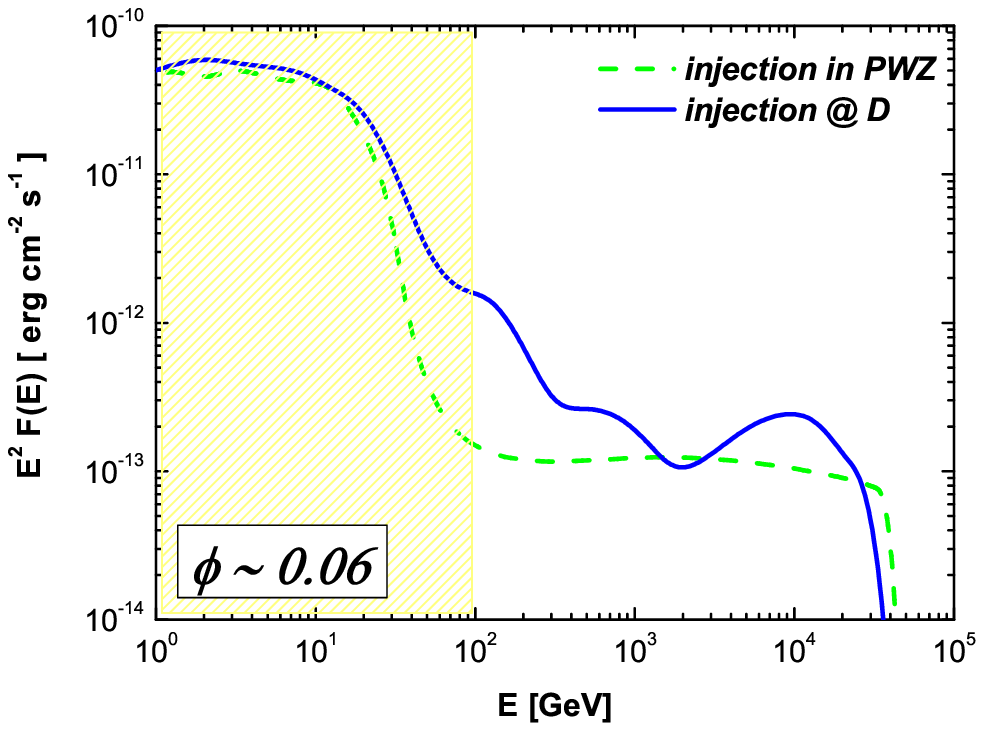}
  \includegraphics*[width=0.49\textwidth,angle=0,clip]{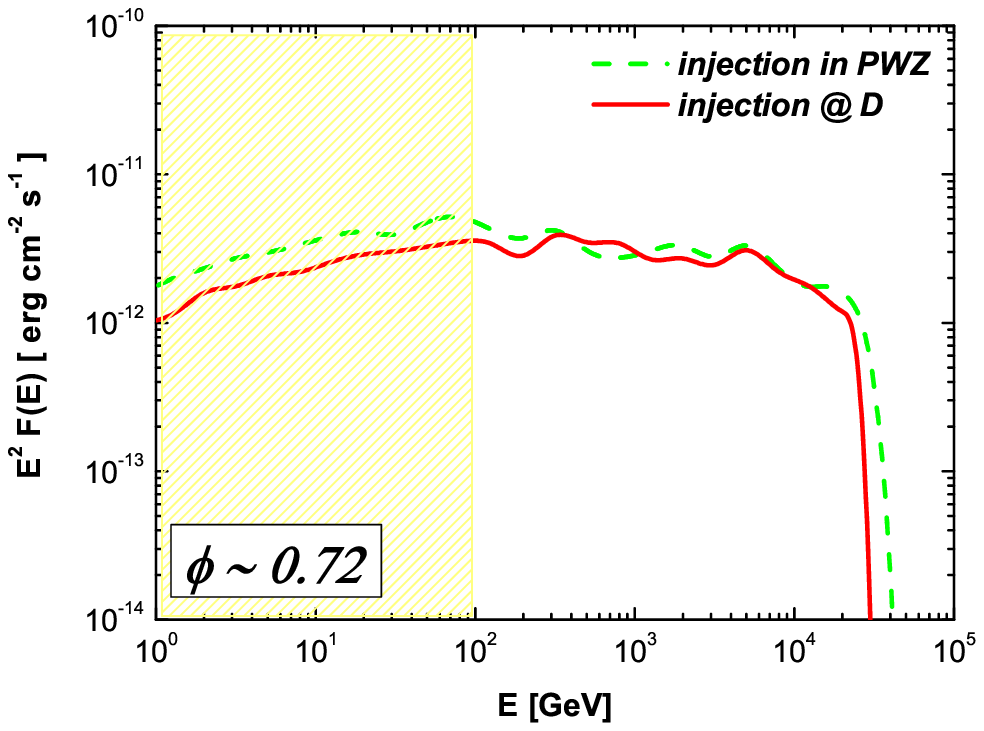} \\
\caption{\label{fig:f_spec_inpwz} {The spectra of photons escaping from the system produced by spectrum of electrons injected at separation distance (solid lines) and inside the PWZ (dashed lines) at SUPC (left figure) and INFC (right). Left: the photon spectrum produced at SUPC ($\phi \sim 0.06$) by electrons injected inside PWZ, at $r_{init} = r_{sh} / 2$ (the spectrum index $\Gamma_e = -2.4$). Right: the photon spectrum produced at INFC ($\phi \sim 0.72$) by electrons injected inside PWZ, at $r_{init} = D-R_s$ ($\Gamma_e = -1.9$) as the shock at INFC is not terminated in the direction to the observer.} }
\end{figure*}

{ 
The model assumes that the initial electron are injected in the vicinity of the pulsar light cylinder. This is actually an assumption  which we have investigated further. Indeed. we have investigate also the changes in the produced photon spectra if injection take place at a further distance inside the PWZ. Results are given in Fig. \ref{fig:f_spec_inpwz}. For phases where the wind is terminated in the direction to the observer (e.g., SUPC) the new injected radius was fixed to $r_{init} = D- r_{sh} / 2$ from the massive star, where $D$ is separation at given phase (i.e., at the middle of the PWZ). For phases such as INFC, where the shock is unterminated, the injection place was shifted by the radius of the star $r_{init} = D-R_s$, that also corresponds to the half-distance of the separation between the pulsar and shock $r_{sh}$ for this phase, if movement is in the direction to the massive star. We can see that the produced photon spectra do not differ significantly in this two scenarios, making our results stable.
}

\subsection{Lightcurve and broad-phase spectra}

\begin{figure*}
  \centering
  \includegraphics*[width=0.49\textwidth,angle=0,clip]{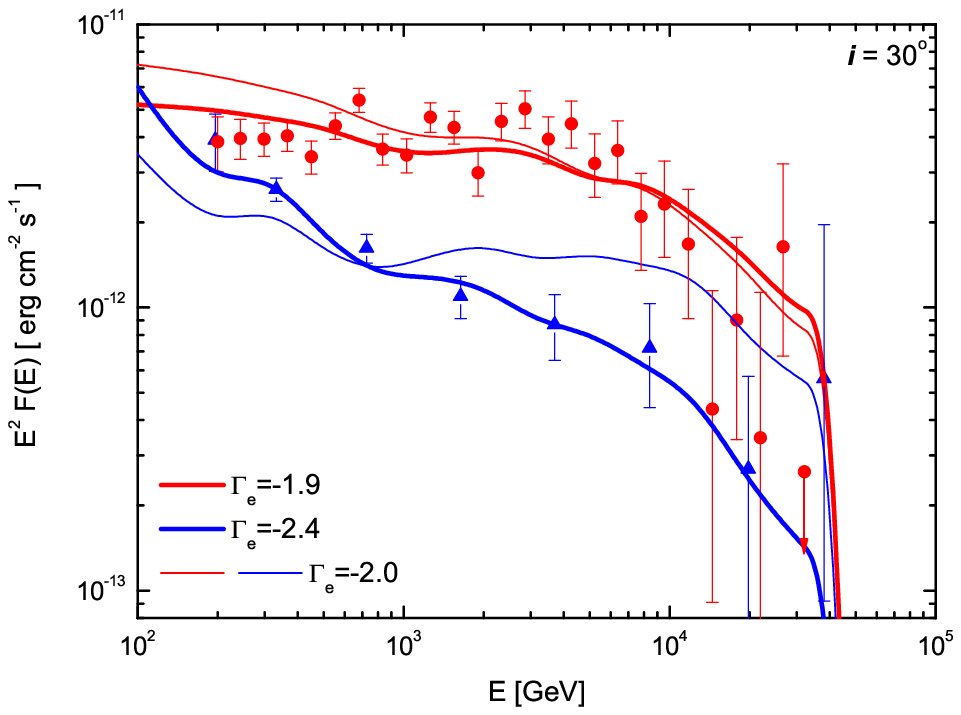}
  \includegraphics*[width=0.49\textwidth,angle=0,clip]{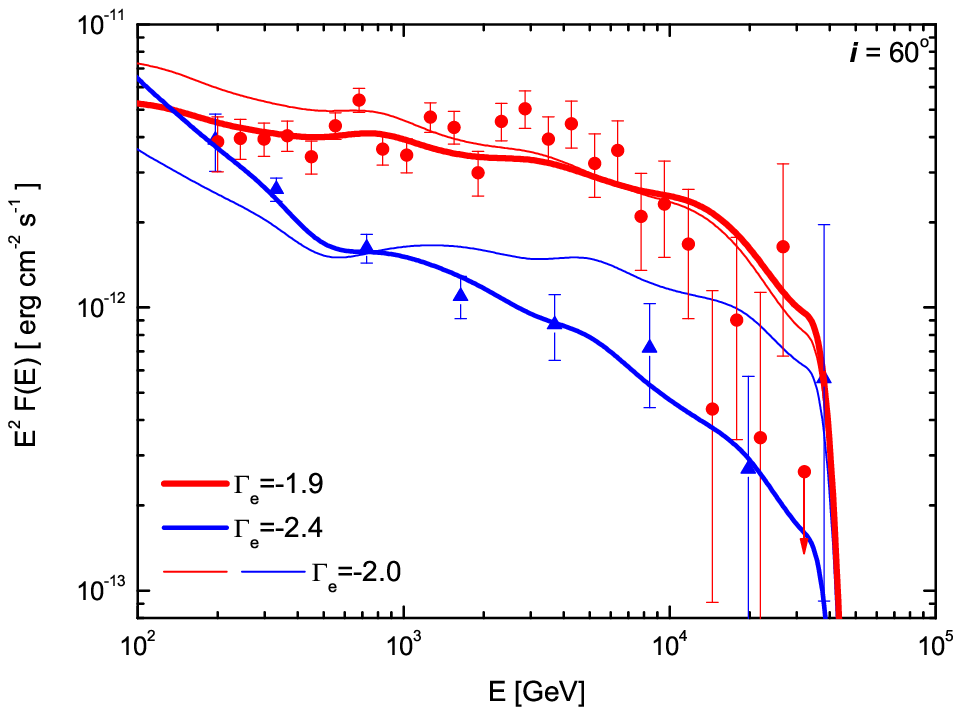} \\
\caption{\label{fig:spec_var} {Very high-energy spectra of LS 5039 around INFC and SUPC, together  with the theoretical predictions in equal phase intervals for power-laws electron distribution and two different inclination angles. The free parameters involved in our model and their assumed values are given in Table \ref{spectrum-values}.  After Sierpowska-Bartosik \& Torres (2008).}}
\end{figure*}

\begin{figure*}
  \centering
  \includegraphics*[width=0.49\textwidth,angle=0,clip]{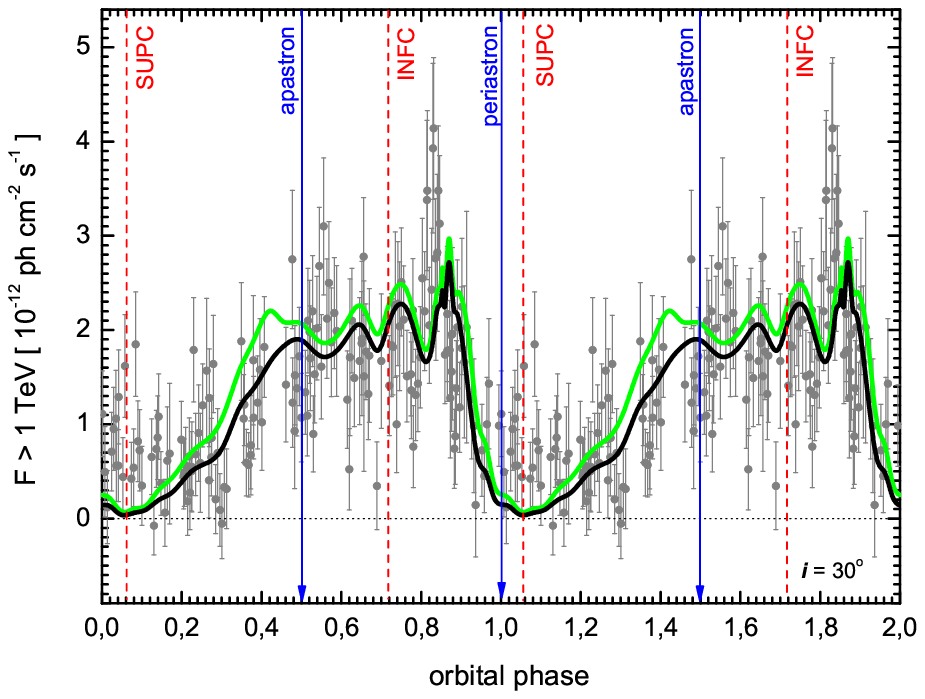}
  \includegraphics*[width=0.49\textwidth,angle=0,clip]{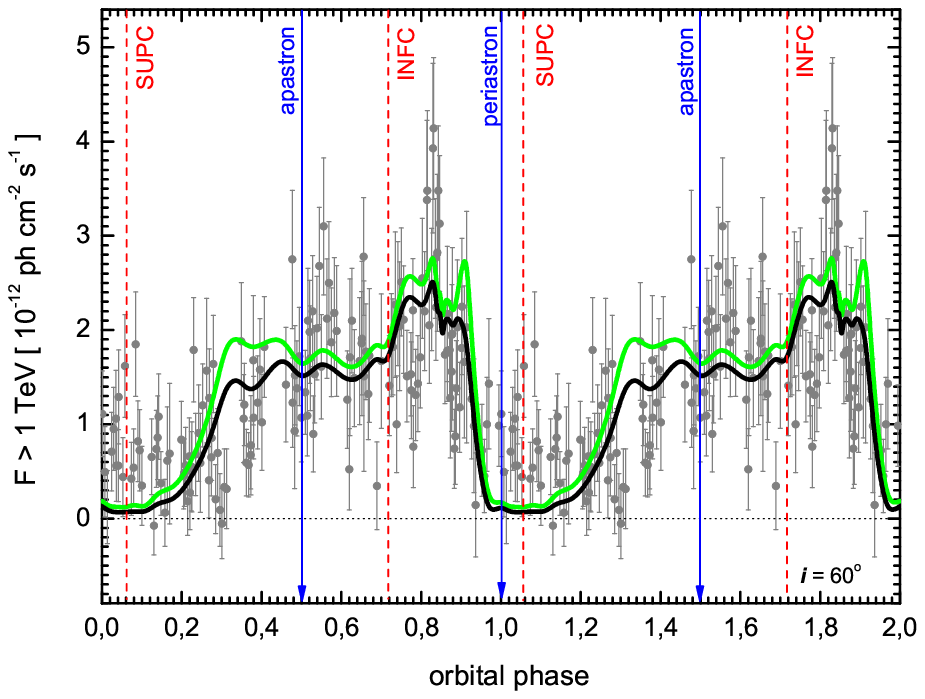} \\
\caption{\label{fig:lc_var} {
H.E.S.S. run-by-run folded (ephemeris of Casares et al. 2005) observations (each point corresponds to 28 min of observations) of LS 5039 with the results of the theoretical model for power-law distribution and two different inclination angles for i=30$^0$ (left) and i=60$^0$ (right). Light (green) lines stand for results obtained with a constant interacting lepton spectrum along the orbit, whereas dark lines (black) correspond to a variable spectrum. After Sierpowska-Bartosik \& Torres (2008).
}}
\end{figure*}

To construct the lightcurve, spectra for over 20 orbital phases were calculated to cover the whole orbit of LS 5039. 
The specific averaged spectra were obtained based on the same orbital intervals as presented in H.E.S.S. data.
The averaged spectra were obtained summing up individual contributions from orbital spectra in given phases, each with a weight ($\delta t/T$) corresponding to the fraction of orbital time that the system spends in the corresponding phase bin.
%
Then, comparing the observational and theoretical averaged spectra, the parameter $\beta$ was estimated. Both for constant and variable injection model, we get that the fraction of the spin-down power in the primary leptons has to be at the level of $\sim 1$ \%. With this parameter in hand each of the single spectra can be equally normalized. 

It is worth noticing how well these lightcurves compare
with those in the work by Bednarek (2007), at least for some of
the specific phases considered by him. Bednarek also included
cascading in his simulations, and the physical input of his model
(although in the case of a microquasar scenario) is similar to
ours. As a result, the anti-correlation phenomena (from GeV
to TeV energies) is also a result of his work. The spectrum
along the orbit with respect to H.E.S.S. datapoints and the possible
short-timescale
variability (see below) was
not provided by Bednarek, so that a comparison with these results is not
possible.
A detailed differentiation between these two models is a
key input for distinguishing (microquasar or $\gamma$-ray
binary) scenarios, even when some assumptions
are intrinsic to each of the models,  isolating the contribution of an equal physical input
can help decide on what object constitute the system LS 5039. As an example:
Bednarek did find in his model that a fixed inclination angle
($i$ = 60$^0$) was needed in order to reproduce the shape of the
H.E.S.S. lightcurve results, whereas in our case, as we see in
Figure \ref{fig:lc_var} the influence of inclination is minor.

\begin{figure*}
\centering
\includegraphics[width=.49\textwidth]{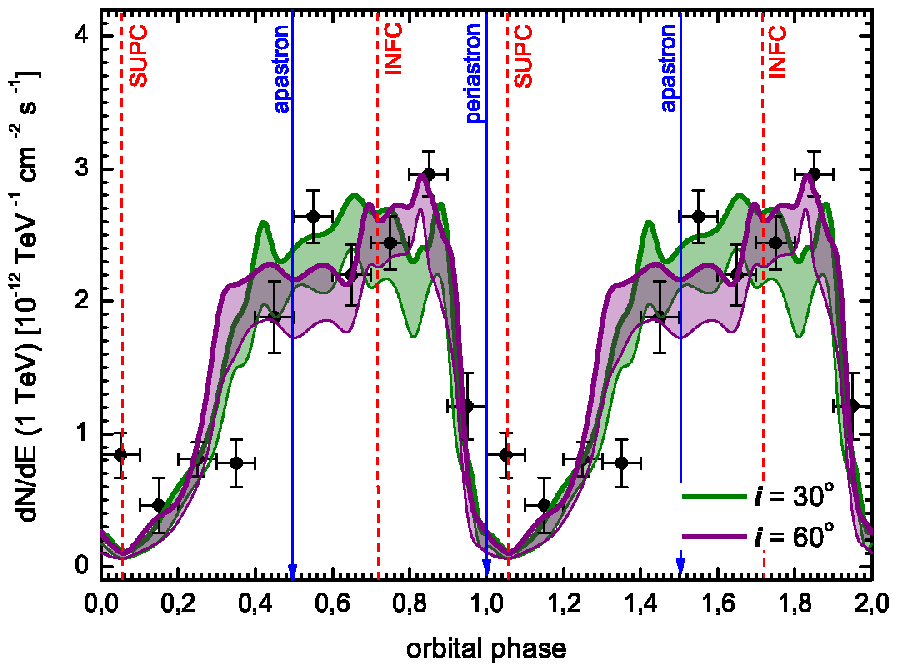}
\includegraphics[width=.49\textwidth]{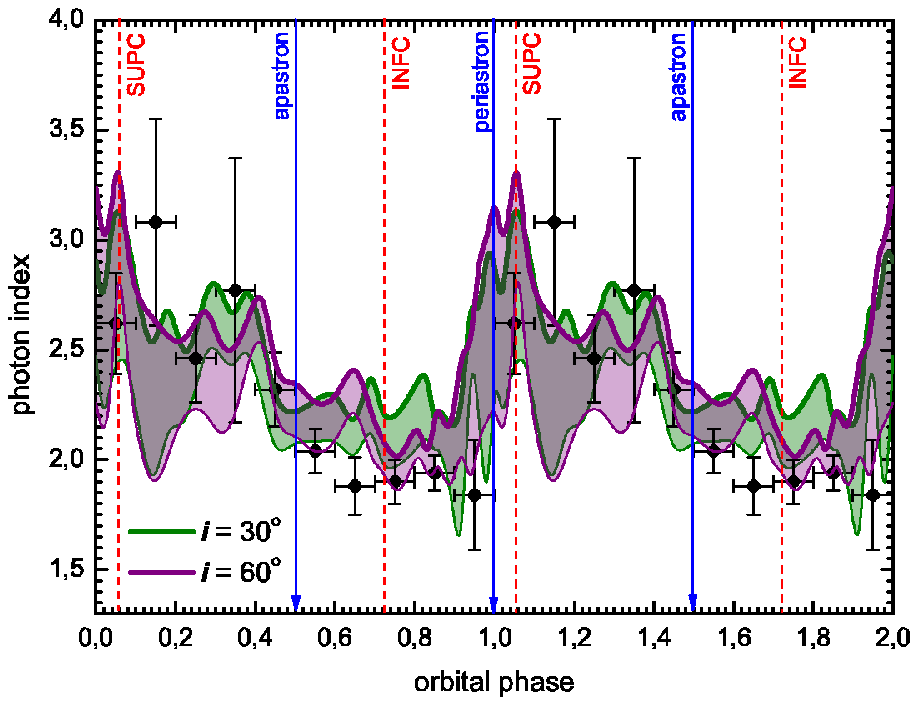}
\caption{Shaded areas in the left (right) panel show the change in the normalization (photon index) of a power-law photon spectra fitted to the theoretical prediction for each of the 0.1 bins of phase in the case of a variable lepton distribution along the orbit. The two different colors of the shading stand for the two inclination angles considered. The size of the shading gives account of the error in the fitting parameters. Data points represent the H.E.S.S. results for a similar procedure: a power-law fit to the observational spectra obtained in the same phase binning. From Sierpowska-Bartosik \& Torres (2008).}
\label{phase-bin}
\end{figure*}

For completeness, we mention that as noted above, H.E.S.S. has also provided the evolution of the normalization and slope of a power-law fit to the 0.2--5 TeV data in 0.1 phase-binning along the orbit (Aharonian et al. 2006). The use of a power-law fit was limited by low statistics in such shorter sub-orbital intervals, i.e., higher-order functional fittings such as a power-law with exponential cutoff were reported to provide a no better fit and were not justified. To directly compare with these results, we have applied the same approach to treat the model predictions, i.e., we fit a power-law in the same energy range and phase binning. We show here this comparison in Fig.  \ref{phase-bin} (taken from Sierpowska-Bartosik and Torres 2008) in the case of a variable lepton distribution along the orbit. We find a rather good agreement between model predictions and data. Results for constant lepton distribution along the orbit can be seen in Sierpowska-Bartosik and Torres (2007a).

In Fig. \ref{fig:spec_var} the SED in H.E.S.S. energy range are shown for both, constant and variable lepton distributions along the orbit, and two inclination of the system $i=30^o$ and $i=60^o$. 
It was shown in the previous Section that in case of the mono-energetic injection, there are photon flux differences in INFC phase due to the change in the angle to the observer. In the power-law  distribution model this effect is less significant due to the contributions of different energies, and the dependence on energy of the processes involved. Based for instance on the injection of constant spectrum with slope $\Gamma_e = -2.0$ we can see that there is no significant differences involving the inclination.

Exploring the more evolved model of the variable injection we see that the steeper the primary spectrum is, the higher the photon flux at lower energy range (SED for lower energies were shown by Sierpowska-Bartosik and Torres 2008). For constant lepton distribution along the orbit, the differences in photon flux below 100 GeV are not so significant, but the difference gets important in the case of a variable lepton distribution (the difference between both models in the lower energy range is about one order of magnitude). Overall, the variable lepton distribution provides a better agreement with all data (note that it has only two extra free parameters when compared with the constant distribution case, see Table \ref{spectrum-values}, but matches more than 10 data points that were missed in the previous case). In particular, it is worth noticing  that there is good agreement with the H.E.S.S. spectra at an energy $\sim 200$ GeV where both SUPC and INFC spectra coincide and above which the photon flux for INFC is higher then the flux for SUPC dominating in the lower energy range; i.e., the energy where the anticorrelation begins (see Sierpowska-Bartosik and Torres 2008). It is also interesting to remind that 
absorption alone would produce strict modulation (zero flux) in the energy range 0.2 to 2 TeV whereas the observations show that the flux at $\sim 0.2$ TeV is stable. Additional processes, i.e. cascading, must be considered to explain the spectral modulation. Our model have these processes consistently included and the lightcurve details arise then as the interplay of the absorption of $\gamma$-rays with the cascading process, in the framework of a varying geometry along the orbit of the system.

\subsection{Testing with future data}

\begin{figure*}
\centering
\includegraphics[width=.32\textwidth]{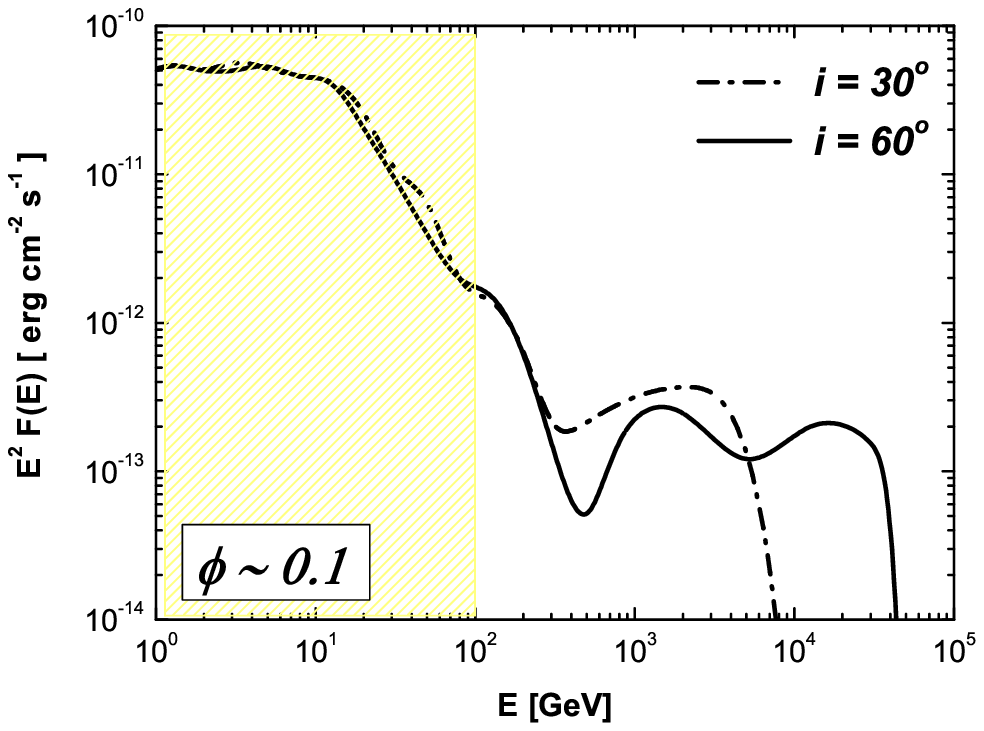}
\includegraphics[width=.32\textwidth]{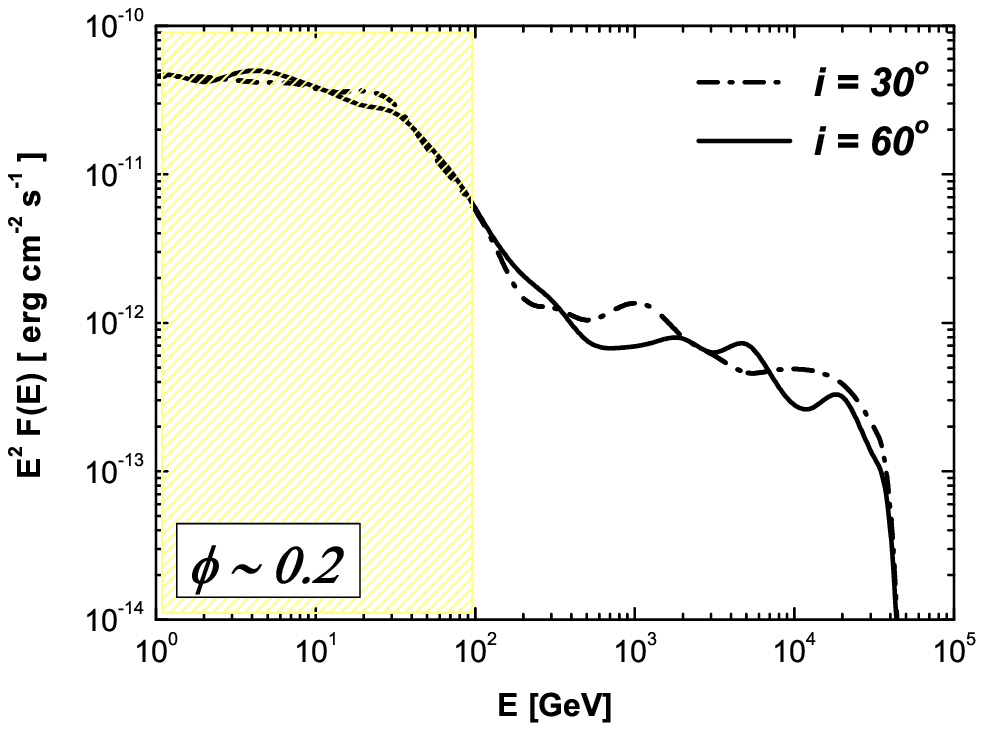}
\includegraphics[width=.32\textwidth]{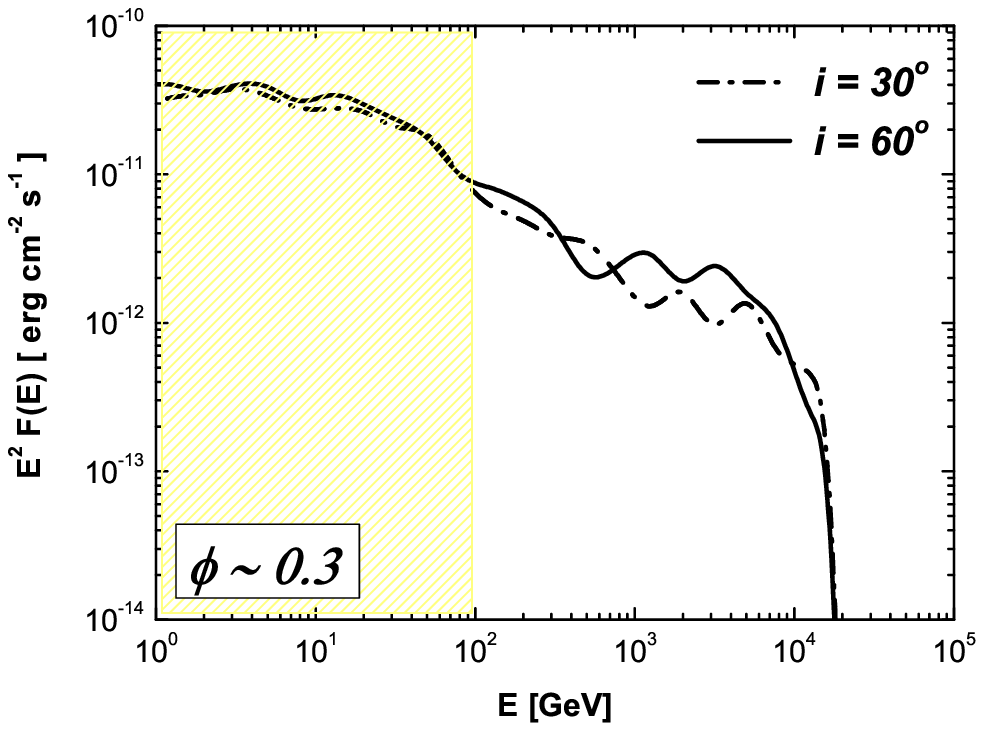}\\
\includegraphics[width=.32\textwidth]{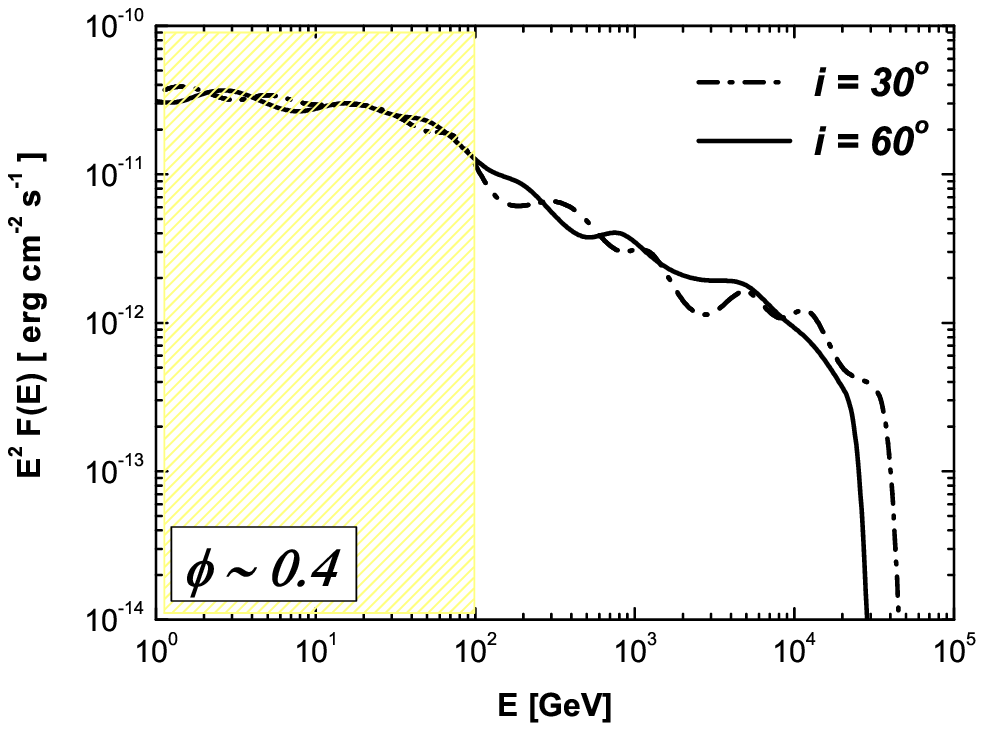}
\includegraphics[width=.32\textwidth]{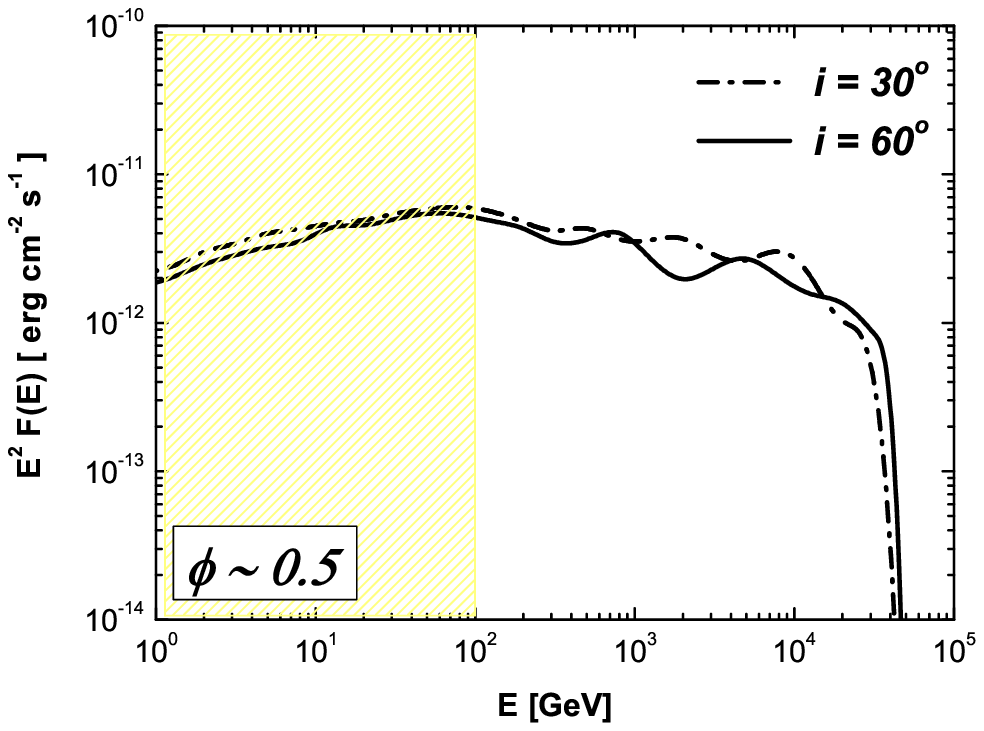}
\includegraphics[width=.32\textwidth]{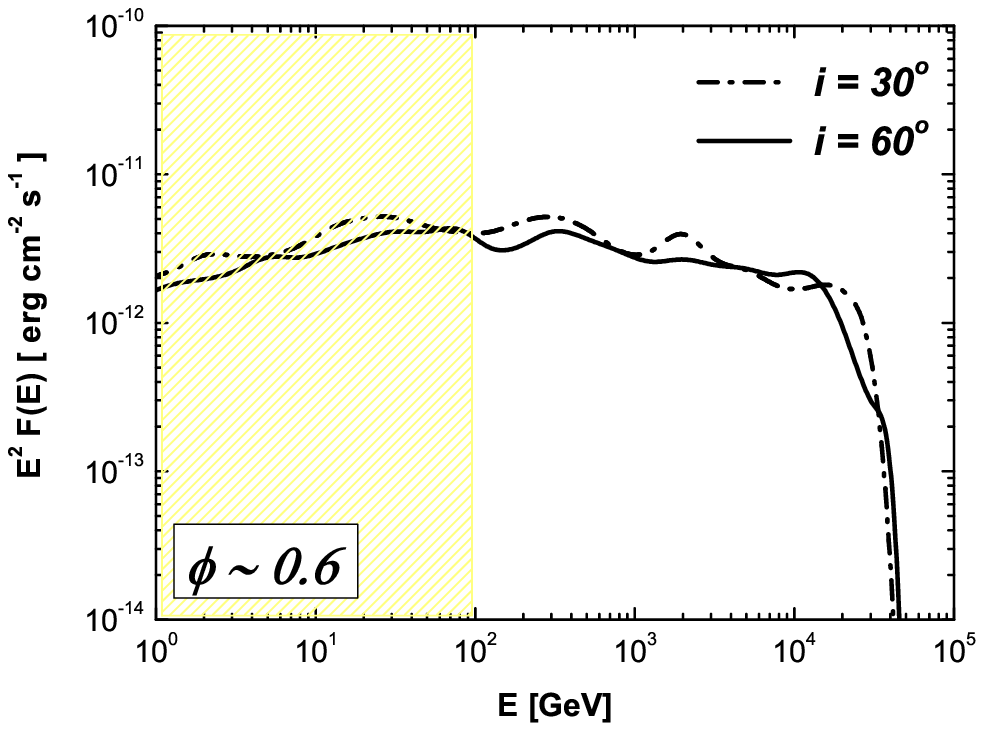}\\
\includegraphics[width=.32\textwidth]{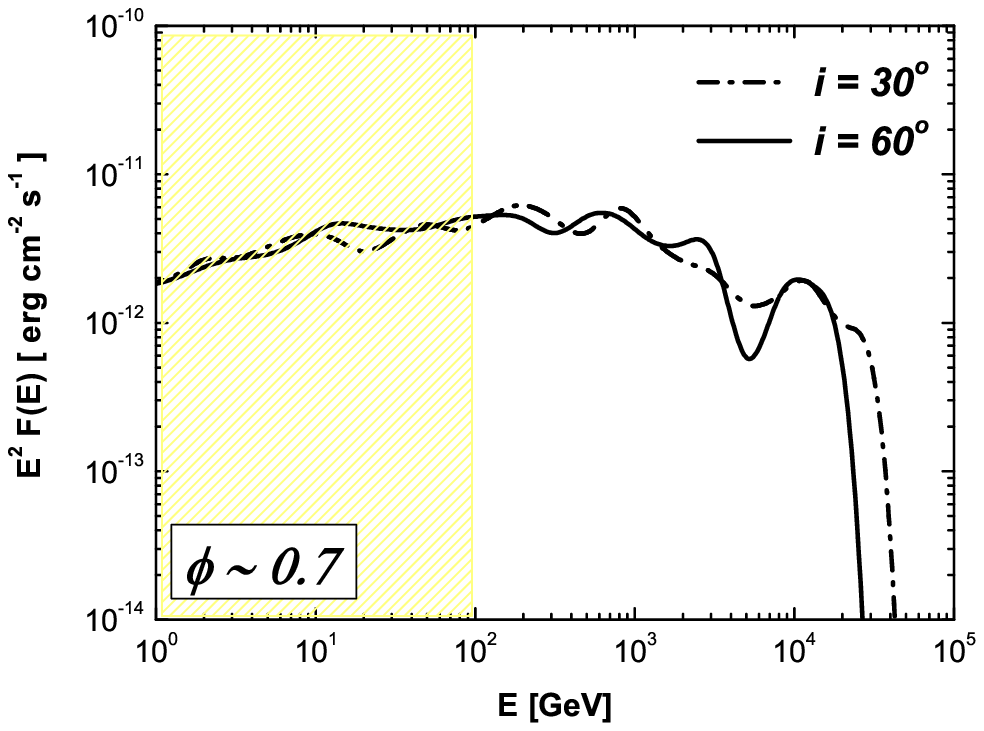}
\includegraphics[width=.32\textwidth]{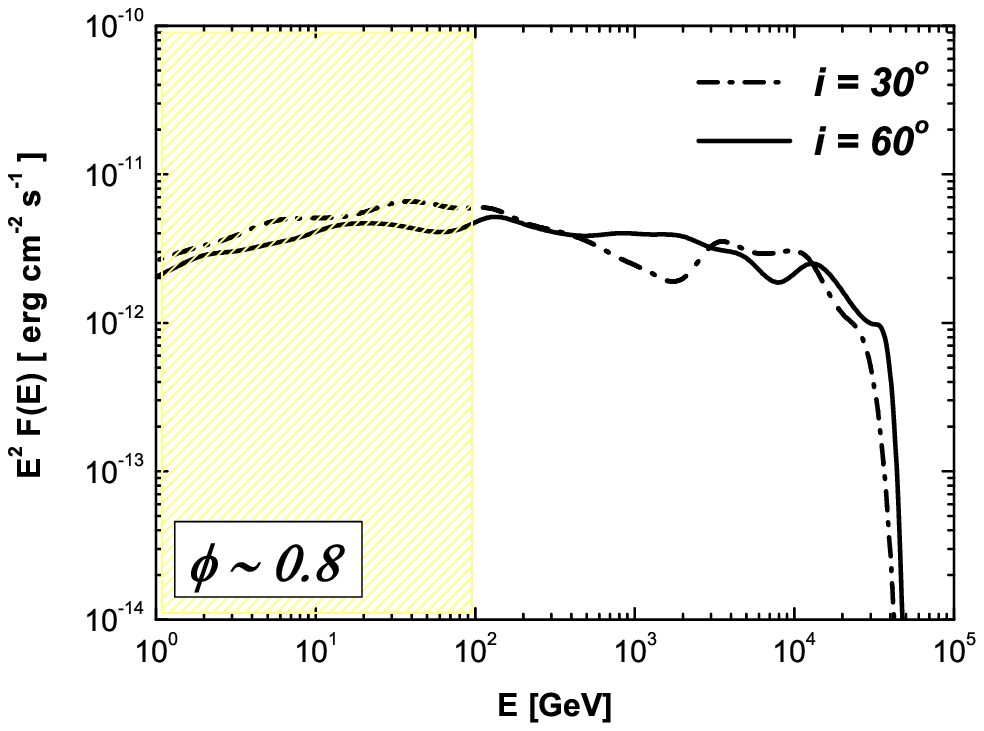}
\includegraphics[width=.32\textwidth]{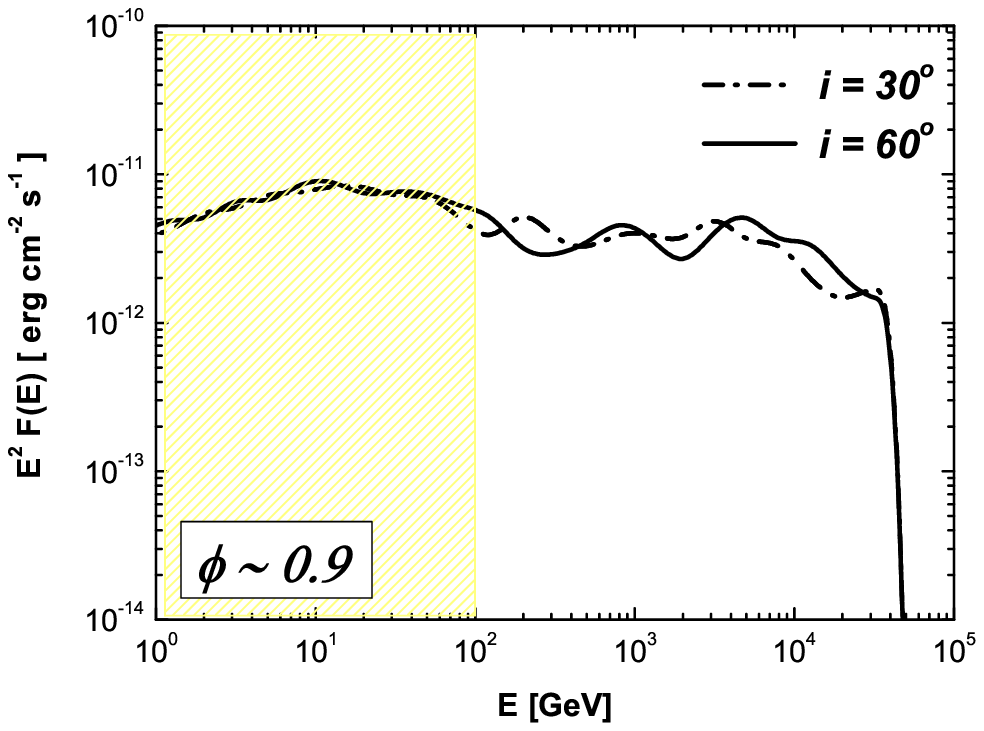}\\
\caption{Evolution of individual spectrum in the case of variable lepton distribution along the orbit, at individual phase bins, from 0.1 to 0.9. The shadow represents the GLAST energy coverage; the rest of the span of the x-axis can be observed by ground-based facilities.}
\label{phase-bin2}
\end{figure*}

Apart from possible testing with GLAST (at the level of lightcurve, spectra, hardness ratios, and differentiation between constant and variable electron distribution, see Sierpowska-Bartosik and Torres 2008), we can provide further possible tests at high $\gamma$-ray energies.
It was already said that power-laws do not always present the best fit to the specific spectra along the orbit. Even when fitting such power-laws to the theoretical predictions provides  agreement with data (see Fig. \ref{phase-bin}), observations with larger statistics (with H.E.S.S., H.E.S.S. II, or CTA) could directly test the model in specific phases. A model failure in specific phases would allow further illumination about the physics of the system.
Figure \ref{phase-bin2} shows the evolution of individual spectrum in the best fitting case of variable lepton distribution along the orbit, at individual phase bins, from 0.1 to 0.9, for testing with future quality of data.

\section{Concluding remarks}

We presented the details of a theoretical model for the high-energy emission from close $\gamma$-ray binaries, and applied it to the particular case of LS 5039. 
The model assumes a pulsar scenario, where either the pulsar or a close-to-the-pulsar shock injects leptons that after being reprocessed by losses to constitute a steady population, are assumed to interact with the target photon field provided by the companion star within the PWZ. 
The model accounts for the highly variable system geometry with respect to the observer, and radiative processes; essentially, anisotropic Klein-Nishina ICS and $\gamma\gamma$ absorption, put together with a Monte Carlo computation of cascading.
The formation of lightcurve and spectra in this model was discussed in detail for the case where the interacting leptons are assumed mono-energetic and described by power-laws. 

Comparing the interacting models of mono-energetic leptons and power-law distributions we can see similar dependencies for the spectral changes along the orbit and the GeV to TeV lightcurves. Thus, the case of a mono-energetic population is useful to understand some of the aspects regarding the formation of the observational features, although it does not match observational data. For the mono-energetic case, we have shown the spectra produced at characteristic phases (INFC, SUPC, periastron, apastron) for primary energies of 1 TeV and 10 TeV. In the power-law distribution model, the specific features of the photon spectra and the lightcurves produced for an specific primary electron energy  overlap. We can still notice the absorption features in the spectra produced at SUPC. We have discussed effects solely based on the optical depths and the general geometrical dependence along the orbit, as well as on the presence of the shock which terminates the pulsar wind in the direction to the observer at SUPC phases. 

A power-law lepton distribution interacting in the PWZ
describes very well the phenomenology
found in the LS 5039 system at all timescales, both flux and spectrum-wise, even at the shortest timescales measured. This latter result is unexpected: we find that there is nothing a priori in the model that allows one to predict that when broad phase spectra data (INFC and SUPC) are reproduced so will be the data at the individual and much shorter phase-binning, less with such a good agreement. This result point perhaps to some 
reliability of the model, at least in its essential ingredients: geometry, cascading, interacting electron population.

However, this model certainly has room for improvement. We emphasize here that we do not have an a priori model for the interacting lepton population itself, although we have discussed the research on dissipation processes in the PWZ which may give raise to such distributions if it results from pulsar injection. In any case, the assumption of power-laws is an approximation to a more complex scenario where the real interacting lepton population is the result of a full escape-loss equation. 
In addition, we are not considering yet the multiwavelength emission at lower energies, since we left out of our description the synchrotron emission of electrons accelerated at the shock and the morphology of the shock along the orbit, what we expect to discuss elsewhere. 
Other than system scalings that are fixed by multiwavelength observations, the model is based on just a handful of free parameters, and it is subject to tests at high and very high-energy $\gamma$-ray observations with both GLAST (described in more detail in Sierpowska-Bartosik \& Torres 2008) and future samples of data at higher energies, where more statistics at finer phase bins can determine better the spectral evolution along the orbit of this interesting system.\\

{\sl 
We acknowledge  extended use of IEEC-CSIC parallel computers cluster.
We acknowledge W. Bednarek for discussions, and the Referee for useful comments.
This work was supported by grants AYA 2006-00530 and CSIC-PIE 200750I029.}

\newpage

\section*{Appendix: Numerical implementation and formulae}

This Appendix introduces further essential details concerning geometry and formulae for the implemented process of Inverse Compton scattering and $e^+e^-$ pair production in the anisotropic radiation field of the massive star.

\subsection*{Monte-Carlo implementation for the cascading process}

A Monte-Carlo procedure is applied to calculate the place of electron interaction, $x_e$, and the energy of the resulting photon in the ICS, $E_{\gamma}^e$; as well as  the place of photon interaction, $x_p$, and the energy of the produced electron/positron, $E_e^p$ (see, e.g., Bednarek 1997). 
The following summarizes the procedure.

When the primary electron is injected, the computational procedure for IC scattering is invoked first in order to get the initial place of interaction (the radial distance from the injection position) and the energy of the up-scattered photon if the interaction takes place. 
For this same electron, the procedure is repeated afterwards up to the moment of electron cooling or when reaching the shock region. The photons produced along the electron path switch on a parallel computational procedure for pair production. This in turn gives the place of photon absorption and the energy of the $e^+e^-$ pair if the process occurs. 
Because the photons can get trough the termination shock if not absorbed in the PWZ, the numerical procedure is not limited to the termination shock radius. If a $e^+e^-$ pair is created within the PWZ, then the procedure for IC scattering is initiated for it, from which a next generation of photons can be produced. 
When photons cross the shock, information about them is separately saved in order to get account of the level of flux  absorption in the MSWZ. The emission spectra are produced from photons which finally escape from the binary, so the photons absorbed in the MSWZ are not included in it.

\begin{figure*}
   \includegraphics[width=0.49\textwidth,angle=0,clip]{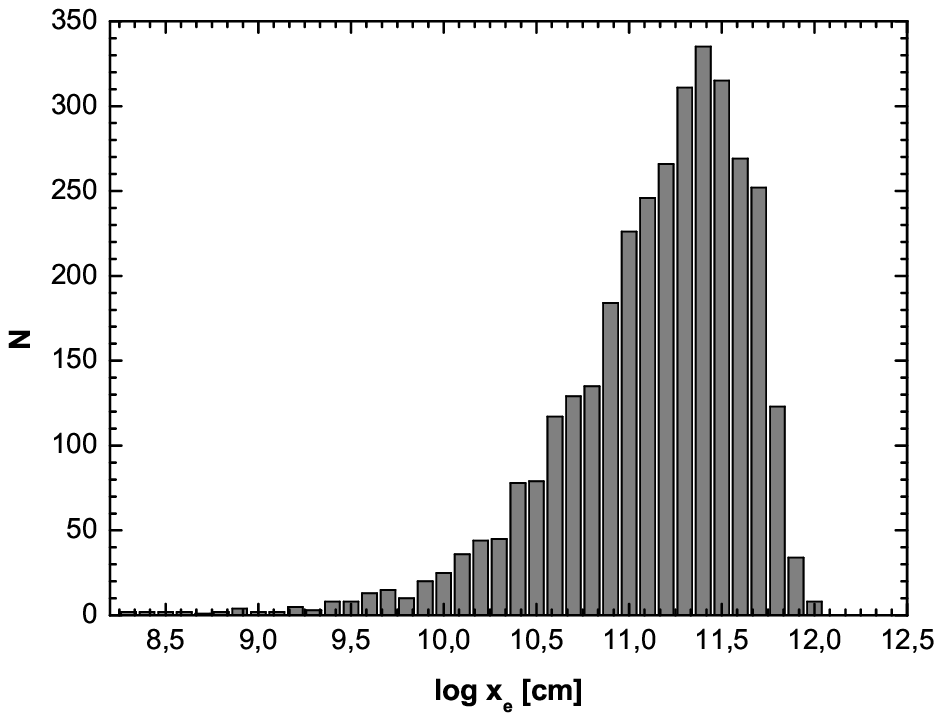}
   \includegraphics[width=0.49\textwidth,angle=0,clip]{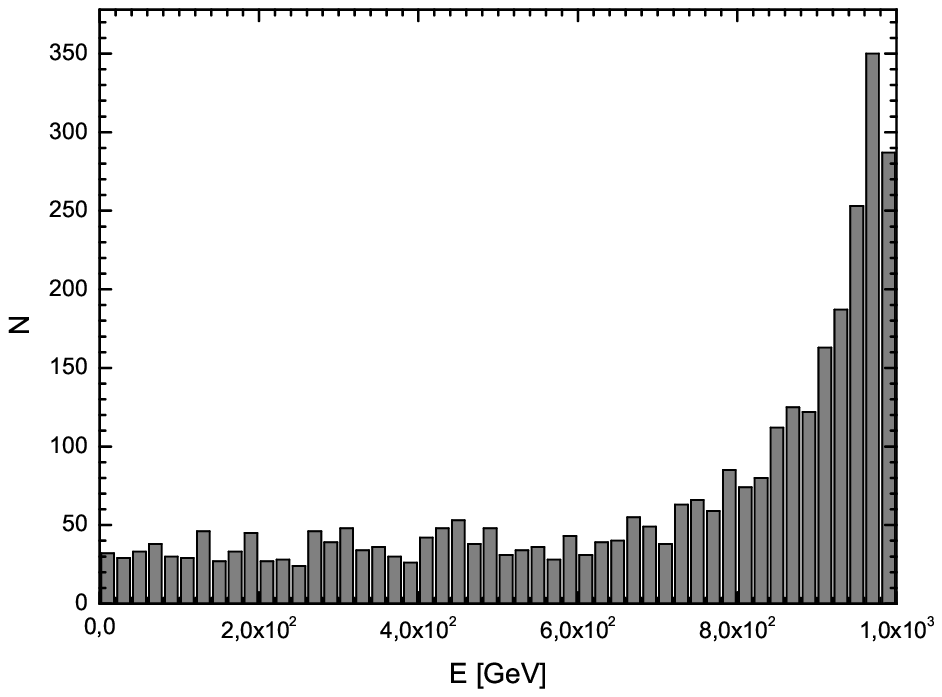}
\caption{\label{fig:ics-x} Left: The distribution of the place of electron interaction due to IC scaterring in the anisotropic radiation field of a massive star. The initial electron energy is here assumed as $E=1\,\rm TeV$,  and the place of its injection is given by the distance to the massive star in the LS 5039 system, $d_s = 2.25\, \rm R_s$, being injected at the angle $\alpha = 150^o$ with respect to the massive star. The number of injected electrons (interactions) is N = 3360. Right: Distribution for the scatterred photon energies for the same simulations.}
\end{figure*}

The place of lepton interaction in IC scattering, that is, the production of a photon at $x_{e1}$, after being injected at a distance $x_i$ from the star, at an angle $\alpha$ 
(we discuss further details of geometry in the next Section and in the Appendix, especially see Fig. \ref{fig:geom0}),
is calculated from  an inverse method. For an specific random number $P_1$ in the range $( 0,1) $, the interaction place $x_{e1}$ is given by the formula:
\begin{equation}
P_1 \equiv e^{-\tau_{ICS}} = \exp \left({- \int_0^{x_{e1}} \lambda_{ICS}^{-1} (E_e, x_i, \alpha, x_e)
\, dx_e} \right),
\label{ics17}
\end{equation}
where $ \lambda_{ICS}^{-1} (E_e, x_i, \alpha, x_e) $ is the rate of electron interaction to IC process (see Eq. \ref{ics16} of the Appendix). 
From this it follows: 
$
-\ln P_1 = \int_0^{x_{e1}} \lambda_{ICS}^{-1} (E_e, x_i, \alpha, x_e) \, dx_e,
$ where the integration is over the propagation path of the electron, $x_e$. Knowing $ \lambda_{ICS}^{-1} (E_e, x_i, \alpha, x_e)$, we keep integrating forward until we find numerical equality of the integral result with the random-generated quantity $-\ln P_1 $, what defines the position of interaction $x_{e1}$. For a large number of simulations, the distribution of interaction distances is presented in Fig. (\ref{fig:ics-x}, left panel).

Once we get the place of electron interaction, $x_{e1}$, simulated in the way described, we get the energy of the scattered photon from an inverse cumulative distribution. 
The energy of the photon $E_{\gamma}^e$  produced in IC is then given by the spectrum of photons after scattering ${dN}/{dE_{\gamma} dt}$ (see Appendix, Eq. \ref{ics14}). 
For  a random number $P_2$ we define:
\begin{equation}
P_2 \equiv \left[ {\int_0^{E_{\gamma}^e} \frac{dN}{dE_{\gamma}dt} \, dE_{\gamma}} \, \right] / \left[
\, {\int_0^{E_{\gamma}^{max}} \frac{dN}{dE_{\gamma}dt} \, dE_{\gamma}} \right],
\label{ics19}
\end{equation}
where $E_{\gamma}^{max}$ is the maximal scattered photon energy 
(see Appendix Eq. \ref{ics4}), 
while the energy $E_{\gamma}^e$ is the Monte-Carlo result for the simulated photon energy. Note that the denominator of the latter expression is the normalization needed for the Monte-Carlo association to succeed. The statistics for the energies of up-scattered photons is shown in Fig. \ref{fig:ics-x}, right panel. 
All relevant formulae and more details of implementation are given in the Appendix.

\begin{figure*}
   \includegraphics[width=0.49\textwidth,angle=0,clip]{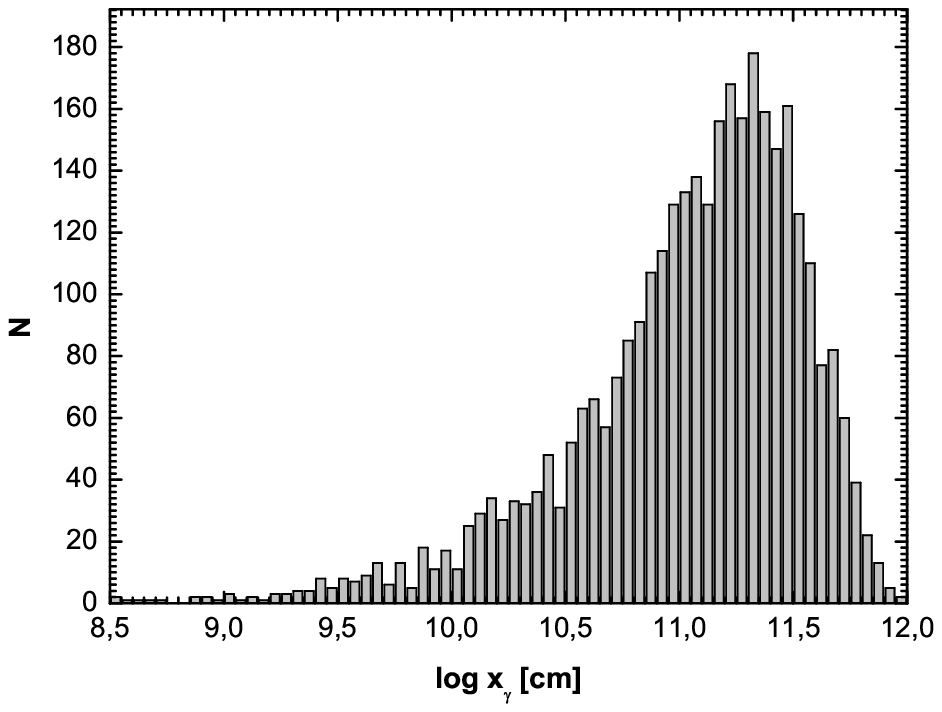}
   \includegraphics[width=0.49\textwidth,angle=0,clip]{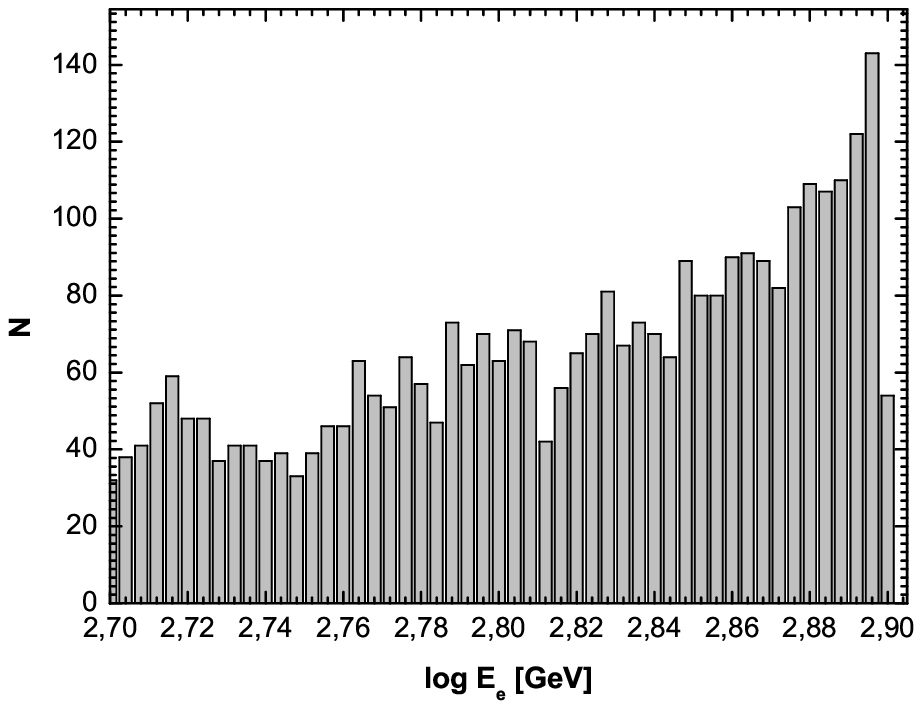}
\caption{\label{fig:gg-run}Left: Distribution of the position of the $\gamma$-photon interaction due to absorption in the anisotropic radiation field. The initial photon energy in this example is assumed as $E=1\,\rm TeV$ and the place of its injection is given by the distance to the massive star  in the LS 5039 system, $d_s = 2.25\, \rm R_s$ at the angle $\alpha = 150^o$ with respect to the massive star. The number of injected photons (interaction) is N = 3358. Right: Distribution of the produced electron energy for the same simulations.}
\end{figure*}

In a similar way we randomize by Monte-Carlo the needed magnitudes for $\gamma\gamma$ absorption. The probability of interaction for a photon with energy $E_{\gamma}$ at a given distance, say $x_{p3}$, from an injection place (i.e., provided $x_i$ and $\alpha$ are known) during the propagation in an anisotropic radiation field is given by expression:
\begin{equation}
P_3 \equiv e^{-\tau_{\gamma \gamma}} = \exp \left({-\int_0^{x_{p3}} \lambda_{\gamma \gamma}^{-1}
(E_{\gamma},x_{i},\alpha, x_{\gamma}) \, dx_{\gamma}} \right) \,.
\label{gp15}
\end{equation}
From this it follows, $\int_0^{x_{p3}} \lambda_{\gamma \gamma}^{-1} (E_{\gamma},x_{i},\alpha,
x_{\gamma}) \, dx_{\gamma} = -\ln P_3,
$
thus, similarly to the process just described, for the random number $P_3$ we can get the specific place of interaction $x_{p3}$. The results of such Monte-Carlo simulations are presented in Fig. \ref{fig:gg-run}, left panel.

The energy of the lepton ($e^+e^-$) produced in the photon absorption process (with the photon having an energy $E_{\gamma}$)  is calculated again from an inverse function. The latter is obtained from the integration of the  $e^+e^-$ spectra produced in the process: 
(see Appendix, Eq. \ref{gp28}):
\begin{equation}
P_4 \equiv \frac{
{\int_{0.5E_{\gamma}}^{E_e^p} \frac{dW(E_{\gamma}, x_0, \alpha,x_{p})}
{dE_e dx_{\gamma}} dE_e} }{
 { { \int_{0.5E_{\gamma}}^{E_e^{max}}} \frac{dW(E_{\gamma}, x_0,
\alpha, x_p)}{dE_e dx_{\gamma}} dE_e } },
\label{gp15b}
\end{equation}
where $P_4$ is a new random number. The integration is over the electron energy $E_e$. For normalizing, as the produced lepton spectra are symmetric with respect to the energy $E_e = E_{\gamma}/2$, the lower limit of this integral is fixed to this latter energy, while the upper limit is equal to maximum energy available in this process $E_e^{max} = E_{\gamma} - m_e c^2$. From the energy of the electron, $E_e^p$, the energy of the associated positron is also obtained as ${E_e^p}' = E_{\gamma} - E_e^p$. Again, note that the denominator of the latter expression is the normalization needed for the Monte-Carlo association to succeed.

\subsection*{Checks for the Monte-Carlo distributions}

\begin{figure*}
   \includegraphics[width=0.49\textwidth,angle=0,clip]{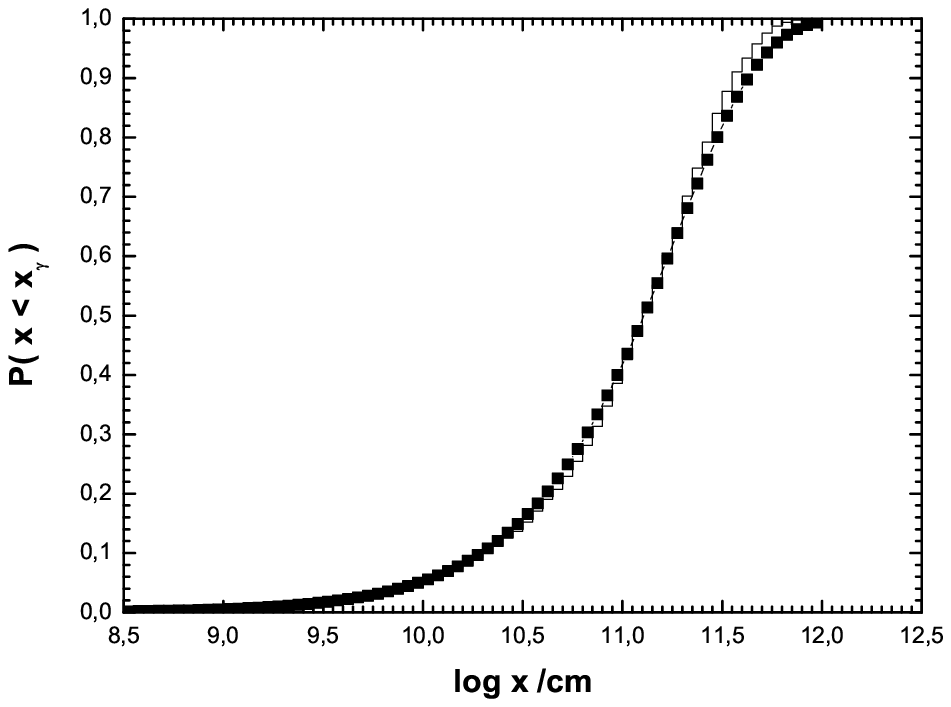}
   \includegraphics[width=0.49\textwidth,angle=0,clip]{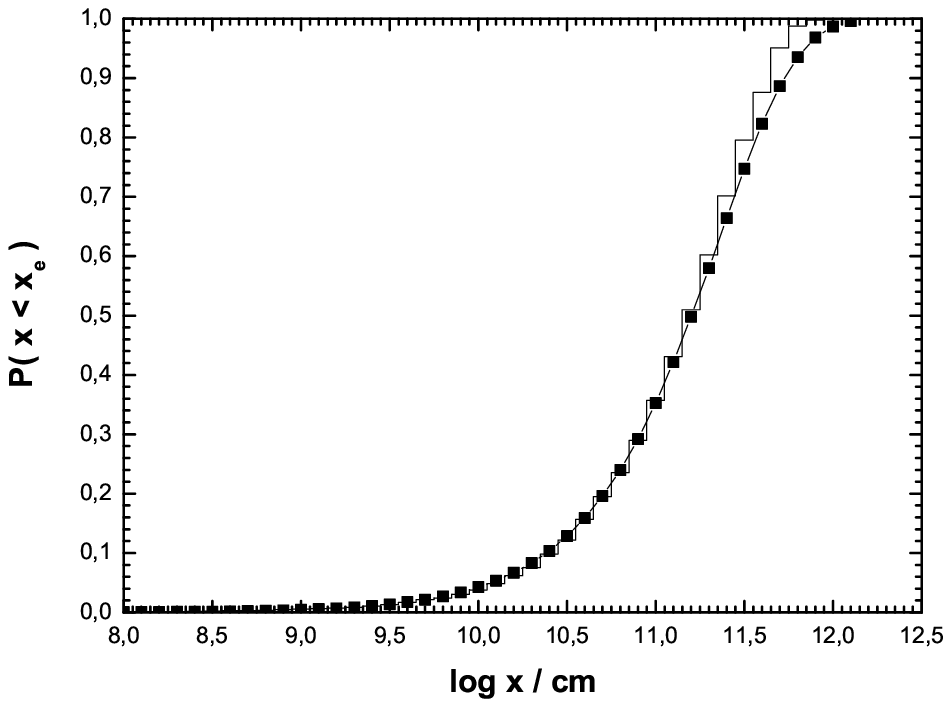}
\caption{\label{fig:MC-check}
Comparison between the event statistics, i.e., $N(x<x_e)/N_{total}$
 with $N_{total}$ being the total number of simulations run, and the analytical computed probability of interaction $1-\exp(-x/\lambda)$, where $\lambda$ is both, the one corresponding to $\gamma\gamma$ absorption (left) and ICS (right).
}
\end{figure*}

Here we discuss the rightness of the simulated random distribution of electrons and photons resulting from the cascading process. In Fig. \ref{fig:MC-check}
we show the comparison between the event statistics, i.e., $N(x<x_e)/N_{total}$
 with $N_{total}$ being the total number of simulations run in these examples, and the analytical computed probability of interaction $1-\exp(-x/\lambda)$, where $\lambda$ is, correspondingly, the one corresponding to $\gamma\gamma$ absorption and ICS.
We see total agreement of the Monte-Carlo and analytical probabilities, i.e., whereas the position of interaction of a single photon or electron,  from which the subsequent cascading process is followed, is obtained through Monte-Carlo and thus is random, the overall distribution maintains the shape provided by the physical scenario: given the target and injection energy, the mean free path defines the distribution for a sufficiently high number of runs.

\begin{table}
\centering
\caption{LS 5039: system parameters }
\vspace{0.2cm}
\begin{tabular}{lll}
\hline
{Parameter} & {Symbol} & {Adopted value}  \\
\hline
Radius of star &  $R_\star$ & $9.3\, R_\odot$\\
Mass of star &  $M_\star$ & $23\, M_\odot$\\
Temperature of star & $T_\star$ & $3.9 \times 10^4$ K\\
Mass loss rate of star & $\dot{M}$ & $10^{-7}\, M_\odot\, \rm yr^{-1}$  \\
Wind termination velocity & $V_{\infty}$ & $2400 \,\rm km\, s^{-1}$ \\
Wind initial velocity & $V_0$ & $4\,\rm km\, s^{-1}$ \\
Distance to the system & $D$ & $ 2.5 \,\rm kpc$\\
Eccentricity of the orbit & $\varepsilon$ & $0.35$ \\
Semimajor axis & $a$ & $0.15\, \rm AU$ $\sim 3.5\, R_\star$  \\
Longitude of periastron & $\omega_{p}$ & $226^o$ \\
\hline
\end{tabular}
\label{orb-param}
\end{table}

\subsection*{Basic geometry }
Parameters such as the viewing angle towards the observer, $\alpha_{obs}$, the separation of the binary, $d$, and the distance to the shock region along the orbit, $r_s$ (Eq. \ref{r0}), are directly connected to $\gamma$-ray observational results and are thus crucial for detailed model discussion. Table \ref{orb-param} gives the set of LS 5039 orbital and binary parameters that are relevant for modeling. These parameters come from the recent work by Casares et al. (2005b), either directly from their measurements or from the values compiled by them.
The inferior conjunction (INFC) is defined as the phase when the binary is viewed from the pulsar side (the smallest $\alpha_{obs}$). Exactly in the opposite side of the orbit we find superior conjunction (SUPC), when the pulsar is behind the massive star (the largest $\alpha_{obs}$), see also Fig. \ref{fig:general}. One can see that for LS 5039, INFC ($\phi \approx 0.72$) is close to apastron phase, while SUPC ($\phi \approx 0.06$) is close to periastron. 

\begin{figure}
  \centering
   \includegraphics*[width=0.5\textwidth,angle=0,clip]{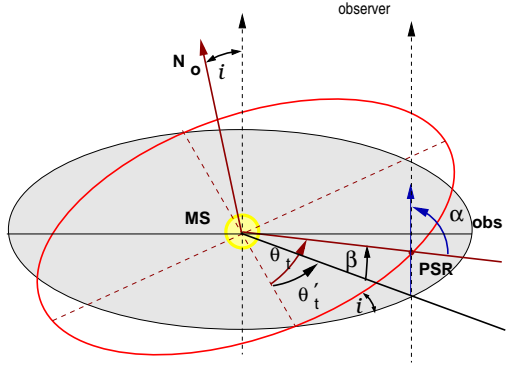}
\caption{\label{fig:alpha_obs_app} The geometry of the orbit of the inclination $i$ in the binary system (with the MS - massive star in the center) where the angle to the observer, $\alpha_{obs}$, is defined. $N_o$ is the normal to the orbital plane, $\theta_t$ is the angle related to the orbital phase and $\theta'_t$ is its projection in the plane of the observer. The angle $\beta$ gives the height of the pulsar for given phase above the observer plane. }
\end{figure}

\begin{figure}
  \centering
   \includegraphics*[width=0.5\textwidth,angle=0,clip]{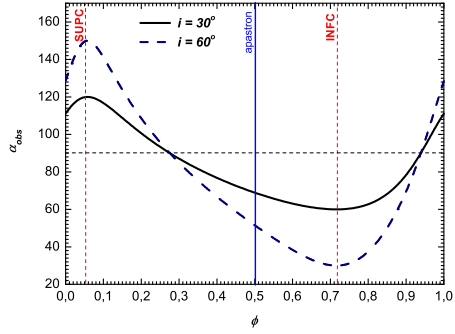}
\caption{\label{fig:alpha_obs} {The LS 5039 pulsar angle to the observer as a function of phase along the orbit for two different values of the binary inclination angle $\textit{i}$. INFC, SUPC, periastron, and apastron phases are marked.}}
\end{figure}

Figs.  \ref{fig:alpha_obs_app} and  \ref{fig:alpha_obs}  shows the angle to the observer, $\alpha_{obs}$, as a function of orbital phase for two considered inclination of the binary, $i=30^o$ and $i=60^o$. In addition, the separation of the binary is marked. The viewing angle changes within the limits $(90^0-i, 90^0+i)$ and depends also on the longitude of the periastron $\omega_{p}$.

The comparison for the two inclinations shows already that $\alpha_{obs}$ varies significantly in the case of the larger angle, as the difference is $\delta \alpha_{obs} = 120^o$, what is crucial for all angle-dependent processes discussed in this paper. The separation of the binary is given by 
\begin{equation} 
r =\frac {p}{1+\varepsilon \cos \theta }, 
\end{equation}
 where $\varepsilon$ is the  eccentricity of the orbit, 
 \begin{equation} 
\varepsilon^2 = (a^2-b^2)/a^2,
 \end{equation}
the value of $p$ is defined as 
\begin{equation} 
p=a(1-\varepsilon^2)
\end{equation}
and where $b$ is the semi-minor axis. As the semi-major axis of LS 5039 is $a = 0.15$ AU $= 3.5\, R_s$ the assumed pulsar in the orbit is at periastron only $2.25\, R_s$ from the massive star, whereas at apastron, it lies at $d = 4.7\, R_s$, about a factor of 2 farther.

To get the relation for the angle to the observer we use the geometry shown in Fig. \ref{fig:alpha_obs_app} where starting from the spherical triangle containing the inclination angle $i$,
we have the relations:
\begin{equation}
\sin \theta_t \cos i = \cos \beta \sin \theta'_t
\end{equation}
and
\begin{equation}
\cos \theta_t = \cos \beta \cos \theta'_t
\end{equation}
which is the cosine theorem for a rectangular spherical triangle.
Then, the angle to the observer defined in Fig. \ref{fig:alpha_obs_app}, is calculated from the formula:
\begin{equation}
\cos^2 \beta  = \sin^2 \theta_t \cos^2 i + \cos^2 \theta_t,
\label{obs_beta}
\end{equation}
from which:
\begin{equation}
\alpha_{obs} = \pi/2 \pm \beta .
\label{obs_beta_prim}
\end{equation}
The sign in this relation depends on the orientation to the observer, i.e., if the pulsar is below or above the orbital plane. Note that Eq. (\ref{obs_beta}) is not fulfilled all around the orbit as in here the angle $\theta_t = 0$ denotes the common point for the orbital and observer plane. The relation can then be used after defining the orientation of the orbital plane with respect to the observer  given by the inclination $i$ and the periastron position $\omega_{per}$. The latter is done when we tilt the inclined orbital plane with $\omega_{per} + \pi/2$. To find the internal phase $\theta_t$ dependent on the true anomaly $\theta$ (with $\theta=0$ at periastron), we have to find the spherical triangle from Fig. \ref{fig:alpha_obs_app} in the system of the orbital plane. Including the tilting of the systems we have the relation: 
\begin{equation}
\theta_t = \theta + \omega_{per} +  \pi / 2. 
\end{equation} 
Then we have to check if the calculated value of $\theta_t$ is within one orbital phase $0 < \theta_t < 2 \pi$. If not the angle have to be replaced by $\theta_t + 2 \pi$ (if $\theta_t < 0$) or $\theta - 2 \pi$ ($\theta_t > 2 \pi$). To use the Eq. (\ref{obs_beta}) and calculate the corresponding angle to the observer we have to recalculate the angle $\theta_t$ once again  as it is only valid  in the range $0 < \theta_t < \pi/2$.  We can divide the orbit in four quarters in which the following transformations are needed:
\begin{enumerate}
\item $\pi/2 - \theta_t$ if  $0 < \theta_t < \pi/2$,
\item $\theta_t - \pi/2$ if  $\pi/2 < \theta_t < \pi$,
\item $3 \pi/2 - \theta_t$ if  $\pi < \theta_t < 3 \pi/2$,
\item $\theta_t - 3 \pi/2$ if  $3 \pi/2 < \theta_t < 2 \pi$.                                                  \end{enumerate}
Inserting such phase into the Eqs. (\ref{obs_beta}) and (\ref{obs_beta_prim}) we get the angle to observer corresponding to the true anomaly $\theta$.

\begin{figure}

\centering
\includegraphics[width=0.5\textwidth, clip]{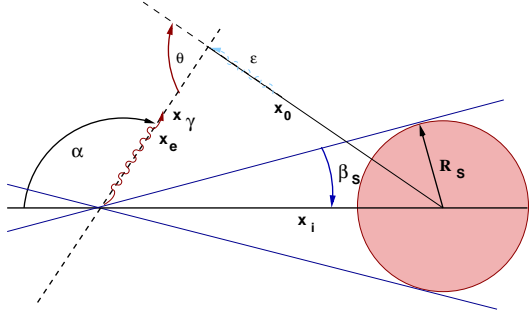} 
\caption{
Basic parameters for photon and electron propagation and their interaction in the anisotropic radiation field of the massive star. The injection is produced at a distance $x_i$ and angle $\alpha$.
$R_{s}$ denotes the massive star radius, and the tangent direction to its surface is given by angle $\beta_{s}$. The propagation path is denoted by $x_{\gamma}$ and $x_e$, while the interaction occur at distance $x_0$. The scattering angle for electron-photon ($\epsilon$) or photon-photon ($\epsilon$) interactions is $\theta$.} 
\label{fig:geom0}
\end{figure}

\subsection*{Thermal radiation}

The source of the thermal radiation field is the massive star of early type (O, Be, WR). The spectrum is described by Planck's law, which differential energy spectrum (the number of photons of given energy $\epsilon$ per unit energy $d\epsilon$, per unit solid angle $\Omega$, per unit volume $V$)  is given by:
\begin{equation}
n(\epsilon) = \frac{dn(\epsilon,\Omega)}{d\epsilon d\Omega dV} = \frac {4
\pi}{(h c)^3} \frac{\epsilon^2}{e^{\epsilon /kT_{s}}-1}, 
\label{gp1}
\end{equation}
where $\epsilon$ is thermal photon energy, $h$ is the Planck constant, and $k$ is Boltzmann constant.

\subsection*{Gamma ray absorption and $e^+e^-$ production: opacity and geometry}
\begin{figure*}
\centering
\includegraphics[width=0.45\textwidth,clip]{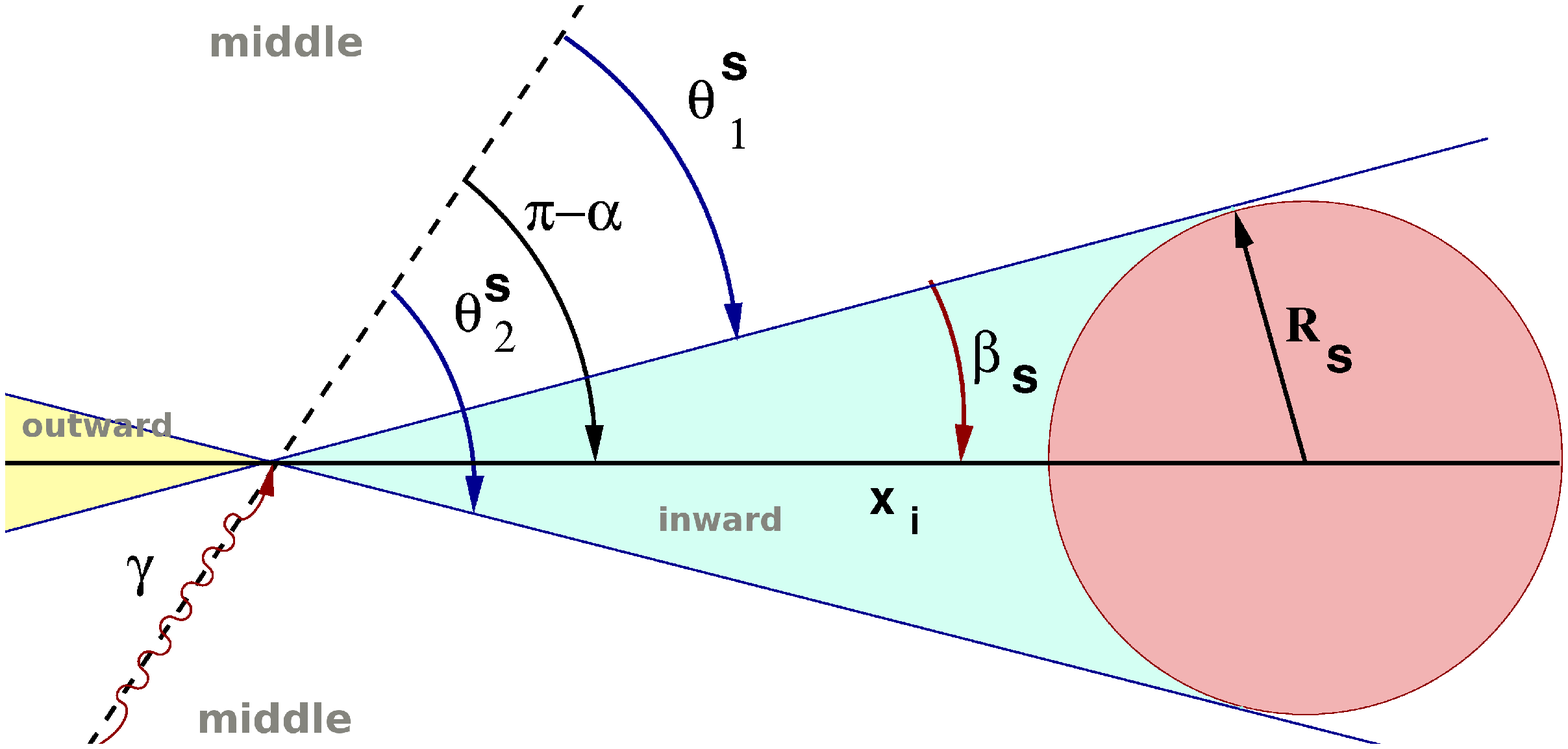}\hfill
\includegraphics[width=0.45\textwidth,clip]{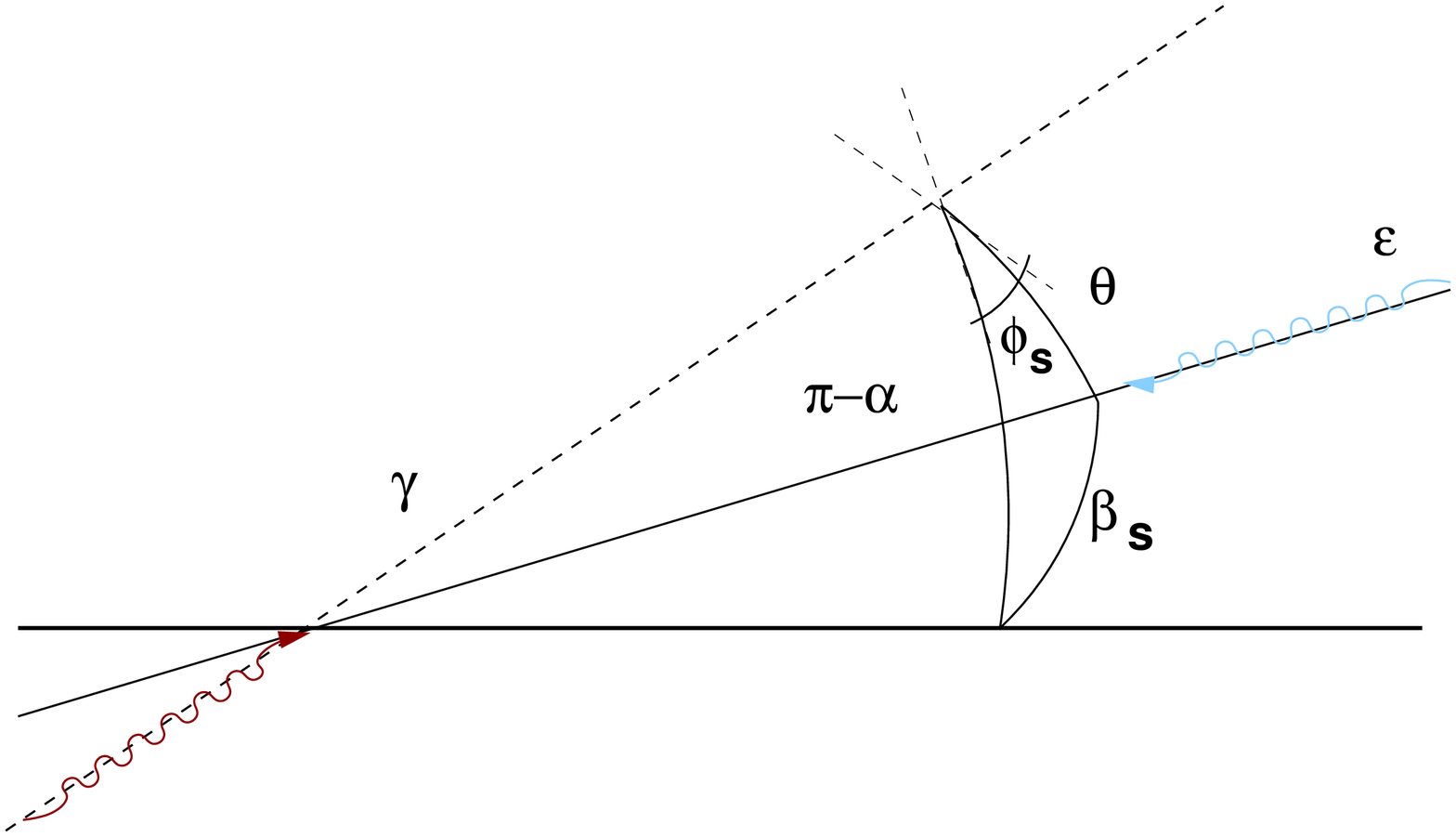} 
\caption{
Left: Spherical geometry for a photon absorption process in the radiation field of the star (a thermal photon is denoted with $\epsilon$). The angle $\beta_{s}$ is defined by the line to the star center and tangent to its surface. The angles $\alpha$ and $\theta$ define the direction of gamma-photon incoming towards the star. The angle $\alpha$ is measured from the central line (joining the star and injection place) whereas the angle $\theta$, from the line of gamma photon propagation. The directions of gamma-photon propagation defined as ``inward'' and ``outward'' with respect to the star are shaded.  Right:  two dimensional geometry for definition of the parameter $\mu = \cos \theta$, where $\mu_1 = \cos \theta_1^{s}$ and $\mu_2 = \cos \theta_2^{s}$ are given by Eq. (\ref{gp13}). The azimuthal angle $\phi_{s}$ is the angle between the direction of gamma propagation and propagation of $\epsilon$..}
\label{fig:geom1-1b}
\end{figure*}

The optical depth to $\gamma$-photon absorption in the radiation field of the massive star up to infinity can then be calculated from the integral: 
\begin{equation}
\tau_{\gamma \gamma} (E_{\gamma},x_{i},\alpha) = 
\int_0^{\infty} \lambda_{\gamma \gamma} ^{-1}  (E_{\gamma},x_{i},\alpha,
x_{\gamma}) \, dx_{\gamma},
\label{gp14}
\end{equation}
where $\lambda_{\gamma \gamma}^{-1}$ is a photon interaction rate to  $e^+e^-$ production in an anisotropic radiation field and $x_{\gamma}$  is its propagation length. When the propagation occur toward the massive star surface the integration is performed up to the stellar surface.
The photon interaction rate, $\lambda_{\gamma \gamma}^{-1}$, is related to a photon of energy  $E_{\gamma}$ injected at a distance $x_{i}$ from the massive star, at angle  $\alpha$ (see Fig. \ref{fig:geom0}),  and is given by the formula:
\begin{equation}
{\lambda_{\gamma \gamma}}^{-1} (E_{\gamma},x_{i},\alpha, x_{\gamma}) = \int
(1+\mu) d\mu
\int d\phi \int n(\epsilon) \sigma_{\gamma \gamma} (\beta) d\epsilon,
\label{gp2}
\end{equation}
where $x_{\gamma}$ is the distance to the interacting photon from the injection place along its propagating path.
The integration limits will be discussed in a different part of the Appendix. For now we focus on the variable $\mu$ in the first integration. It is the cosine of the photon-photon scattering angle $\mu = \cos \theta$ (see Fig. \ref{fig:geom0}). The angle $\phi$ is the azimuthal angle between the photon ($\gamma$-ray) propagation direction and the direction to the massive star, while  $\phi_s$ gives its limit value for the direction tangent to the massive star surface (see Fig. \ref{fig:geom1-1b}). 
The cross section to $e^+e^-$ production is denoted as $\sigma_{\gamma \gamma} (\beta)$, where the parameter $\beta$ in the center of mass system is
\begin{equation}
\beta = \sqrt{\omega^2-m^2}/\omega, 
\end{equation}
with 
\begin{equation}
\omega^2 = \frac{1}{2} E_{\gamma} \epsilon (1+\cos{\theta}) 
\end{equation}
being the photon energy squared  (in this notation it is assumed also that $c=1$). 
From this we get 
\begin{equation}
\beta^2 = 1-{2m^2}/{E_{\gamma}\epsilon(1+\mu)}. 
\end{equation}
The kinematic condition for the angle $\theta$ which defines the threshold for the $e^+e^-$ creation is given by expression:
\begin{equation}
\mu \geq \mu_{lim} = \frac{2m^2}{E_{\gamma}\epsilon}-1.
\label{gp4gr}
\end{equation}
To simplify the equations we rewrite the internal integration in Eq. (\ref{gp2}) making use of $I_1(\mu)$, such that,
\begin{equation}
I_1(\mu) = \int n(\epsilon) \sigma_{\gamma \gamma} (\beta) d\epsilon.
\label{gp3}
\end{equation}
With the replacement 
\be 
\beta^2 = 1 - a/\epsilon , 
\ee
where 
\be
a=2 m^2/E_{\gamma}(1+\mu),
\ee 
and defining the constant $S=8 \pi /(hc)^3$, the spectrum of thermal photons is now given by the formula:
\begin{equation}
n (\epsilon) = S \frac{a^2}{(1-\beta^2)^2}
\frac{1}{e^{(a/[(1-\beta^2)kT_{s})]}-1}.
\label{gp4}
\end{equation}
Substituting in the internal integral $I_1(\mu)$, and introducing the integration variable to $\beta$, via 
\be d\epsilon = 2a \, \beta/(1-\beta^2)^2 \, d\beta, 
\ee
yields to the integral:
\begin{equation}
I_1(a) = 2S \int_0^1  \frac{a^3}{(1-\beta^2)^2}
\frac{1}{e^{(a/(1-\beta^2)kT_{s})}-1} \sigma_{\gamma \gamma}(\beta) d\beta.
\label{gp5}
\end{equation}
The lower limit of integration is from the energy condition for the process $\gamma + \gamma \rightarrow e^+e^-$, i.e., it follows from the threshold condition $E_{\gamma} \epsilon (1+\mu) = 2 m^2$, where $\omega = m$. . 
The upper integration limit comes from the relativistic limit $\omega \gg m$, where we get $\beta \approx 1$. 
To proceed forward, we introduce a dumb variable, $b$, by $a = kT_{s} b$, to get:
\begin{equation}
I_1(b) = 2S \int_0^1 (kT_{s})^3 b^3 \sigma_{\gamma \gamma}(\beta)
\frac{\beta}{(1-\beta^2)^4} \frac{1}{e^{(b/(1-\beta^2))}-1} \, d\beta.
\label{gp6}
\end{equation}

\begin{figure}
\centering
\includegraphics[width=0.35\textwidth,clip]{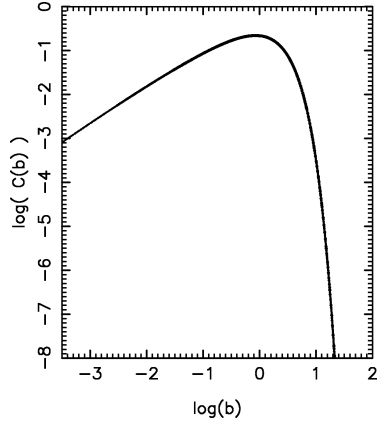}
\caption{The plot of the function $C(b) = I_1(\beta)/C_1$, where the integral $I_1$ is given by Eq. (\ref{gp8}). } 
\label{fig:tabwew1}
\end{figure}

The cross section for $e^+e^-$ pair production is (Jauch and Rohrlich, 1980):
\begin{equation}
\sigma_{\gamma \gamma}(\beta) = \frac{1}{2} r_0^2 \pi (1-\beta^2) \lbrack
(3-\beta^4) \ln \frac{1+\beta}{1-\beta} - 2\beta(2-\beta^2)\rbrack,
\label{gp7}
\end{equation}
where $r_0$ is the classical electron radius, and $\sigma_{T} = \frac{8}{3}
\pi r_0^2$ is the Thomson cross section. When putting this expression into the integral $I_1(b)$ (Eq. \ref{gp6}) we get finally,
\begin{eqnarray}
I_1(b) = C_1 \frac{3}{16} b^3 \int_0^1 \left[(3-\beta^4) \ln \frac{1+\beta}{1-\beta}
- 2\beta (2-\beta^2)\right]
\times \nonumber \\
 \frac{\beta}{(1-\beta^2)^3}
\frac{1}{e^{(b/(1-\beta^2))}-1} d\beta, 
\label{gp8}
\end{eqnarray}
with $C_1 = 16 \pi (kT_{s}/hc)^3 \sigma_T$.

Weparametrizee the internal integral and write it as a function of $b$, 
\be C(b) \equiv  I_1(\beta)/C_1.\ee 
The plot for this function is shown in Fig. (\ref{fig:tabwew1}). Then, the integral we are after can be written as $I_1(b) = C_1 C(b)$ and 
\begin{equation}
{\lambda_{\gamma \gamma}}^{-1} (E_{\gamma},x_{i},\alpha, x_{\gamma})  = C_1 \int
(1+\mu) d\mu \int d\phi \, C(b). 
\label{gp9}
\end{equation}
In the second internal integral, we have to fix the limits (having in mind that the angle $\phi$ depends on the angle $\theta$, then also on the variable $\mu$), 
\begin{equation}
I_2 = \int_{-\phi_{s}}^{\phi_{s}}C(b) d\phi = 2 \phi_{s} C(b)  =
C(b) \Phi(\mu). \label{gp10}
\end{equation}
{The angle $\phi_{s}$ determines the maximal azimuthal angle of $\gamma$-ray photon propagation with respect to the direction of the thermal photon, so that it gives the directions tangent to the star surface. This condition for $\phi_{s}$ can be determined from the spherical triangle} (shown in Fig. \ref{fig:geom1-1b}). 
From the cosine theorem of spherical trigonometry 
\be
\cos{\beta_{s}} = \cos\theta \cos{(\pi - \alpha)} + \sin \theta \sin{(\pi - \alpha)} \cos \phi_{s}, \ee
where $\beta_{s}$
is defined as in Fig. (\ref{fig:geom1-1b}). 
After some algebra we get:
\begin{equation}
\cos \phi_{s} = \frac{\cos \beta_{s} + \mu \cos \alpha}{\sqrt{1-\mu^2}
\sin \alpha}, \label{gp11}
\end{equation}
where $\sin \beta_{s} = R_{s}/x_i$, while $0 \leq \phi_{s} \leq \pi$. The angle $\alpha$ determines the direction of incoming $\gamma$-ray photon (see Fig.  \ref{fig:geom0}).

The limits of integration with respect to the parameter $\mu$ (see Fig. \ref{fig:geom1-1b}), are the range of angles for soft photons coming from the star, 
\begin{equation}
{\lambda_{\gamma \gamma}}^{-1}  (E_{\gamma},x_i,\alpha, x_{\gamma}) = C_1
\int_{\mu_1}^{\mu_2} (1+\mu) \Phi(\mu) \, d\mu \, C(b) , \label{gp12}
\end{equation}
where $\mu_1 = \cos \theta_1^{s}, \, \mu_2 = \cos \theta_2^{s}$.
Using basic trigonometrical dependencies between the angles,  $ \theta_1^{s} = \pi - \alpha - \beta$ and  $ \theta_2^{s} = \pi - \alpha + \beta$ (Fig. \ref{fig:geom1-1b}, left), we get two expressions:
\begin{eqnarray}
\mu_1 &=& -\cos \alpha \cos \beta_{s} + \sin \alpha \sin \beta_{s}, \nonumber \\
\mu_2 &=& -\cos \alpha \cos \beta_{s} - \sin \alpha \sin\beta_{s}. \label{gp13}
\end{eqnarray}

Depending on the angle of $\gamma$-photon propagation $\alpha$ we can precise the conditions for the above integration limits. We distinguish three regimes:
\begin{enumerate}
\item the range for $\alpha$ towards the star surface:
$(\pi - \beta_{s} \leq \alpha \leq \pi)$,
\item the ``middle'' range:
$(\beta_{s} < \alpha < \pi - \beta_{s})$,
\item the range outwards of the massive star surface:
$(0 \leq \alpha \leq \beta_{s}) $.
\end{enumerate}
Integration over $\mu$ can be simplified by separating the main integral, Eq.
(\ref{gp12}), in the sum of the corresponding integrals for these regimes, each depending on their limits $\mu_1, \mu_2, \mu_{lim}$ (see Eq. \ref{gp13} and \ref{gp4gr}). 
A simplification occurs for the specific range of the angle $\theta$ for which the function $\Phi(\mu) = 2 \phi_{s}$ (Eq. \ref{gp10}) is maximal, i.e., $\Phi(\mu) = 2 \pi$, what corresponds to the scenario when the photon propagate toward the star. Another simplification happens when $\Phi(\mu)$ is  minimal, i.e., 
$\Phi(\mu) = 0$, for an outgoing photon. 
{To simplify the formulae for numerical implementation we introduce the function $f(\mu,b) = C_1 C(b) (1 + \mu)$ such that}
\be {\lambda_{\gamma \gamma}}^{-1} (E_{\gamma},x_{i},\alpha, x_{\gamma})  = \int_{\mu_1}^{\mu_2} f(\mu,b) \Phi(\mu) \, d\mu, \ee 
where we introduced internal function $f(\mu,b) = C_1 (1+\mu) \, C(b)$. To fasten the simulations the integration limits can be fixed following the rules (note that  relation $(\mu_1 \geq \mu_2)$ is always  fulfilled):

\begin{enumerate}

\item for ``inward'' propagation :
\begin{itemize}

\item if $(\mu_{lim} \leq \mu_2 \leq \mu_1)$:\\
$ {\lambda_{\gamma \gamma}}^{-1} (E_{\gamma},x_{i},\alpha, x_{\gamma}) = \int_{\mu_2}^{\mu_1} f(\mu,b) \Phi(\mu) \, d\mu + \int_{\mu_1}^{1}
f(\mu,b) 2\pi \, d\mu $,

\item if $(\mu_2 \leq \mu_{lim} \leq \mu_1)$:\\
$ {\lambda_{\gamma \gamma}}^{-1} (E_{\gamma},x_{i},\alpha, x_{\gamma}) = \int_{\mu_{lim}}^{\mu_1} f(\mu,b) \Phi(\mu) \, d\mu +
\int_{\mu_1}^{1} f(\mu,b) 2\pi \, d\mu $,

\item if $(\mu_2 \leq \mu_1 \leq \mu_{lim})$:\\
$ {\lambda_{\gamma \gamma}}^{-1} (E_{\gamma},x_{i},\alpha, x_{\gamma}) = \int_{\mu_{lim}}^{1} f(\mu,b) 2\pi \, d\mu$.
\end{itemize}
Apart from these,  the case of $\alpha = \pi$ can also be separately defined (incoming at a central spot of the star) when $(\mu_1= \mu_2)$ and $ {\lambda_{\gamma \gamma}}^{-1} (E_{\gamma},x_{i},\alpha, x_{\gamma}) = \int_{\mu_{min}}^{1} f(\mu,b) 2\pi \, d\mu $, where $\mu_{min}$ is the smaller value from  $(\mu_1, \mu_{lim})$. \\

\item for the ``middle" range of propagation:
\begin{itemize}
\item if $(\mu_{lim} \leq \mu_2 \leq \mu_1)$:\\
$ {\lambda_{\gamma \gamma}}^{-1} (E_{\gamma},x_{i},\alpha, x_{\gamma}) = \int_{\mu_2}^{\mu_1} f(\mu,b) \Phi(\mu) \, d\mu $,
\item if $(\mu_2 \leq \mu_{lim} \leq \mu_1)$:\\
$ {\lambda_{\gamma \gamma}}^{-1} (E_{\gamma},x_{i},\alpha, x_{\gamma}) = \int_{\mu_{lim}}^{\mu_1} f(\mu,b) \Phi(\mu) \, d\mu $,
\item if $(\mu_2 \leq \mu_1 \leq \mu_{lim})$:\\
$ {\lambda_{\gamma \gamma}}^{-1} (E_{\gamma},x_{i},\alpha, x_{\gamma}) = 0$.
\end{itemize}

\item for ``outwards" directions:
\begin{itemize}
\item if $(\mu_{lim} \leq \mu_2 \leq \mu_1)$:\\
$ {\lambda_{\gamma \gamma}}^{-1} (E_{\gamma},x_{i},\alpha, x_{\gamma}) = \int_{\mu_2}^{\mu_1} f(\mu,b) \Phi(\mu) \, d\mu +
\int_{\mu_{lim}}^{\mu_2} f(\mu,b) 2\pi \, d\mu $,
\item if $(\mu_2 \leq \mu_{lim} \leq \mu_1)$:\\
$ {\lambda_{\gamma \gamma}}^{-1} (E_{\gamma},x_{i},\alpha, x_{\gamma}) = \int_{\mu_{lim}}^{\mu_1} f(\mu,b) \Phi(\mu) \, d\mu $,
\item if $(\mu_2 \leq \mu_1 \leq \mu_{lim})$:\\
$ {\lambda_{\gamma \gamma}}^{-1} (E_{\gamma},x_{i},\alpha, x_{\gamma}) =0 $.
\end{itemize}
Apart from these, as before, the case of $\alpha =0$ can also be defined (escaping radially from a central spot of the star) when $(\mu_1= \mu_2)$ i if $(\mu_1 > \mu_{lim}) $, and we have $ {\lambda_{\gamma \gamma}}^{-1} (E_{\gamma},x_{i},\alpha, x_{\gamma}) =
\int_{\mu_{lim}}^{\mu_1} f(\mu,b) 2\pi \, d\mu $, otherwise ${\lambda_{\gamma \gamma}}^{-1} (E_{\gamma},x_{i},\alpha, x_{\gamma}) = 0$.

\end{enumerate}

\subsection*{The energy spectrum for $e^+e^-$ pairs produced by $\gamma\gamma$ absorption process}

To simulate the energy of an electron (positron) produced in the process of gamma absorption we follow the energy distribution of $e^+e^-$ pairs (see Eq. \ref{gp15b}).
The $e^+e^-$ pair spectrum produced by a photon of energy $E_{\gamma}$ at a specific distance $x_p$ 
is given by the expression:
\begin{equation}
\frac{dW(E_{\gamma}, x_i, \alpha, x_{\gamma})}{dE_e dx_{\gamma}} = \int (1+\mu)
d\mu \int d\phi \int n(\epsilon) \sigma(E_{\gamma}, E_e, \mu) d\epsilon,
\label{gp17}
\end{equation}
where $E_{\gamma}$ is the energy of the interacting photon, $x_{\gamma}$ is the propagation length, $x_i$ the distance of the injection place form the massive star, $\alpha$ is the angle of propagation, and $E_e$ is the energy of produced lepton ($e^+e^-$). The limits of integration are discussed in the following paragraph.

The energy spectrum of thermal photons $\epsilon$ is again given by the Planck's spectrum $n(\epsilon)$ (Eq. \ref{gp1}). The cross section for pair production $\sigma(E_{\gamma}, E_e, \mu)$ depends on the photon energy $E_{\gamma}$ and energy of electron $E_e$. The variable $\mu = \cos \theta$, angles $\theta$ and $\phi_{s}$ are defined in Fig. \ref{fig:geom1-1b}).

The cross section is given by the formula (Akhieser \& Berestezki, 1965):
\begin{eqnarray}
\sigma(E_{\gamma}, E_e, \mu) &=& \frac{r_0^2}{2} \frac{m^2}{E_{\gamma}}
\frac{1}{\varepsilon_c \beta_c \gamma_c} \times [  \frac{2\varepsilon_c^2 - m^2
+(\varepsilon_c^2- m^2)A^2(E_e,\varepsilon_c)} {m^2 B^2
(E_e,\varepsilon_c)+\varepsilon_c^2 A^2(E_e,\varepsilon_c)} 
 \nonumber \\ 
&-& \frac{2(\varepsilon_c^2- m^2)^2 A^4(E_e,\varepsilon_c)} {(m^2
B^2(E_e,\varepsilon_c)+ \varepsilon_c^2 A^2(E_e,\varepsilon_c))^2} ],
\label{gp18}
\end{eqnarray}
where $\varepsilon_c$ is the photon energy in centrum of momentum (CM) of the interacting photons $\varepsilon_c^2 = \frac{1}{2} E_{\gamma} \epsilon (1+\mu)$. The parameter $\gamma_c = (E_{\gamma}+\epsilon)/2\varepsilon_c$ is Lorentz factor in CM system and $m = m_e c^2$ is the electron mass. 
In Eq. (\ref{gp18}), the following functions are introduced: 
\begin{eqnarray}
B^2(E_e,\varepsilon_c) &=& \frac{(E_e-\gamma_c \varepsilon_c)^2}{(\gamma_c
\beta_c)^2 (\varepsilon_c^2 - m^2)}, \\
A^2(E_e,\varepsilon_c) &=& 1-B^2(E_e,\varepsilon_c),
\label{gp19}
\end{eqnarray}
where $\beta_c$ is the velocity of the CM system in units of the speed of light, $\gamma_c \beta_c = \sqrt{\gamma_c^2-1}$. 
The internal integration in Eq. (\ref{gp17}), i.e., that performed over the thermal photons spectrum, according to Eq. (\ref{gp18}), is: 
\begin{eqnarray}
I_1(E_{\gamma}, E_e, \mu) = \frac{8 \pi}{(h c)^3}
\int_{\epsilon_{min}}^{\infty}
\frac{\epsilon}{(e^{\epsilon/kT_{s}} - 1)} \frac{1}{\varepsilon_c \beta_c
\gamma_c} \times \nonumber \\
\left[ \frac{2\varepsilon_c^2-m^2+(\varepsilon_c^2-m^2)A^2(E_e,\varepsilon_c)}
{m^2B^2(E_e,\varepsilon_c)+\varepsilon_c^2 A^2(E_e,\varepsilon_c)} 
-  \right.
\nonumber \\ \left.
\frac{2(\varepsilon_c^2- m^2)^2 A^4(E_e,\varepsilon_c)} {(m^2
B^2(E_e,\varepsilon_c)+ \varepsilon_c^2 A^2(E_e,\varepsilon_c))^2} \right]  \,
d\epsilon.
\label{gp20}
\end{eqnarray}
The lower integration limit is given by the condition:
\begin{equation}
\epsilon_{min} = \frac{m^2 E_{\gamma}}{2 E_e (E_{\gamma} - E_e)(1 + \mu)} =
\frac{F}{\kappa(1-\kappa)}.
\label{gp21}
\end{equation}
In the last expression, we have introduced the new variables $\kappa = E_e/E_{\gamma}$ and $F =m^2/2E_{\gamma}(1+\mu)$, where $\kappa \in (0,1)$.

\begin{figure}
\centering
   \includegraphics[width=0.35\textwidth,clip]{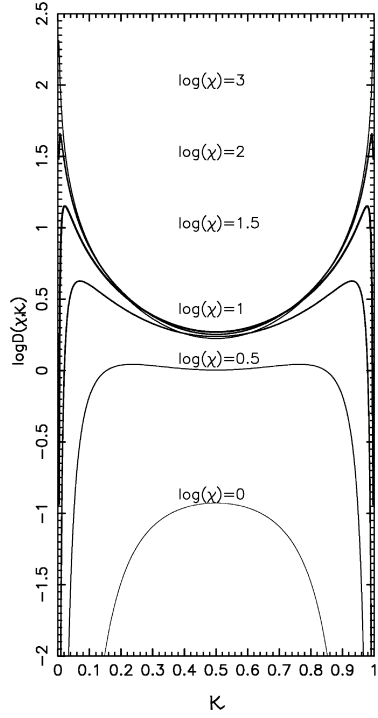}
\caption{The integral  function $D(\chi, \kappa)$ (Eq. \ref{gp27}), where $\chi = 2 kT_{s} E_{\gamma}(1+\mu)/m^2$ and $\kappa = E_e/E_{\gamma}$ are the parameters.  }
\label{fig:tabwew2}
\end{figure}

For a photon energy $E_{\gamma} \gg \epsilon$, we have $\gamma_c \gg 1$, $\beta \simeq 1$, and $\varepsilon_c \beta_c \gamma_c = E_{\gamma}/2$. After rewriting  Eq. (\ref{gp20}) we get 
\begin{equation}
I_1(E_{\gamma}, E_e, \mu) = \frac{8 \pi}{(h c)^3} \frac{2}{E_{\gamma}}
\int_{\epsilon_{min}}^{\infty} \frac{\epsilon}{(e^{\epsilon/kT_{s}} - 1)}
f(\epsilon) \, d\epsilon, \label{gp22}
\end{equation}
where, to simplify the formula, we have defined 
\begin{eqnarray}
f(\epsilon) = \frac{2 \varepsilon_c^2-m^2 + (\varepsilon_c^2-m^2) A^2(E_e,
\varepsilon_c)} {m^2B^2(E_e,\varepsilon_c) + \varepsilon_c^2
A^2(E_e,\varepsilon_c)} - \nonumber \\
\frac{2(\varepsilon_c^2- m^2)^2
A^4(E_e,\varepsilon_c)} {(m^2 B^2(E_e,\varepsilon_c)+ \varepsilon_c^2
A^2(E_e,\varepsilon_c))^2}.
\label{gp23}
\end{eqnarray}

To proceed forward, we introduce the variable $z = \epsilon /kT_{s}$, so that 
$kT_{s} dz = d
\epsilon$, and the integral is now given by the expression:
\begin{equation}
I_1(E_{\gamma}, E_e, \mu)  = \frac{8 \pi}{(h c)^3}
\frac{2(kT_{s})^2}{E_{\gamma}}
\int_{z_{min}}^{\infty} \frac{z}{e^z - 1} f(z) \, dz,
\label{gp24}
\end{equation}
where the lower limit of the integral is given by $z_{min} = \epsilon_{min}/kT_{s}$, 
\be
z_{min} = m^2/(2E_{\gamma} kT_{s} (1+\mu) \kappa(1-\kappa)) =  m^2/\chi
\kappa(1-\kappa) \ee 
and a new parameter, 
\be
\chi = 2E_{\gamma} kT_{s} (1+\mu),\ee
was introduced. The parameter $\chi$ is limited by the condition $\varepsilon_c^2 > m^2$, so then  
\be \chi > 4m^2/z_{max}
= 4kT_{s}m^2/\epsilon_{max}, \ee
where $\epsilon_{max} $ is the maximal thermal photon energy from the blackbody spectrum (we assume $\epsilon_{max} = 30kT_{s}$). 
By using the variable 
\be
\varrho = \varepsilon_c^2 = \frac{1}{2} E_{\gamma} \epsilon (1+\mu) = z\chi/4, 
\ee
the additional needed functions are given by expressions:
\begin{eqnarray}
B^2(\chi, \kappa) &=& \frac{\varrho(2\kappa-1)^2}{\varrho - m^2}, \\
A^2(\chi, \kappa) &=& 1-B^2(\chi, \kappa).
\label{gp25}
\end{eqnarray}
Function (\ref{gp23}) can now be rewritten as:
\begin{eqnarray}
f(z) = \frac{2\varrho - m^2+(\varrho-m^2) A^2(\chi, \kappa)}{m^2 B^2(\chi,
\kappa)+ \varrho A^2(\chi, \kappa)} - \nonumber \\
\frac{2(\varrho- m^2)^2 A^4(\chi,
\kappa)}{\left[ m^2 B^2(\chi, \kappa)+\varrho A^2(\chi, \kappa)\right]^2}.
\label{gp26}
\end{eqnarray}

We can further conveniently replace 
the parameter $\chi$ with a dimensionless form $\chi = 2 kT_{s}
E_{\gamma}(1+\mu)/m^2$ to get finally the full set of needed magnitudes in a useful way for integrating:
\be
I_1(E_{\gamma}, E_e, \mu) =  \frac{16 \pi (kT_{s})^2}{(hc)^3}
\frac{1}{E_{\gamma}}\, D(\chi, \kappa), 
\label{gp27}
\ee
\be
D(\chi, \kappa) =  \int_{z_{min}}^{\infty} \frac{z}{e^z - 1} f(z) \, dz, 
\label{57}
\ee
\be
f(z) = \frac{2\varrho - 1+(\varrho-1) A^2(\chi, \kappa)}{B^2(\chi, \kappa)+
\varrho A^2(\chi, \kappa)} - \frac{2(\varrho- 1)^2 A^4(\chi, \kappa)}{\left[
B^2(\chi, \kappa)+\varrho A^2(\chi, \kappa)\right]^2}, 
\ee
\be
B^2(\chi, \kappa) = \frac{\varrho(2\kappa-1)^2}{\varrho -1}, 
\ee
\be
A^2(\chi, \kappa) = 1-B^2(\chi, \kappa).
\ee
To produce the electron of the energy $\kappa$ the process is limited by the condition $z>z_{\kappa}$, where  
\be z_{\kappa} = \frac{1}{\chi \kappa (1-\kappa)}.
\ee  On the other hand, from the kinematics of the process we have limitation $z>z_{\chi}$, where
\be z_{\chi} = \frac{4}{\chi} =  \frac{2m^2}{E_{\gamma} kT_{s} (1+\mu)}.\ee 
The lower limit of the sought integral is the greater value of the two expressions quoted above. 
The integral given in Eq. (\ref{57}) can be tabulated with respect to the parameters $\chi$ and $\kappa$. The plot of the function $D(\chi, \kappa)$ (Eq. \ref{57}) is presented in Fig. (\ref{fig:tabwew2}).

The remaining internal integral in Eq. (\ref{gp17}) is given by the function $\Phi(\mu) =  \int_{-\phi_{s}}^{\phi_{s}} d\phi= 2 \phi_{s}$, where the angle $\phi_{s}$ is defined by Eq. (\ref{gp11}). With this in mind, the formula for the spectrum of pairs is given by:
\begin{equation}
\frac{dW(E_{\gamma}, x_i, \alpha,x_{\gamma})}{dE_e dx_{\gamma}} =
\frac{C_2}{E_{\gamma}^{2} } \int (1+\mu) \Phi(\mu) D(\chi, \kappa)\, d\mu,
\label{gp28}
\end{equation}
where the constant $C_2 = 3 \sigma_{T} m^2 (kT_{s})^2/(hc)^3$. The integral limits, $(\mu_1, \mu_2)$ can be calculated according to Eq. (\ref{gp13}), following the rules described in previous paragraphs.

\begin{figure*}
\centering
\includegraphics[width=0.35\textwidth,clip]{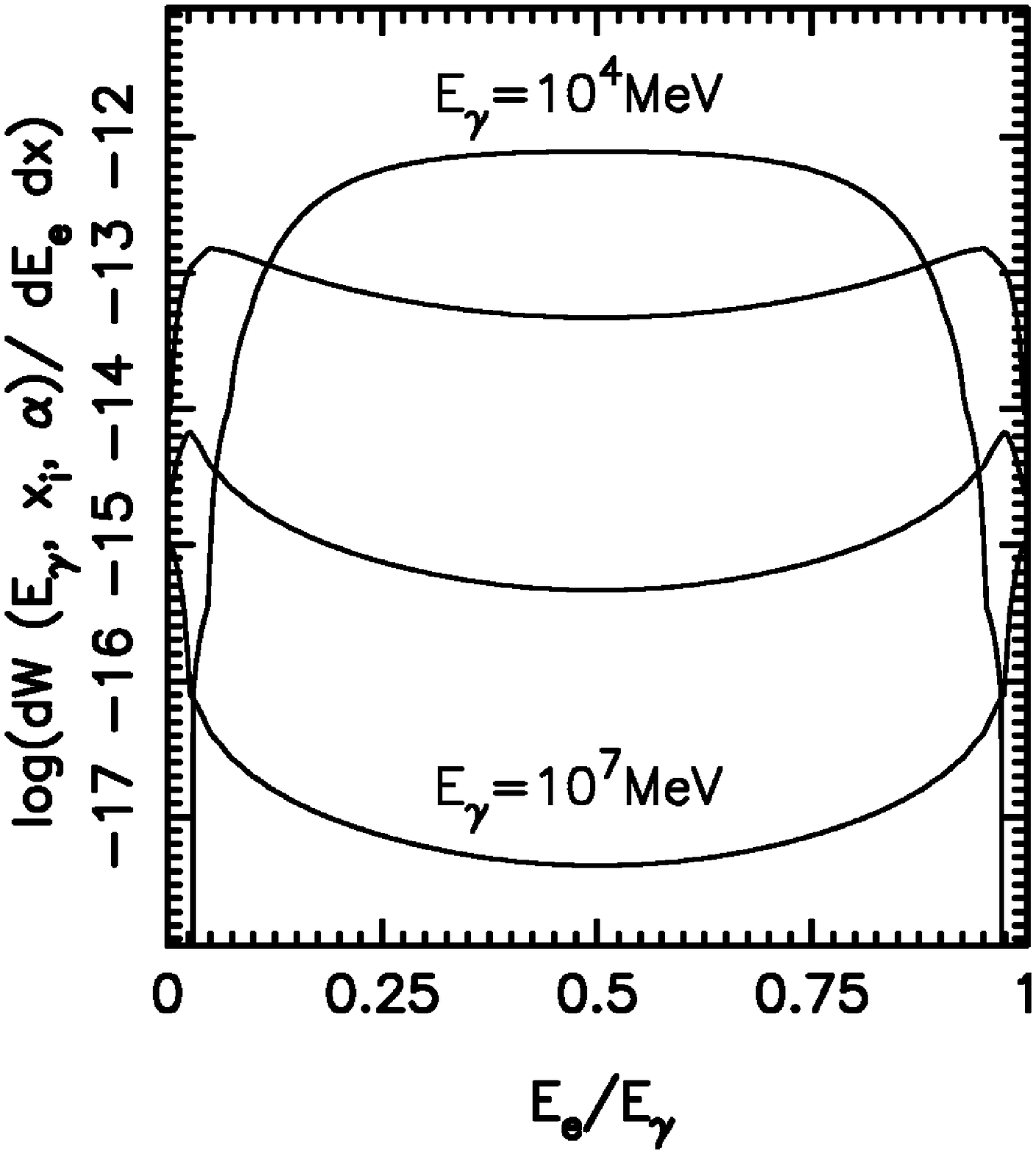}\hspace{0.2cm}
\includegraphics[width=0.35\textwidth,clip]{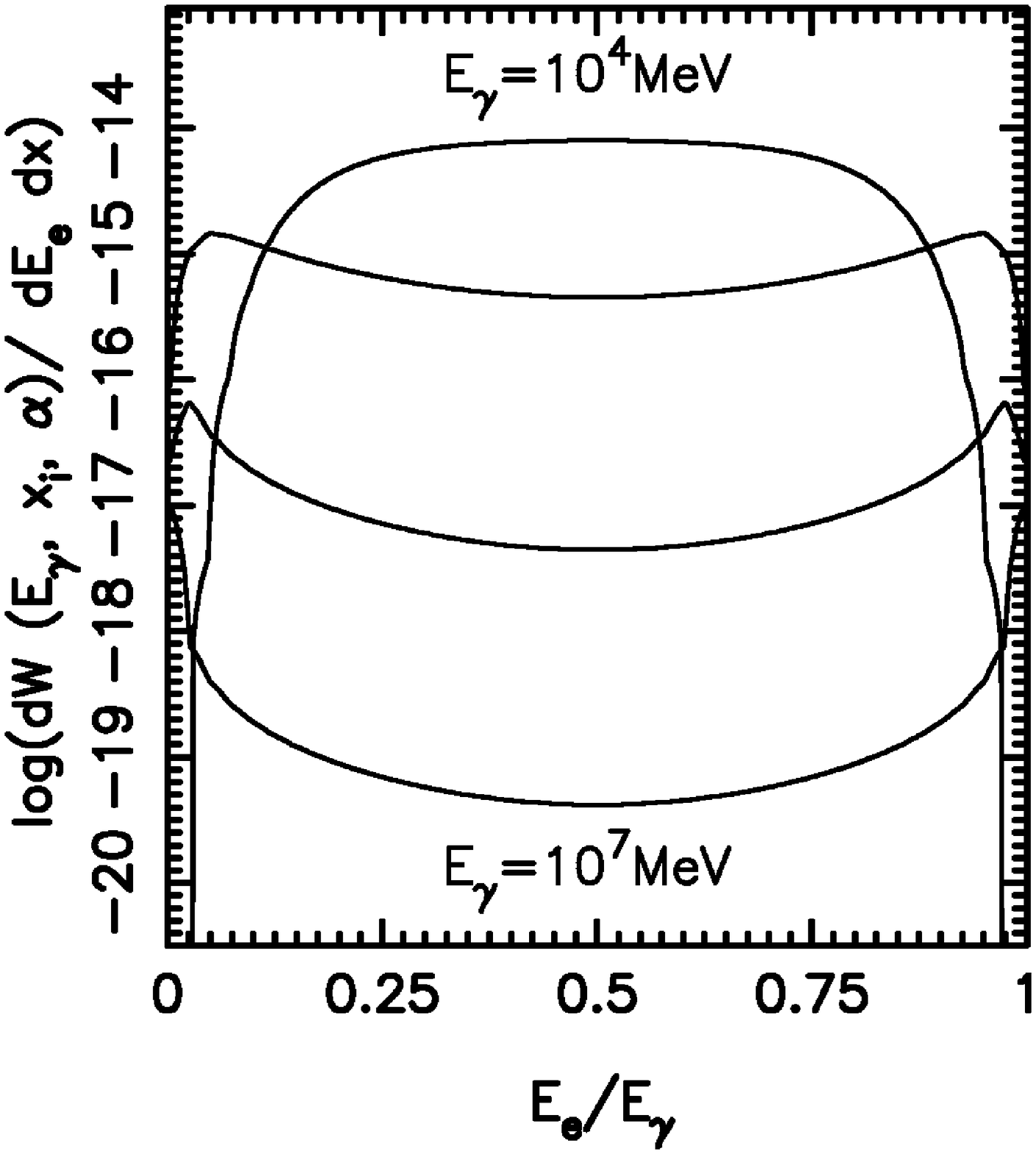}\\
\includegraphics[width=0.35\textwidth,clip]{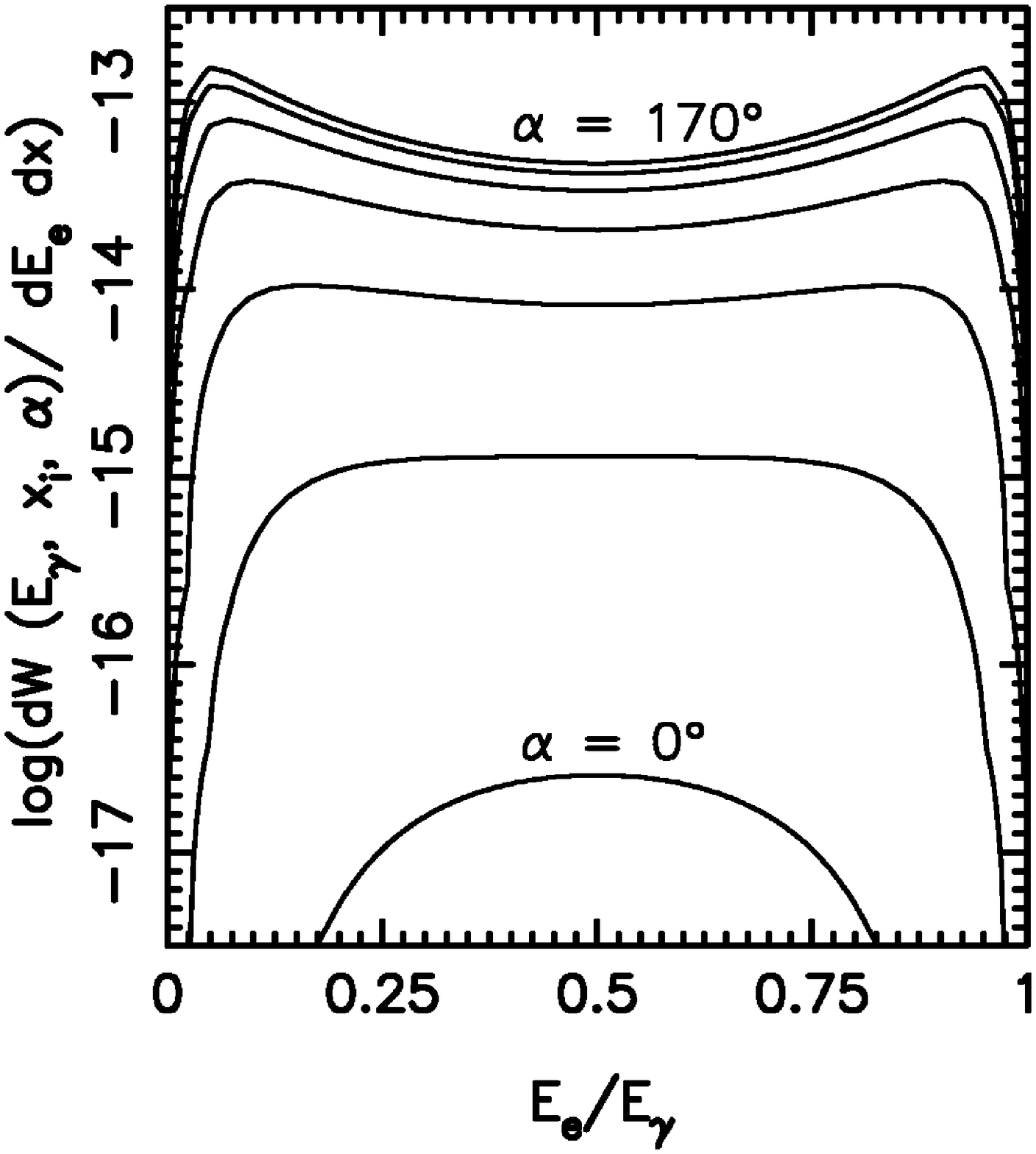}\hspace{0.2cm}
\includegraphics[width=0.35\textwidth,clip]{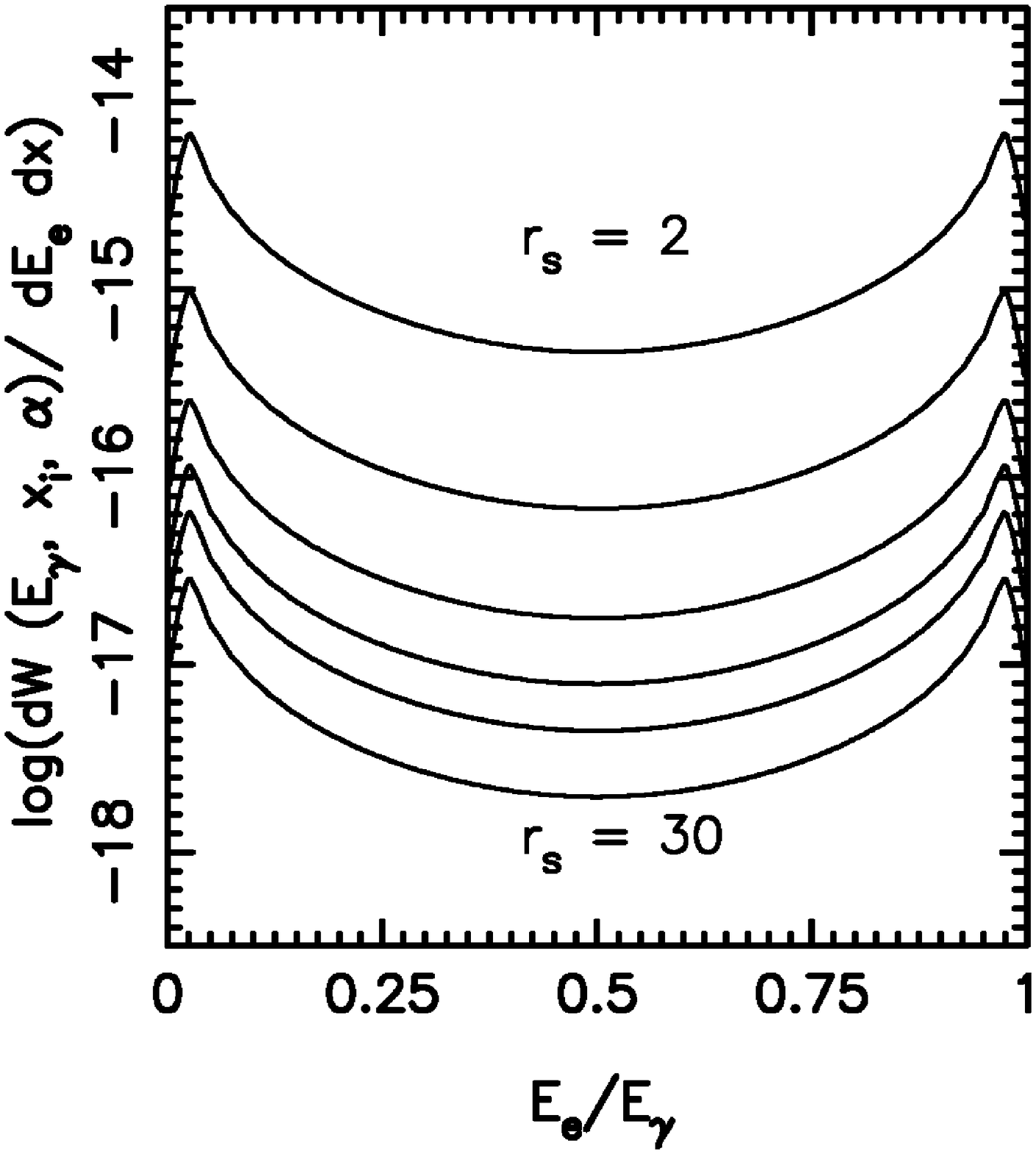}
\caption{\label{fig:wid-ee}
The differential spectra for $e^+e^-$ pairs produced in the absorption process of a photon with energy $E_{\gamma}$. The massive star has here a surface temperature $T_{s}=10^5\,\rm K$, and radius $R_{s} = 10\, R_{\odot}$. The spectra for a fixed place of photon injection, defined by $x_i = 2R_{s}$, and $\alpha = 170^o$ for different energies of photon are shown at the left in the top panels. The energies of gamma photon are $E_{\gamma} = 10^4, 10^5, 10^6, 10^7 \,\rm MeV$, from top to bottom curves. The same dependence but for $x_i = 10R_{s}$ is shown to the right in the top panels. 
The $e^+e^-$ spectra from different angles $\alpha$ at the same injection distance $x_i =2R_{s}$ and energy of photon $E_{\gamma} = 10^5\,\rm MeV$ are presented in the left, bottom panels. The spectra are calculated for $\alpha=30^o, 60^o, 90^o, 120^o, 150^o$. The dependency on the distance of the photon injection place, for $x_i = 2,5,10,15,20,30 \,
R_{s}$, and fixed $E_{\gamma} = 10^6\,\rm MeV$ and $\alpha= 170^o$ are shown in figure to the right, bottom panels. The energy of leptons are normalized to the injected photon energy.}
\end{figure*}

The $e^+e^-$ pair spectra as a function of initial parameters for the process, $x_i$, $\alpha$, (see Fig. \ref{fig:geom0}), and the energy of interacting photon, $E_{\gamma}$, are shown in Fig. (\ref{fig:wid-ee}). 
{The produced $e^+e^-$ pair energy spectra are symmetric with respect to the electron energy $E_e/E_{\gamma} = 0.5$, what is characteristic for the gamma absorption process. Figures show also that the process of pair production strongly depends on the assumed geometry of interacting particles so that the efficiency is decreasing for gamma photons propagating in the outward direction and if injected farther from the star surface. Note that the presented spectra are calculated for specific parameters of the massive star, but serve as a general example.}

\subsection*{The production of high-energy photons in Inverse Compton scattering }

\begin{figure}
  \includegraphics[width=9cm,clip]{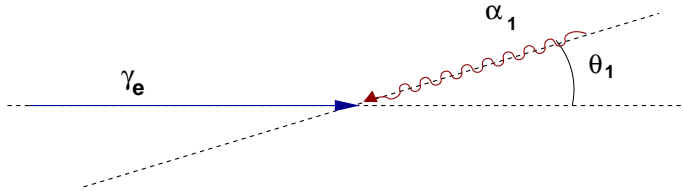}
 \caption{\label{fig:geom2a} The geometry for the Inverse Compton scattering in the observer  (LAB) frame. The Lorentz factor of an electron is $\gamma_e$, the energy of a photon before scattering is $\alpha_1 $, and angle of scattering is $\theta_1$ (the angle between the directions of propagation of the low energy photon and the relativistic electron).}
\end{figure}

\begin{figure}
  \centering
  \includegraphics[width=9cm,clip]{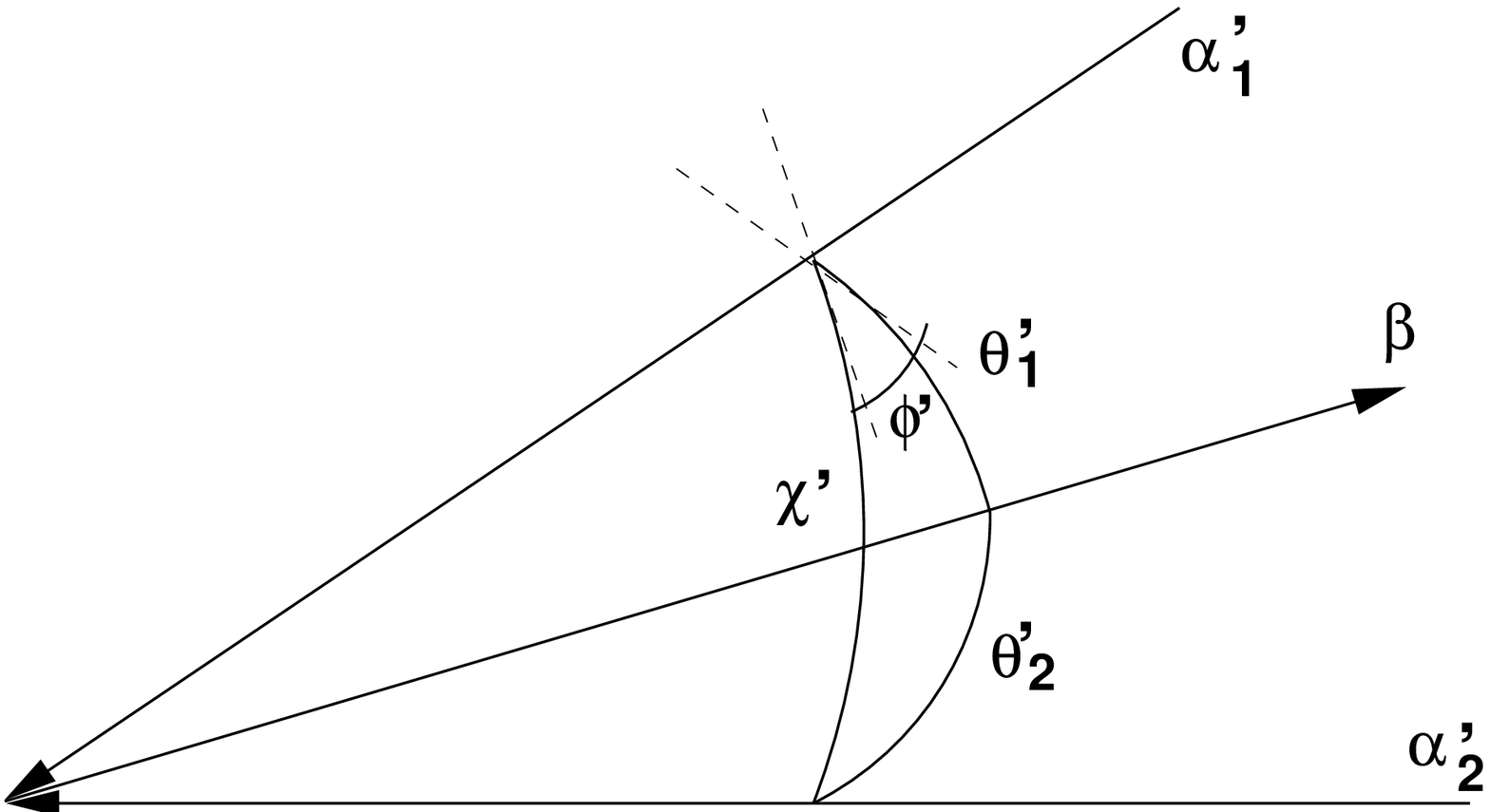}
  \caption{The kinematics of IC scattering in the center of mass system (CMS). The photon before scattering is denoted by $\alpha_1' = \epsilon /  m_e c^2 $, and after it, by $\alpha_2' = E_{\gamma} / m_e c^2 $. The propagation of the photon before and after interaction are given by angles $\theta_1'$ and $\theta_2'$, both are polar angles measured with respect to the velocity vector of the electron $\vec{\beta} = \vec{v}/c$. The scattering angles are the polar angle $\chi'$,  and the azimuthal angle $\phi'$.}
  \label{fig:geom2}
\end{figure}

The electron propagates in the radiation field of density $n(\epsilon)$, and the electron injection place is defined by the distance $x_i$ and angle $\alpha$ with respect to the massive star (see Fig. \ref{fig:geom0}). The thermal photons are again described by a blackbody energy spectrum (Eq. \ref{gp1}).

The rate of interaction of a high-energy electron of energy $E_e$, at the distance  $x_e$ from its injection place, is given by the integral:
\begin{equation}
\lambda_{ICS}^{-1} (E_e,\, x_i, \, \alpha,\, x_e) = \frac{1}{c}
\int_0^{E_{\gamma}^{max}}
\frac{dN(E_e, x_i, \alpha, x_e)}{dE_{\gamma}dt} \, dE_{\gamma}.
\label{ics1}
\end{equation}
The integration of the scattered photon spectra $dN(E_e,\, x_i,\, \alpha,\,
x_e)/dE_{\gamma}dt$ is taken over the energies of the produced photons, $E_{\gamma}$, and is limited by the maximal 
photon energy scattered $E_{\gamma}^{max}$.

We calculate the optical depth to IC scattering, at the propagation path $x_e$, integrating the rate of electron interactions in the anisotropic radiation field:
\begin{equation}
\tau(E_e,\, x_i, \, \alpha,\, x_e) = 
\int_0^{\infty} \lambda_{ICS}^{-1} (E_e,\, x_i, \, \alpha,\, x_e)\, dx_e.
\label{ics2}
\end{equation}
If the electron propagates directly towards the massive star, the path is terminated by the massive star surface. 

\subsection*{The kinematics of Inverse Compton scattering}
We adapt here the formulas from Jones (1968) (therein Eq. 1 through 10) and Blumenthal and Gould (1970) (their chapters 2.6-2.7), wherein the scattered photon spectra are calculated in the isotropic radiation case. In our work, we integrate over the angles of incoming low energy photons (the solid angle which defines the surface of the massive star) what differs from integration over whole solid angle in the isotropic radiation field.

The energy of photon before scattering in the LAB system (see Fig. \ref{fig:geom2a}) is denoted by  $\alpha_1 = \epsilon /  m_e c^2 $, while that after scattering, by $\alpha_2 = E_{\gamma} / m_e c^2 $. In the CMS (see Fig. \ref{fig:geom2}) all quantities are marked with prime variables, so that the energy of the photon before and after scattering are written as $\alpha_1'$ and $\alpha_2'$. 
The angle between the direction of the electron propagation and the incoming stellar photon is denoted as $\theta_1$ ($\theta_1'$ in the CMS), while the angle between the electron direction and the photon after scattering, is $\theta_2$ ($\theta_2'$ in the CMS). 
These are polar angles measured with respect to the electron velocity vector $\vec{\beta} = \vec{v}/c$. The angles of scattering in the electron rest frame generate a spherical triangle, so that 
\be \cos\theta_2' = \cos \theta_1' \cos \chi' + \sin \theta_1' \sin \chi' \cos \phi' 
\ee 
(see Fig. \ref{fig:geom2}). The energy of the scattered photon in the CMS is given by the formula: 
\begin{equation}
\alpha_2' = \frac{\alpha_1'}{1+\alpha_1'(1-\cos\chi')},
\label{ics3}
\end{equation}
where $\chi'$ is the photon scattering angle, defined as the azimuthal angle between the propagation direction of the photon after scattering, $\alpha_2'$, and the photon propagation direction before it, $\alpha_1'$. 

The energy of photon in the LAB system can be found from a Lorentz transformation. In a simplified case, when the photon propagate head-on with respect to the electron direction, and after interaction, it has the energy:
\begin{eqnarray}
\alpha_2 = \gamma_e \alpha_2' (1+\beta \cos(\pi - \chi'))  \approx \gamma_e
\alpha_2' (1-\cos \chi').
\label{ics4}
\end{eqnarray}
In the electron's rest frame (see Fig. \ref{fig:geom2}), for the polar angle $\theta_1' = 0$ we have $\chi' = \theta_2'$ (see Eq. \ref{ics4}), what means that the photon energy after scattering (see Eq. \ref{ics4}) does not depend on the azimuthal scattering angle $\phi'$.

We can calculate  the maximal energy of the scattered photon from relations (\ref{ics3}) and (\ref{ics4}), such that $\alpha_{max}=2\gamma_e \alpha_2'$. In the Thomson limit ($\alpha_1' \ll 1$), the maximal photon energy  depends on the electron energy and equals $\alpha_{max}^T \approx 4\gamma_e^2 \alpha_1$. In the Klein-Nishina limit, which is the highly relativistic case, this energy is proportional to the initial electron energy and it is $\alpha_{max}^{KN} \approx \gamma_e \alpha_1$.

\subsection*{The IC scattered photon spectrum in an anisotropic target field} 

\begin{figure}
\centering
\includegraphics[width=0.35\textwidth,clip]{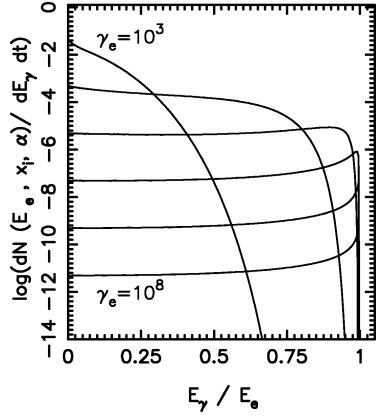}
\caption{The photon spectra (Eq. \ref{ics16}) for fixed initial parameters (the electron injection place): the distance to the massive star $x_i = 10 R_{s}$ and angle $\alpha = 180^o$ (see Fig. (\ref{fig:geom0})), as a function of electron energy $E_e$. The temperature of the star is herein taken as $T_{s}=10^5\, K $ and the radius $R_{s} = 10\, R_{\odot}$. The electron Lorentz factors shown are : $\gamma_e = 10^3, 10^4, 10^5, 10^6, 10^7, 10^8$.} 
\label{fig:wg1}
\end{figure}

In the case of mono-energetic isotropic photon background (in the LAB system) the angular spectrum of incoming low energy photons in the electron rest frame is given by the  formula (Jones 1968):
\begin{equation}
n' ( \theta_1' ) d \cos \theta_1' =
\frac{d \cos \theta_1' }{2 \gamma_e (1 - \beta \cos \theta_1' )^2}.
\label{ics7}
\end{equation}
For relativistic electrons ($\gamma_e \gg 1$), the incoming target photon polar angles are in the range 
$0\leq \theta_1' \leq \theta_{1/2}'$, where $ \theta_{1/2} \approx 1/ \gamma_e $ is the angle of a photon `cone' apex.

The photon energy spectrum before scattering in the electron rest frame is:
\begin{equation}
n'( \alpha_1' ) d \alpha_1' = \frac {\alpha_1'}{2 \gamma_e \alpha_1^2}
\Theta(\alpha_1'; \frac{\alpha_1}{2\gamma_e}, \alpha_1 2 \gamma_e) d \alpha_1',
\label{ics8}
\end{equation}
for $\gamma_e \gg 1$, where function $\Theta(x; a,b)$ is the characteristic function defined in a range $(a,b)$ and takes the values:
\[ \Theta(x; a,b) = \left\{ \begin{array}{ll}
                    1 & \mbox{$a \leq x \leq b$}, \\
                    0 & \mbox{$b < x < a$}.
                    \end{array} \right. \]
The cross section to IC scattering in the Klein-Nishina range is given by the formula:
\begin{eqnarray}
\sigma(\alpha_2', \alpha_1', y')  =  &&
\frac{r_0^2(1+y'^2)}{2[1+\alpha_1'(1-y')]^2}  \times  \nonumber \\ &&
\left\lbrace 1+ \frac{\alpha_1'^2(1-y')^2}{(1+y'^2)[1+\alpha_1'(1-y')]}
\right\rbrace \times \nonumber \\  &&
\delta(\alpha_2' - f(\alpha_1', y')), 
\label{ics9}
\end{eqnarray}
where $y' = \cos \chi' $, $r_0 = e^2/mc^2$ is the classical electron radius and the internal function is given by 
\be
f(\alpha_1', y') = \alpha_1'/[1+\alpha_1'(1-y')].
\ee
The number of scatterings per unit time $t'$ is 
$ n' c \sigma$. 
If $dN/dt = \gamma_e^{-1} dN/dt'$, then after integration over $\phi'$ one gets:
\begin{eqnarray}
\frac{d^4 N}{dt d\alpha_1' d\alpha_2' dy'} & = & \frac{\pi r_0^2 c}{2 \alpha_1^2
\gamma_e^2} \frac{ 1 + y'^2 } 
{[ 1 + \alpha_1' (1 - y') ]^2} \times \nonumber \\ && \left(
1+\frac{\alpha_1'^2(1-y')^2}{(1+y'^2)[1+\alpha_1'(1-y')]}  \right) \times
\nonumber \\ && \alpha_1' \delta(\alpha_2' - f(\alpha_1', y')) \Theta(\alpha_1';
\frac{\alpha_1}{2 \gamma_e}, \alpha_1^2 2 \gamma_e).
\label{ics10}
\end{eqnarray}
From the relation 
\be d\alpha_2' d\alpha_1' dy' = [1+\alpha_1'(1-y')]^2 d\alpha_2' dy' df
\ee 
and after integration over $f$ we get:
\begin{eqnarray}
\frac{d^3N}{dt d\alpha_2' dy'} & = & \frac{\pi r_0^2 c}{2 \alpha_1^2 \gamma_e^2}
\left[ (1+y'^2) + \frac{(\alpha_2')^2 (1-y')^2}{1-\alpha_2'(1-y')} \right] \times
\nonumber \\ && \frac{\alpha_2'}{1-\alpha_2'(1-y')} \times  \nonumber \\
&& \Theta(\frac{\alpha_2'}{1-\alpha_2'(1-y')}; \frac{\alpha_1}{2 \gamma_e}, \alpha_1^2 2
\gamma_e ).
\label{ics11}
\end{eqnarray}
Applying the Doppler-shift formula to the energy   $\alpha_2'$ and LAB frame energy  $\alpha_2$ yields to relation 
\be \alpha_2' = \alpha_2/\gamma_e(1-\beta y'). 
\ee 
Then, by means of the replacement $\rho = 1-\beta y'$:
\begin{eqnarray}
\frac{d^3N}{dt d\alpha_2 d\rho} &=& \frac{\pi r_0^2 c \alpha_2}{2 \alpha_1^2
\gamma_e^4 (1-\alpha_2/\gamma_e)} \times \nonumber \\ && \left[ \rho^2 -2\rho + 2 +
\frac{(\alpha_2/\gamma_e)^2}{1-\alpha_2/\gamma_e} \right] \frac{\Theta(\rho; \rho_1,
\rho_2)}{\rho^2}, \label{ics12}
\end{eqnarray}
which is obtained under the assumption that $(1-y') \approx \rho$. The limiting values in the argument of the function $\Theta(\rho; \rho_1, \rho_2)$ are:
\begin{eqnarray}
\rho_1 &=& \frac{\alpha_2}{2 \alpha_1 \gamma_e^2 (1- \alpha_2 / \gamma_e)}, \\
\rho_2 &=& \frac{2 \alpha_2}{ \alpha_1 (1- \alpha_2 / \gamma_e)},
\label{ics13}
\end{eqnarray}
where the parameter $\rho$ is limited by conditions $\rho_l  \leq \rho \leq \rho_p$, with
\begin{eqnarray}
\rho_l &=& \frac{1}{2\gamma_e^2}, \nonumber \\
\rho_p &=& 2
\label{rholp}
\end{eqnarray}

To calculate the spectra of scattered photons from the spectrum (Eq. \ref{ics12})
as a function of our parameters, $dN/dt dE_{\gamma}(E_e, x_i, \alpha, x_e)$, we replace $E_{\gamma} = \alpha_2 m_e c^2$ and $\epsilon = \alpha_1 m_e c^2$ and integrate over the background photon spectra $n(\epsilon)$ and parameter $\rho$,
\begin{eqnarray}
\frac{dN}{dt dE_{\gamma}} (E_e, x_i, \alpha, x_e) & = & \frac{r_0^2 c
E_{\gamma}}{2 \gamma_e^4 (mc^2)^3(1-E_{\gamma}/E_e)}
\int \frac{n(\epsilon_1)}{\epsilon_1^2} d\epsilon_1 \times \nonumber \\
&& \int \Phi(\mu(\rho)) \left[ \rho^2 -2\rho + 2 +
\frac{(E_{\gamma}/E_e)^2}{1-E_{\gamma}/E_e} \right] \times \nonumber \\ &&
\frac{\Theta(\rho; \rho_1,
\rho_2)}{\rho^2} \, d\rho,
\label{ics14}
\end{eqnarray}
where 
\be \mu = E_{\gamma} mc^2 / (\epsilon \gamma_e(E_e-E_{\gamma})\rho) -1,\ee
and the function $\Phi(\mu)$ has been already defined in previous paragraphs as $\Phi(\mu) =  \int_{-\phi_{s}}^{\phi_{s}} d\phi= 2 \phi_{s}$.

\begin{figure*}
\centering
\includegraphics[width=0.35\textwidth,clip]{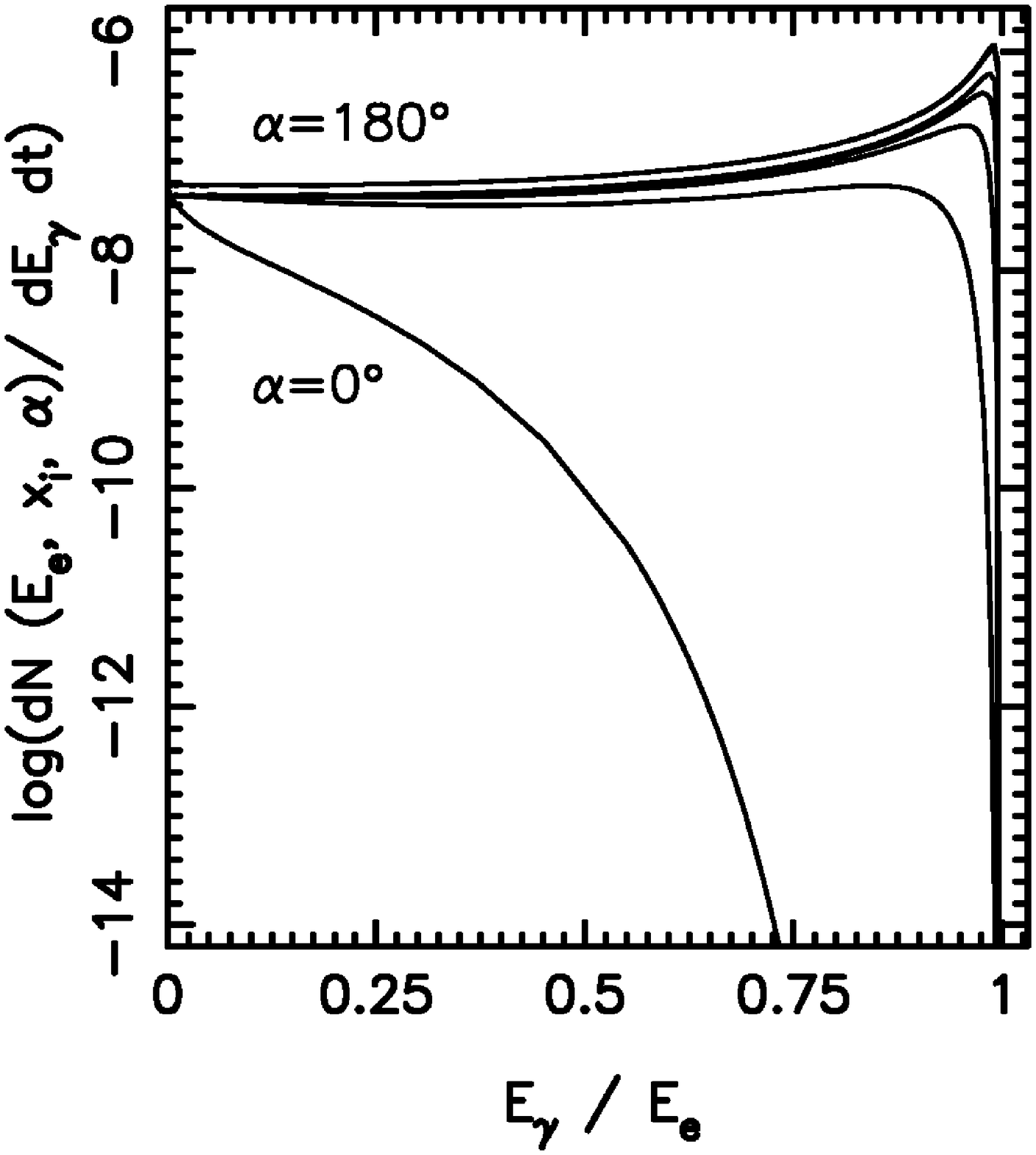} \hspace{0.2cm}
\includegraphics[width=0.35\textwidth,clip]{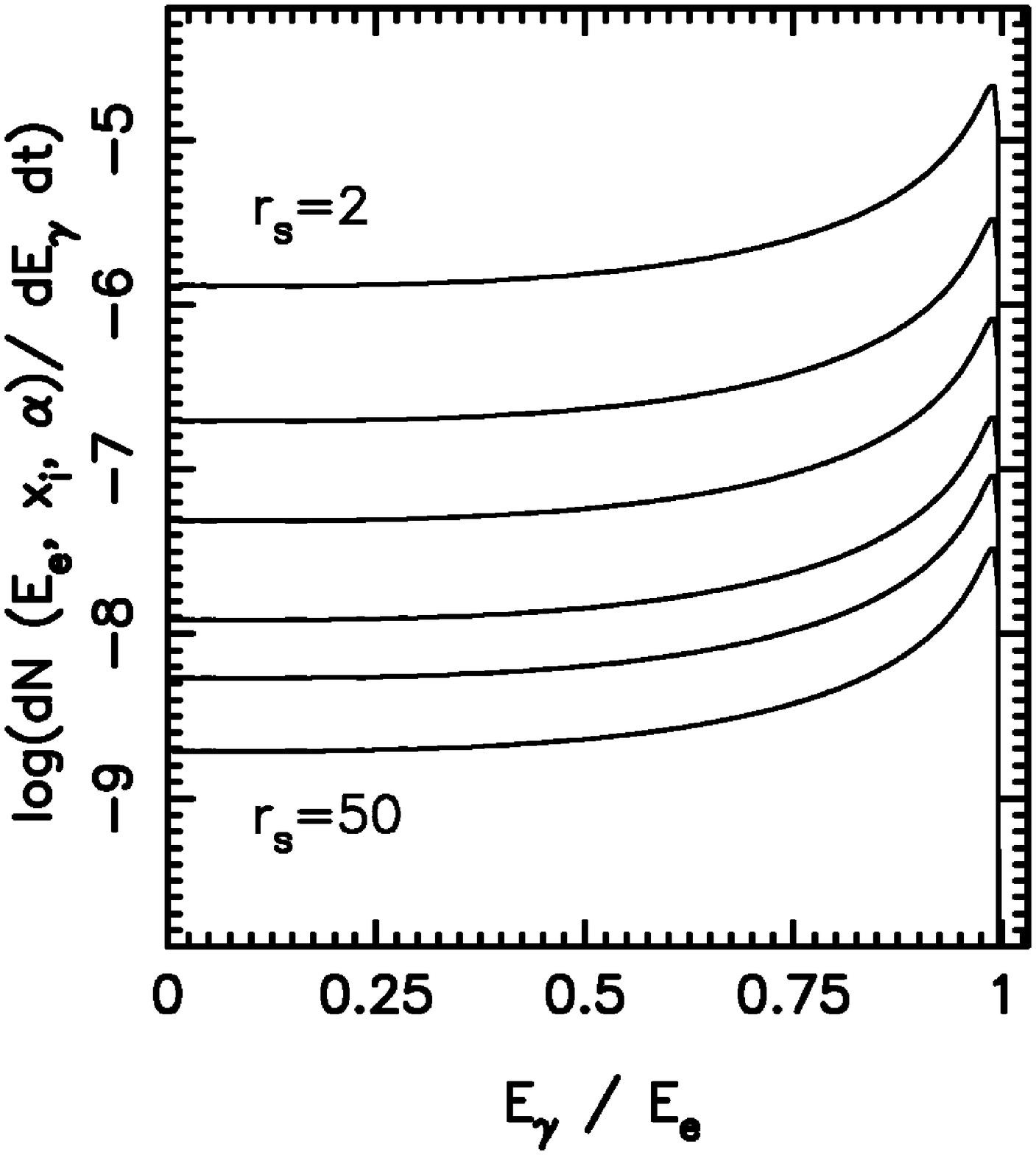}
\caption{Left: 
The photon spectra (Eq. (\ref{row:num86})) for a 
specific electron energy $\gamma_e = 10^6$ and fixed distance to the massive star $x_i = 10 R_{s}$ ($x_e=0$), calculated for different electron propagation angles: $\alpha = 0^o, 30^o, 60^o, 90^o, 120^o, 150^o, 180^o$. Right: The photon spectra depending on the electron energy $E_e$ and the initial injection parameters. The spectra are calculated for  $\gamma_e = 10^6$, the angle of propagation $\alpha = 180^o$ and subsequent distances to the massive star: $x_i =$
5, 10, 20, 30, 50\, $R_{s}$ ($r_s = x_i/R_{s}, x_e = 0 $) (right panel). The temperature of the star is herein taken as $T_{s}=10^5\, K $ and its radius is  $R_{s} = 10\, R_{\odot}$. }
\label{fig:wg2}
\end{figure*}

For convenience, let us rewrite the internal integral on $\rho$ as follows: 
\begin{eqnarray}
G(E_{\gamma}, \epsilon, \mu ) = &&
\int_{\rho_{min}}^{\rho_{max}}
\Phi(\mu)
\left[ \rho^2 -2\rho + 2 + \frac{(E_{\gamma}/E_e)^2}{1-E_{\gamma}/E_e} \right] \times
\nonumber \\ &&
\frac{\Theta(\rho; \rho_1, \rho_2)}{\rho^2} \, d\rho.
\label{ics14a}
\end{eqnarray}
If the range of parameter for $\rho$ from Eq. (\ref{ics14a}) is consistent with the range $(\rho_l, \rho_p)$ (Eq. \ref{rholp}), then the integration limits $(\rho_{min}, \rho_{max})$ correspond to the common part: $(\rho_1, \rho_2) \cap (\rho_l, \rho_p)$, and the integral over $\rho$ is given by function:
\begin{eqnarray}
G(E_{\gamma}, \epsilon, \mu ) = 2 \pi
\left[ \rho - 2 \log \rho -
\frac{2}{\rho} - \frac{(E_{\gamma}/E_e)^2}{(1-E_{\gamma}/E_e) \rho}
\right]_{\rho_{min}}^{\rho_{max}}.
\label{ics14b}
\end{eqnarray}
From the Doppler formula and transformations between the systems we get, 
\be \rho =
E_{\gamma}/\gamma_e^2 \epsilon (1+\beta \cos \theta_2)(1-E_{\gamma}/E_e), \ee
so that the integration limit over $\rho$, for angles $\theta_1^{s}$ and $\theta_2^{s}$
(see Fig. \ref{fig:geom1-1b}) are given by the formulas:
\begin{eqnarray}
\rho(\theta_1) & = &
\frac{E_{\gamma}}{\epsilon \gamma_e (1+\beta \cos
\theta_1^{s})(1-E_{\gamma}/E_e)},
\label{ics15a}\\
\rho(\theta_2) & = &
\frac{E_{\gamma}}{\epsilon \gamma_e (1+\beta \cos
\theta_2^{s})(1-E_{\gamma}/E_e)},
\label{ics15}
\end{eqnarray}
where the angles $\theta_1^{s},\, \theta_2^{s}$ are the angles which defined the directions of low energy photons coming from the star. 
In addition, the parameter $\rho$ is limited by kinematic condition for the scattering to happen, \begin{equation} \cos \theta_{kin} > 2m_e^2/E_{\gamma}\epsilon -1 . \end{equation} The integration in respect to $\rho$ can be separated to get the sum of integrals, as for $\rho > \rho(\theta_{kin})$ the internal function $\Phi(\mu)=2 \pi$, and the function $G(E_{\gamma}, \epsilon, \mu )$ is given by (\ref{ics14b}).
If the function $\Phi(\mu)=2 \pi$ we can integrate the function $G(E_{\gamma}, \epsilon, \mu )$ (Eq. \ref{ics14a}). Otherwise, if 
$\Phi(\mu)=2 \phi_{s}$ we can rewrite the function $G(E_{\gamma}, \epsilon, \mu )$ in a compact way as follow. Thus, these cases are summarized by
\begin{eqnarray}
G(E_{\gamma}, \epsilon, \mu ) = &&
\int_{\rho_{min}}^{\rho_{max}}
2 \phi_{s} 
G_{\rho}(E_{\gamma}, \epsilon, \mu )\, d\rho, \\
G(E_{\gamma}, \epsilon, \mu ) = && 
2 \pi 
\left[  G_{int}(E_{\gamma}, \epsilon, \mu )  \right]_{\rho_{min}}^{\rho_{max}}.
\label{ics14sim}
\end{eqnarray}
where $G_{\rho}(E_{\gamma}, \epsilon, \mu ) = \left[ \rho^2 -2\rho + 2 +(E_{\gamma}/E_e)^2/(1-E_{\gamma}/E_e) \right] \times
\Theta(\rho; \rho_1, \rho_2)/\rho^2$ and $G_{int}(E_{\gamma}, \epsilon, \mu ) =  \rho - 2 \log \rho - {2}/{\rho} - {(E_{\gamma}/E_e)^2}/({(1-E_{\gamma}/E_e) \rho})$.
The limits of integration depend on combinations of $(\rho(\theta_1), \rho(\theta_2))$ and  $(\rho_l, \rho_p) $, and respecting the constraints put by $\rho(\theta_{kin})$. 
The final integration $G(E_{\gamma}, \epsilon, \mu )$ is given by the formulas:
\begin{enumerate}

\item for ``outwards'' propagation :
\begin{itemize}
\item if $(\rho_{max} < \rho_1)$: $G(E_{\gamma}, \epsilon, \mu ) = 0$,\\

\item if $(\rho_{max} > \rho_1)$ and $(\rho_{min} < \rho_1)$ :

\begin{description}
 \item and  $(\rho_{max} < \rho_2)$ :\\
 $G(E_{\gamma}, \epsilon, \mu ) = \int_{\rho_1}^{\rho_{max}} 2 \phi_{s} G_{\rho}(E_{\gamma}, \epsilon, \mu )\, d\rho $,
\item or $(\rho_{max} > \rho_2)$ :\\
 $G(E_{\gamma}, \epsilon, \mu ) = \int_{\rho_1}^{\rho_{2}} 2 \phi_{s} G_{\rho}(E_{\gamma}, \epsilon, \mu )\, d\rho + 2 \pi 
\left[  G_{int}(E_{\gamma}, \epsilon, \mu )  \right]_{\rho_{2}}^{\rho_{max}}$,
\end{description}

\item if $(\rho_{max} > \rho_1)$ and  $(\rho_{min} > \rho_1)$:

\begin{description}
 \item if $(\rho_{min} < \rho_2)$ and  $(\rho_{max} <  \rho_2)$:\\
 $G(E_{\gamma}, \epsilon, \mu ) = \int_{\rho_{min}}^{\rho_{max}} 2 \phi_{s} G_{\rho}(E_{\gamma}, \epsilon, \mu )\, d\rho $,
\item if $(\rho_{min} < \rho_2)$ and  $(\rho_{max} > \rho_2)$ :\\
 $G(E_{\gamma}, \epsilon, \mu ) = \int_{\rho_{min}}^{\rho_{2}} 2 \phi_{s} G_{\rho}(E_{\gamma}, \epsilon, \mu )\, d\rho + 2 \pi 
\left[  G_{int}(E_{\gamma}, \epsilon, \mu )  \right]_{\rho_{2}}^{\rho_{max}}$,
\item if $(\rho_{min} > \rho_2)$:\\
 $G(E_{\gamma}, \epsilon, \mu ) = 2 \pi \left[  G_{int}(E_{\gamma}, \epsilon, \mu )  \right]_{\rho_{2}}^{\rho_{max}}$.
\end{description}

\end{itemize}

\item for the ``middle" range of propagation:
\begin{itemize}
\item if $(\rho_{max} < \rho_1)$ or $(\rho_{min} > \rho_2)$:\\
$G(E_{\gamma}, \epsilon, \mu ) = 0$,
\item if $(\rho_{max} > \rho_1)$ and $(\rho_{max} < \rho_2)$:\\
$G(E_{\gamma}, \epsilon, \mu ) = \int_{MAX(\rho_{min}, \rho_1)}^{\rho_{max}} 2 \phi_{s} G_{\rho}(E_{\gamma}, \epsilon, \mu )\, d\rho $,
\item if $(\rho_{max} > \rho_1)$ and $(\rho_{max} > \rho_2)$:\\
$G(E_{\gamma}, \epsilon, \mu ) = \int_{MAX(\rho_{min}, \rho_1)}^{\rho_{2}} 2 \phi_{s} G_{\rho}(E_{\gamma}, \epsilon, \mu )\, d\rho $.
\end{itemize}
where $MAX(a,b)$ gives the larger number from the brackets.

\item for ``inward" directions:

\begin{itemize}
\item if $(\rho_{min} > \rho_2)$ : $G(E_{\gamma}, \epsilon, \mu ) = 0$, 

\item if $(\rho_{min} < \rho_2)$ and $(\rho_{min} > \rho_1)$:

\begin{description}
 \item if $(\rho_{max} > \rho_2)$ :\\
 $G(E_{\gamma}, \epsilon, \mu ) = \int_{\rho_{min}}^{\rho_{2}} 2 \phi_{s} G_{\rho}(E_{\gamma}, \epsilon, \mu )\, d\rho $,
\item if $(\rho_{max} < \rho_2)$ :\\
 $G(E_{\gamma}, \epsilon, \mu ) = \int_{\rho_{min}}^{\rho_{max}} 2 \phi_{s} G_{\rho}(E_{\gamma}, \epsilon, \mu )\, d\rho $,
\end{description}

\item if $(\rho_{min} < \rho_2)$ and  $(\rho_{min} < \rho_1)$:

\begin{description}
 \item if $(\rho_{max} > \rho_2)$ :\\
 $G(E_{\gamma}, \epsilon, \mu ) = \int_{\rho_1}^{\rho_{2}} 2 \phi_{s} G_{\rho}(E_{\gamma}, \epsilon, \mu )\, d\rho + 2 \pi 
\left[  G_{int}(E_{\gamma}, \epsilon, \mu )  \right]_{\rho_{min}}^{\rho_{1}}$,
\item if $(\rho_{max} < \rho_2)$ and $(\rho_{max} > \rho_1)$ :\\
 $G(E_{\gamma}, \epsilon, \mu ) = \int_{\rho_1}^{\rho_{max}} 2 \phi _{s} G_{\rho}(E_{\gamma}, \epsilon, \mu )\, d\rho + 2 \pi 
\left[  G_{int}(E_{\gamma}, \epsilon, \mu )  \right]_{\rho_{min}}^{\rho_{1}}$,
\item if $(\rho_{max} < \rho_2)$ and $(\rho_{max} < \rho_1)$ :\\
$G(E_{\gamma}, \epsilon, \mu ) = 2 \pi \left[  G_{int}(E_{\gamma}, \epsilon, \mu )  \right]_{\rho_{min}}^{\rho_{max}}$.
\end{description}
 
\end{itemize}
\end{enumerate}

Finally the rate of electron scattering on its propagation way to IC process is given by expression:
\begin{eqnarray}
\lambda_{ICS}^{-1} (E_e,\, x_i,\, \alpha,\, x_e) & = &
\frac{1}{c} \int_0^{E_{\gamma}^{max}} \frac{dN }{dE_{\gamma}dt} (E_e,\, x_i,\,
\alpha, \, x_e) \, dE_{\gamma}, 
\label{ics16}
\end{eqnarray}
where,
\begin{eqnarray}
\frac{dN}{dt dE_{\gamma}} (E_e,\, x_i,\, \alpha, \, x_e)  =  && C \,
\frac{1}{\gamma_e^4} \frac{E_{\gamma}}{ (1 - E_{\gamma}/E_e)} 
\times \nonumber \\ &&
\int_{ \epsilon_{min}}^{ \epsilon_{max}} \frac{d\epsilon }{ e^{ \epsilon /
kT_{s}} - 1 } \, G(E_{\gamma},\, \epsilon,\, \mu ),
\label{row:num86}
\end{eqnarray}
with the constant $C = r_0^2 c /2 (mc^2)^3$. 
The function $G(E_{\gamma}, \epsilon, \mu )$ is described by Eqs. (\ref{ics14a}) and (\ref{ics14b}), which have to be chosen with respect to the value of the function $\Phi(\mu)$ and the limits to the parameter $\rho$, as explained. The lower limit for the integral in Eq. (\ref{row:num86}) corresponds to the lowest energy of the scattered photon 
\be \epsilon_{min} = E_{\gamma}/4\gamma_e^2,\ee while the upper one in the integral of Eq. (\ref{ics16}) is the maximal scattered photon energy 
\be E_{\gamma}^{max} = 4 \gamma_e^2 \epsilon . \ee 
The photon spectra calculated from Eq. (\ref{row:num86}) are presented in Figs. (\ref{fig:wg1}) and (\ref{fig:wg2}). They are calculated with  respect to the electron energy $E_e$ and its initial parameters of injection close to the massive star (see Fig. \ref{fig:geom0}). 
{The photon energy spectra reflect the features of the cross section for IC scattering, with characteristic peaks at the high energy range (for $E_{\gamma}/E_e \sim 1$) -- Klein-Nishina regime if the energy of the incoming electron is relativistic. These plots shows also the dependence on the propagation angle (with the highest production rate for direction toward the massive star) and place of the electron injection. }

\end{document}